\documentclass[12pt,preprint]{aastex}
\usepackage{color}
\usepackage[below]{placeins,epsfig}
\usepackage{natbib}

\def\gtorder{\mathrel{\raise.3ex\hbox{$>$}\mkern-14mu
    \lower0.6ex\hbox{$\sim$}}}
\def\ltorder{\mathrel{\raise.3ex\hbox{$<$}\mkern-14mu 
    \lower0.6ex\hbox{$\sim$}}}

\shorttitle{Cosmological Evolution of Galaxies}
\shortauthors{Isaac Shlosman}

\begin{document}

\title{Cosmological Evolution of Galaxies}

\author{Isaac Shlosman\\  
        Department of Physics \& Astronomy, University of Kentucky\\
        Lexington, KY 40506-0055, USA\\
        shlosman@pa.uky.edu}

%

\begin{abstract}
{I review the subject of the cosmological evolution of galaxies, including
different aspects of growth in disk galaxies, by focussing on the angular
momentum problem, mergers, and their by-products. I discuss the alternative to
merger-driven growth -- cold accretion and related issues. In the follow-up, I
review possible feedback mechanisms and their role in galaxy evolution. Special
attention is paid to high-redshift galaxies and their properties. In the next
step, I discuss a number of processes, gas- and stellar-dynamical, within the
central kiloparsec of disk galaxies, and their effect on the larger spatial
scales, as well as on the formation and fuelling of the seed black holes in
galactic centres at high redshifts.}
\end{abstract}

%
%

\section{Introduction: the paradigm of galaxy evolution}
\label{sec:intro}

The aim of these lectures is to review the main aspects of secular galaxy
evolution, i.e., galaxy evolution on cosmological scales. Two issues stand out
in this attempt. First, the subject is immense and cannot be covered within the
scope of this chapter. And second, one cannot ignore the fact that galaxy
evolution is part of the overall evolution in the Universe -- from the largest
spatial scales ruled by dark matter (DM) to the smallest ones taken over by
dissipative baryons that can form stars and grow supermassive black holes (SMBHs). In
other words, the process of galaxy formation can be influenced strongly by a
huge range of spatial scales. Hence, our attempt to discuss secular galaxy
evolution will be rather modest in depth, only highlighting those issues which
appear to lie at the forefront of current research. Lastly, we shall focus on
disk galaxy evolution and only briefly mention elliptical galaxies.

Galactic morphology is largely a reflection of underlying dynamical and secular
processes on relevant scales. That morphological evolution does indeed take
place has been established fairly well, e.g., a recent quantitative analysis
comparing galaxy populations at redshifts $z$\,$\sim$\,0.6 and $z$\,=\,0
(Delgado-Serrano {\it et al.} 2010). While the fraction of ellipticals has
barely changed over the last $\sim$5\,Gyr, the fraction of peculiar galaxies
grew in favour of spirals by a factor of $\sim$2--3. So peculiar morphology
increases dramatically with $z$ (e.g., Brinchmann \& Ellis 2000).

As the galaxy population does indeed exhibit evolutionary trends, the question
is what drives this process and how can we analyse and quantify it. Overall, our
goal lies in explaining the origin of the contemporary Hubble sequence and in
describing changes in this sequence over $z$. In this context, two alternative
views exist. Firstly, the Hubble fork is determined by the initial conditions,
i.e., by Nature. That means, for example, that the massive galaxies form in the
highest overdensities, which themselves resulted from initial conditions. 
Alternatively, it is the environment, i.e., Nurture, not Nature, that determines
the galaxy properties. Within this framework, evolution is driven solely by
interactions, e.g., between galaxies, between galaxies and the intergalactic
medium (IGM), etc. 

In the past couple of decades, the issue of structure formation in the Universe,
in terms of the dichotomy of top-to-bottom versus bottom-up scenarios, has been
resolved in favour of the latter, and of the cold dark matter (CDM) paradigm.
Unfortunately, even within the bottom-up framework, it remains unclear when and
where the baryons matter. Clearly, on large scales the baryons follow the DM.
But where and when do the baryons run {\it amok}{\footnote{From the Malay
meaning `mad with uncontrollable rage'.}? Inside the DM haloes? In the cold
filaments? As the baryons collapse into the haloes, where do stars form -- in
disks or spheroids? Is gas fragmentation encouraged or suppressed during this
infall? To what degree is the angular momentum conserved during collapse?

These questions open a Pandora's box of dissipative baryon dynamics partially
decoupled from the DM. They underline pressing problems of structure formation
on galactic and subgalactic scales. Most importantly, they emphasise the old/new
dichotomy of what is primarily responsible for disk evolution: internal or
external factors. Keeping this in mind, it is possible to construct a follow-up
list of outstanding problems according to anyone's taste. Why do disks form
inside triaxial haloes? (The haloes appear universally triaxial in numerical
simulations, e.g., review by Shlosman 2008.) What is the prevailing morphology
of the early galaxies: disk, elliptical, or some other unspecified morphology?
Can archaeology help to uncover the details of disk formation and evolution?
Important issues here are: can disks survive the epoch of major mergers? and do
geometrically thick disks come from mergers or from supernova (SN) feedback?

A separate set of problems is related to disk-halo dynamical and secular
interactions, which can have profound effects on both components -- a kind of a
dynamical feedback. To what extent does the disk evolution at high $z$ differ
from that at low $z$? What is the origin of bulges and galactic bars? And what
does this tell us about internally versus externally-driven disk evolution?

Beyond star formation, what is the role of baryons in disk evolution? What are
the dynamical corollaries of gas presence, e.g., in maintaining the disk spiral
structure? More specifically, does the gas (and the baryon fraction) vary
systematically with the halo mass, $M_{\rm h}$? When do baryons form central
SMBHs? Does feedback from stellar and SMBH evolution
encourage or suppress further star formation?  

Finally, are disk galaxies doomed, in the sense that ultimately they will fall
into high-density regions, and what does that tell us about the
morphology-density relation at high $z$? When did the current morphology-density
relation form, and what regulates star formation in disks, stellar mass loss or cold
infall? 

Much of the theoretical progress understanding the drivers of galaxy evolution
is due to numerical simulations of collisionless and dissipative processes in
the Universe. Over the last few decades, astronomy has acquired precious support
from an experiment, albeit a virtual one. Because the dominant processes are so
nonlinear, the synthesis of theoretical, observational and experimental
components has contributed much of our current understanding of structure
formation in the Universe on all spatial scales. If our goal is to be able to
`produce' realistic galaxies that can be directly compared with observations, we
are well underway.

The challenges in understanding galaxy formation and evolution are amplified by
the unknown physics of the dominant processes (e.g., star formation, mechanical
and radiative physics, turbulence), supplemented by often counter-intuitive
nonlinear dynamics and by insufficient observational constraints. The numerical
approach also suffers from the large dynamic range to be addressed by
simulations -- from $\sim$10--100\,Mpc down to $\sim$AU. Gravity is the source
of this difficulty.

The cornerstone of the current galaxy formation paradigm was established and
refined in 1970s--1990s. It has been very successful in predicting and
explaining disk galaxy properties. A two-stage process has been suggested, based
on hierarchical clustering and DM halo formation in the first stage, and gas
cooling and collapse into pre-existing potential wells in the second one (e.g.,
White \& Rees 1978; Fall \& Efstathiou 1980; Mo {\it et al.} 1998). Prior to
gravitational collapse, the DM haloes acquire angular momentum ($J$) via
gravitational torques, while baryons follow the DM and have the same specific
angular momentum, $j = J/M$. In the next step, while the DM `warms' up during
virialisation, the baryons can cool down and continue the collapse, with $j$,
and hence its distribution, roughly constant. Finally, the low-$j$ baryons accrete
onto the inner regions and cause central starbursts, resulting in the formation
of galactic bulges in disk galaxies and SMBHs in their centres. Meanwhile
high-$j$ baryons form galactic disks with self-regulated star formation. In this
framework, $j$ determines the disk size, and the disk surface density fixes the
timescale for star formation. 

Another milestone has been passed with the understanding that pure DM structure
formation leads to a universal density profile in virialised objects (e.g.,
Navarro {\it et al.} 1996, hereafter NFW). This density profile has been
approximated by a power law with a slope of $-2$ (in log\,$\rho$--log\,$R$) at
some characteristic radius $R_{\rm s}$. It becomes shallower and tends to a
slope of $-1$ toward the centre, and steepens to a slope of $-3$ at large radii.
Over a range of radii, the NFW density profile leads to a flat circular velocity
curve. The dissipative baryon influx into the halo is accompanied by a
substantial adiabatic contraction of the non-dissipative DM in the central
region, which is being dragged in, modifying the mass distribution there (e.g.,
Blumenthal {\it et al.} 1986). The adiabatic contraction of the DM halo has been
invoked in order to explain the `conspiracy' of so-called maximum disks (e.g.,
Burstein \& Rubin 1985). However, additional processes that may have been
omitted in the original estimate of the adiabatic contraction can complicate
this picture quite substantially (e.g., Primack 2009).

This galaxy formation paradigm has received substantial support from both
observations and high-resolution numerical simulations, although a long list of
caveats exists. The theoretical background for DM mass and angular momentum
distributions is still unclear and remains at the forefront of astrophysical
research. 

In this conjecture, the value of the angular momentum, its distribution and
conservation, emerge as one of the dominant parameters, if not the dominant one
determining galaxy evolution, along with the halo mass. To understand the
caveats associated with $J$, we define the dimensionless angular momentum
parameter which characterises the DM haloes (e.g., Bullock {\it et al.} 2001):

\begin{equation}
\lambda' = \frac{J}{\sqrt{2}M_{\rm h}R_{\rm h}v_{\phi}} \sim 0.03 - 0.05 ,
\label{eq:lambda}
\end{equation}
where $M_{\rm h}$ and $R_{\rm h}$ are the halo's virial mass and radius, respectively, 
and $v_{\phi}$ is its circular velocity. Numerical simulations point to the 
universality of the $\lambda'$ log-normal distribution

\begin{equation}
P(\lambda')d\lambda' = \frac{1}{\sqrt{2\pi\sigma^2}} {\rm exp} \biggl[-
     \frac{{\rm ln}^2\left(\lambda'/\bar{\lambda'}\right)}{2\sigma^2}\biggr] 
        \frac{d\lambda'}{\lambda'},
\label{eq:pdf}
\end{equation}
where $\bar{\lambda'} = 0.035\pm 0.005$ is the best-fit value, and
$\sigma(\lambda')\sim 0.5$ is the width of the log-normal distribution.
Equation\,\ref{eq:lambda} is a slightly modified version of the original spin
parameter $\lambda$ introduced by Peebles (1969), with
\hbox{$\lambda$\,$\sim$\,0.01--0.1} being the range found for DM haloes, and its median
$\lambda$\,$\sim$\,0.035. The simple relation between $\lambda$ and $\lambda'$
is given by $\lambda' = \lambda |{\rm ESIS/ENFW}|^{1/2}$, where ESIS and ENFW
are the energies of isothermal and NFW haloes.

The collapsing baryons will be stopped by the centrifugal barrier if $J$ is
conserved. In the absence of DM, $\lambda'$\,$\sim$\,$(R_{\rm d}/R_{\rm
h})^{1/2}$, where $R_{\rm d}$ is the radius of a disk embedded in the halo. To
reach rotational support, $\lambda'$ must increase by a factor of ten, to
$\lambda'$\,$\sim$\,0.5. This means that a collapse will proceed over two
decades in $R$ and the resulting $R_{\rm d}/R_{\rm h}$\,$\sim$\,0.01 will be
uncomfortably small. On the other hand, baryon collapse within a DM halo leads
to  $\lambda'$\,$\sim$\,$R_{\rm d}/R_{\rm h}$. This requires only a collapse by
a factor of 10 in $R$ in order to be in agreement with observations, resulting
in $R_{\rm d}$\,$\sim$\,8\,($\lambda'/0.035$)($H_{\rm 0}/H_{\rm
z})(v_{\phi}/200\,{\rm km\,s^{-1}})$\,kpc (e.g., Mo {\it et al.} 1998). Here
$H_0$ and $H_{\rm z}$ are the Hubble constants at $z$\,=\,0 and at an arbitrary
redshift, and $v_\phi$ is the maximum rotational velocity in the disk (which is
also the halo circular velocity for maximum disks). The corollaries are that,
e.g., haloes with higher $\lambda'$ lead to lower surface brightness disks, and
that the resulting disk size distribution originates in the $\lambda'$
distribution. 

However, the observed spread in $R_{\rm d}$ appears to be larger than the spread
in $\lambda'$ (e.g., de Jong \& Lacey 2000). Furthermore, an analysis of the
Courteau (1997) sample of local massive disks has shown that gas may lose some
of its angular momentum (Burkert {\it et al.} 2009). Comparison of these disks
with the half-light radii of $z$\,$\sim$\,2 SINS galaxies (e.g.,
F\"orster-Schreiber {\it et al.} 2009; Cresci {\it et al.} 2009), reveals a
similar deficiency of the spin parameter for the latter. We return to this issue
in Section\,\ref{sec:angmom}. While the above discrepancies must be clarified, a
number of other caveats threatening the current paradigm require much more
serious attention, and will be discussed below.   

%
%

\section{Disk growth: the angular momentum problem}
\label{sec:angmom}

Within the galaxy formation paradigm of White \& Rees (1978) and Fall \&
Efstathiou (1980), the DM haloes acquire their spin from tidal interactions and
baryons increase their rotational support by falling into the potential wells of
these haloes and by conserving their angular momentum. To what degree the baryon
angular momentum is preserved during this process has emerged as one of the key
issues in our understanding of disk galaxy formation.\looseness-1

\subsection{\textit{\textbf{The angular momentum catastrophe}}}
\label{sec:catast}

So is there a problem with $J$ when forming galactic disks? We start with the
so-called angular momentum `catastrophe' which has emerged from the comparison
of observations with numerical simulations of disk formation. While the observed
disks have shown a deficiency of specific angular momentum $j$ by a factor of
two, modelled disks  appear to have radial scalelengths smaller by a factor of
10 compared with observations, resembling bulges rather than disks (e.g.,
Navarro \& Steinmetz 2000). In other words, the actual disks rotate too fast for
a given luminosity $L$.  At the same time, the slopes of the modelled disks
turned out to be in agreement with the $I$-band Tully-Fisher (TF) relation for
late-type galaxies, $L$\,$\sim$\,$v_{\rm max}^\alpha$, where
$\alpha$\,$\sim$\,2.5--4, increasing to $\sim$3 in $I$. In our context, this is
important because TF defines the relation between the dynamical mass and $L$.
Simulations have shown much more compact endproducts of baryonic collapse, where
the gas has lost a substantial fraction of its $j$. As we shall see below, the
possible remedy to this problem involves a combination of numerical and physical
factors, but no universal solution has been found so far. 

What are the possible contributors to the angular momentum catastrophe? The
first suspect is lack of numerical resolution. Indeed, increasing the resolution
by adding smooth particle hydrodynamics (SPH) particles (e.g., Lucy 1977;
Monaghan 1982) has improved the situation by helping to resolve the maxima of
rotation curves, resulting in flatter and lower curves, thus having an effect on
the central and outer regions of modelled galaxies (e.g., Governato {\it et al.}
2004; Naab {\it et al.} 2007; Mayer {\it et al.} 2008).

Numerical non-conservation of $j$ is another suspect. In the outer disks, it can
cause an excess of angular momentum transfer to the halo (e.g., Okamoto {\it et
al.} 2005). In the inner disks, this effect could smear the radial gradients of
rotation velocity, and artificially stabilise the disks against bar instability.
Modelling an isolated galaxy formed from a non-cosmological Milky Way-type DM
halo has demonstrated that low-resolution disks heat up and lose $j$ to the
halo. Colder disks remain larger and have smaller bulge-to-disk ratios. A
possible solution lies in the increase of the number of SPH particles and the
introduction of a multi-phase interstellar medium (ISM). 

Next, the disk ISM cooling below the Toomre (1964) threshold of
$Q$\,$\equiv$\,$c_{\rm s}\kappa/\pi G\Sigma$\,=\,1 leads to an excessive
fragmentation (we return to this problem when discussing over-cooling).
Here $c_{\rm s}$ is the sound speed in the gas, $\kappa$ -- the epicyclic
frequency, and $\Sigma$ -- the disk (gas) surface density. The multi-phase ISM
introduced by Springel \& Hernquist (2003) through a modified equation of state
clearly alleviated the problem. The gas cooling must be compensated by some kind
of feedback to avoid fragmentation and the follow-up runaway star formation.
This feedback from stellar evolution, e.g., from SNe and OB stars,
has a clear effect on the bulge-to-disk ratio (Robertson {\it et al.} 2004;
Heller {\it et al.} 2007b).   

Finally, it was pointed out that baryons can hitchhike inside DM halo
substructures which spiral in to the centre as a result of dynamical friction
(e.g., Maller \& Dekel 2002). This process is similar to that in a disk, where
in the absence of star formation, dense gas clumps would heat up the disk,
spiral in to the centre and possibly contribute to the bulge growth (e.g.,
Shlosman \& Noguchi 1993; Noguchi 1999; Bournaud {\it et al.} 2007). This
process, related to overcooling, can be easily over- or underestimated because
it depends on a number of additional processes, e.g., gas-to-stars conversion,
ablation, etc. (see Section~\ref{sec:bulge}). Also, one cannot neglect the
contribution of clumpy DM and baryons to the mechanical feedback (i.e., by
dynamical friction), and, consequently, to the fate of the central DM cusp  and
the formation of a flat density core in the halo (e.g., El-Zant {\it et al.}
2001; Tonini {\it et al.} 2006; Romano-D\'iaz {\it et al.} 2008a).    

\subsection{\textit{\textbf{The cosmological spin distribution and individual haloes}}}
\label{sec:spin}

We now turn to the distribution of angular momentum in DM haloes. Early
considerations of the baryon collapse assumed an initial solid-body rotation for
uniform-density gas spheres. The angular momentum distribution (AMD) can be
written as $M(<j)$\,=\,$M_{\rm h}[1-(1-j/j_{\rm max})^{3/2}]$, where $j_{\rm
max}$ is some maximum $j$, if baryons preserve $j$ during this process, which
results in exponential disks (e.g., Mestel 1963).  More generally, what is the
shape of an AMD which leads to exponential disks? The initial solid-body
rotation law is soon replaced, following assumptions that baryons are mixed well
with the DM and share its $j$. Assuming $j$ conservation, $M_{\rm d}(<j)/M_{\rm
d} = M_{\rm h}(<j)/M_{\rm h}$, where $M_{\rm d}$ is the disk mass (e.g., Fall \&
Efstathiou 1980; Bullock {\it et al.} 2001; Maller \& Dekel 2002), the new AMD
can be written as
\begin{equation}
P(j) = \frac{\mu j_0}{(j+j_0)^2}    \hskip .5truein {\rm if}\, j\ge 0
\label{eq:specmom}
\end{equation}
and $P(j) = 0$ if $j < 0$. This leads to a mass distribution with $j$
\begin{equation}
M(<j) = M_{\rm h}\frac{\mu j}{j+j_0},
\label{eq:massj}
\end{equation}
with two parameters, $\mu$ -- the shape, and $j_0 = (\mu-1) j_{\rm max}$ --
whose effect on $M(<j)$ has been displayed in Bullock \textit{et al.} (2001,
Fig.~4) for four haloes. When using the best-fit $j_0$, and normalising by
$M_{\rm h}$, all haloes follow the same curve nicely -- an indication of the
universality of the AMD. The resulting (computed) disk surface density is nearly
exponential, but with an over-dense core and an overextended tail.

Using Swaters' (1999) sample of 14 dwarf disk galaxies, van den Bosch {\it et
al.} (2001) have demonstrated that the specific AMDs (i.e., $M(<j)$) of these
disks differ from halo AMDs proposed by Bullock {\it et al.} (2001). But the
total disk $j$'s, i.e., $P(\lambda)$, are similar to those of haloes. Although
the pre-collapse baryon and DM AMDs are expected to be similar, the disks formed
have been found to lack low and high $j$ -- in direct contradiction to the
simple theoretical expectations discussed above. Moreover, the mass fraction of
Swaters' disks appears to be significantly smaller than the universal baryon
fraction in a $\Lambda$CDM universe. Does this mean that disks form from a small
fraction of baryons and, in spite of this, attract most of the available angular
momentum? One can argue that a redistribution of baryon $j$ should occur (at a
stage earlier than anticipated), or that the high-$j$ baryons avoid the disks.
What processes are responsible for modifying the baryon AMDs? As argued by van
den Bosch {\it et al.} (2001), a plausible explanation can lie in the
selective loss of the low-$j$ gas, leading to essentially bulgeless disks.

The closely-related issue of the existence of bulgeless disk galaxies is
discussed in Section~\ref{sec:feedback}.

\subsection{\textit{\textbf{The tidal torque theory and the origin of the universal 
angular momentum distribution}}}
\label{sec:ttt}

Rotation of galaxies results from gravitational torques near the time of the
maximum expansion (e.g., Hoyle 1949; Doroshkevich 1970; White 1978). If $J$ is
calculated using the spherical shell approximation (i.e., linear tidal torque
theory, hereafter TTT) further redistribution during the non-linear stage
contributes little (Porciani {\it et al.} 2002a,b). Alternatively, $J$ can be
acquired in the subsequent non-linear stage of galaxy evolution via mergers,
accounting for the orbital angular momentum and its redistribution (e.g., Barnes
\& Efstathiou 1987). 

According to the TTT, $J$ is gained mostly in the linear regime of growing
density perturbations due to the tidal torques from neighbouring galaxies. The
maximum contribution to $J$ comes from the times of the maximum cosmological
expansion of the shell. Little $J$ is exchanged in the non-linear regime after
the decoupling and virialisation of the DM. Baryons are well mixed with the DM
and acquire similar $j$ (see Section~\ref{sec:intro}). The TTT follows the
evolution of the angular momentum of the DM halo based on the moment of inertia
tensor and the tidal tensor. Within this framework, galactic spins can be
estimated using the quasi-linear theory of gravitational instability. The TTT
has been extensively tested by numerical simulations.

\begin{figure}[ht!!!]  
\begin{center}
\includegraphics[angle=0,scale=0.83]{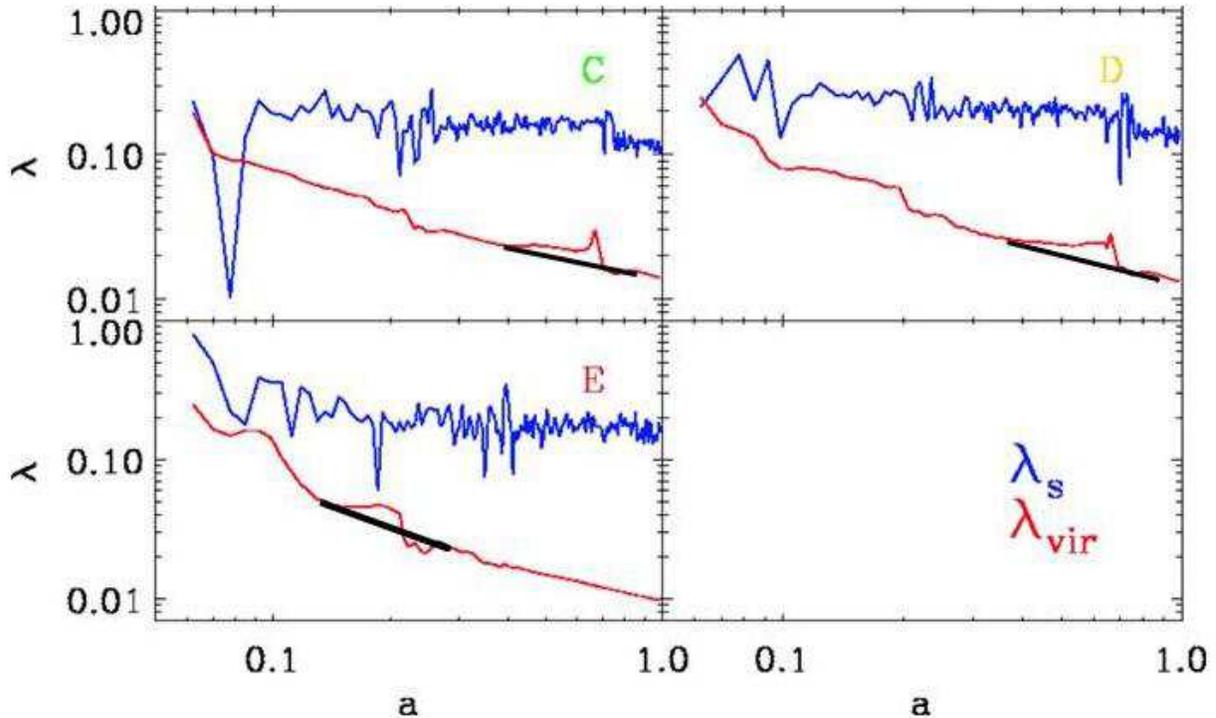}
\end{center}
\caption{\small Evolution of DM halo spin $\lambda$ as a function of $a$, the
cosmological expansion parameter, during major mergers in three models, C--E. In
all frames: the blue (upper) curves show the spin $\lambda_{\rm s}$ within the
characteristic NFW radii. The red (lower) curves show the evolution of the spin
$\lambda_{\rm vir}$ within the halo virial radii. The thick (black) line has
been applied to show the `normal' evolution during major mergers (from
Romano-D\'iaz {\it et al.} 2007).}
\label{fig:fig01}
\end{figure}

So far we have not considered the effect of the mergers on the evolution of $J$
(or $\lambda$). We define mergers as major if their mass ratio is larger than
$1:3$. Wechsler (2001) has argued that major mergers dominate the $j$-balance in
galaxies. Do they? The crucial point here is to account for the redistribution
of the angular momentum during and after the merger. Hetznecker \& Burkert
(2006) have separated the main contributors to $\lambda$\,$\sim$\,$J|E|^{1/2}
M^{-5/2}$, i.e., $J$, $E$ and $M_{\rm h}$, and followed them during and after
the merger event. While the DM halo $J$ experiences a jump during a major
merger, the subsequent mass and energy redistribution in the halo washes out the
gain in $J$. A similar conclusion has been reached by Romano-D\'iaz {\it et al.}
(2007) and is displayed in Fig.~\ref{fig:fig01}. The corollary of
high-resolution DM simulations is that there is no steady increase in the halo
cosmological spin $\lambda$ with time due to major mergers. 

To summarise, hydrodynamical simulations show that the angular momentum of
baryons is not conserved during their collapse. A number of factors can
contribute to this process, but there is no definitive answer yet. The proposed
solutions often amount to fine tuning.  There is also a mismatch in the
$j$-profiles. The distribution of the modelled specific angular momentum does
not agree with the observations.

\begin{figure}[ht!!!]  
\begin{center}
\includegraphics[angle=0,scale=0.83]{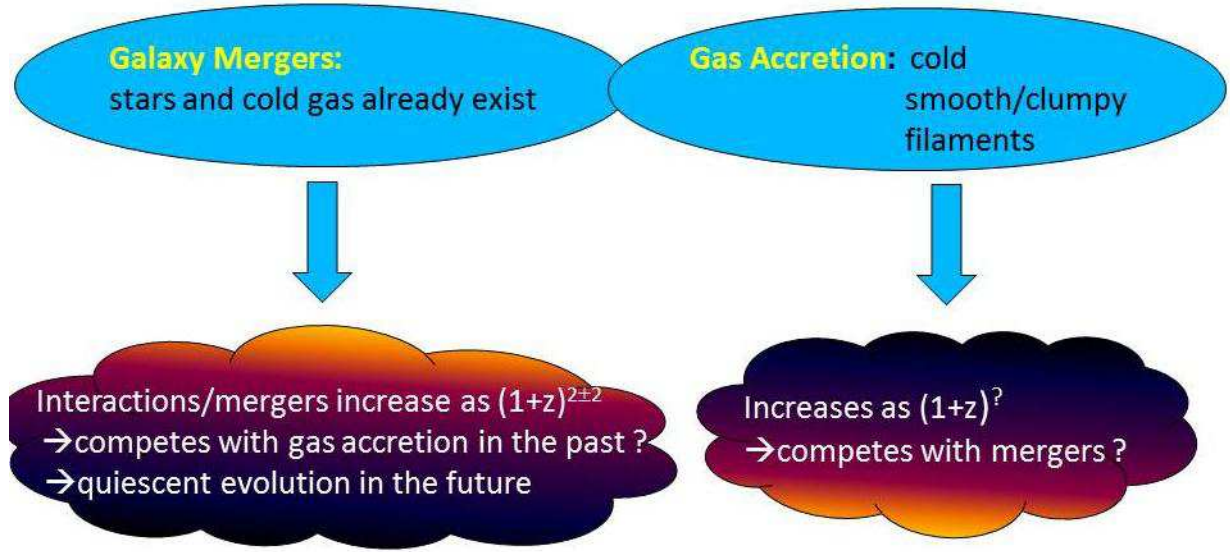}
\end{center}
\caption{\small Disk growth mechanisms involve mergers and/or gas accretion. Both options
depend on the environment, and their effects are gradually quenched in clusters of
galaxies. The dependence on redshift is $(1+z)^{2\pm 2}$ for mergers, and seems to
have a peak around $z\sim 2$, but this is yet to be verified for the gas accretion.}
\label{fig:fig02}
\end{figure}

%
%

\section{Disk growth: mergers}
\label{sec:mergers}

In this and the following sections we shall review the details of secular and
dynamical growth in disk galaxies which involve merging and/or the accretion of
cold gas (Fig.\,\ref{fig:fig02}). In the simplest (unrealistic) case of an
isolated collapsing cloud, the angular momentum is conserved because the
escaping radiation is isotropic and carries little $J$ anyway. It is possible
that some of the gas will rebound and escape in the equatorial plane taking away
some $J$. 

The lowest energy state for a rotating self-gravitating cloud occurs when almost
all of the mass is found in the central compact accumulation (i.e., SMBH) and a
negligible mass and all of the angular momentum escape to infinity. Such a state
is never achieved, however, because the efficiency of $J$-transfer is declines
rapidly as the process goes on. Amusingly, to quote Colgate \& Petschek (1986),
`there seems to be too much angular momentum in the Universe to allow the
formation of stars [...]. This fundamental problem begs for solution. Since net
$J$ appears to be zero over very large scales [...] our problem is restricted to
local patches of the Universe where matter collapses to form relatively dense
rotating objects [...].'  We note that in order to understand how galaxies
grow, one needs to resolve the issue of what is the most efficient angular
momentum loss mechanism in particular circumstances.

Disk galaxies form inside DM haloes, and it seems clear that Nature's choice is
the formation of self-gravitating disks rather than a complete and efficient
collapse to the very centre. Disks can form slowly from a weakly rotating gas
cloud, in an adiabatic process, and their properties in this case will be linked
to the initial conditions. Or they can form on a short dynamical timescale,
forgetting the initial conditions in the process. In such a process, how much
$J$ is transferred to the DM and how much is redistributed among the baryons?

What factors affect the secular growth of galactic disks? The following list
of physical factors and associated processes was not compiled in order of 
significance, and are not mutually exclusive:
\begin{itemize}
\renewcommand{\theenumi}{(\alph{enumi})}

\item Environment: determined by high/low density fields and their corollary --
supply of cold gas (from filaments, etc.), major and minor mergers (in the low
velocity dispersion field), interactions/harassment (in the high velocity
dispersion field), stripping (tidal, ram, strangulation).

\item Rate of star formation, feedback from stellar evolution and active
galactic nucleus (AGN), galactic winds.

\item Intrinsic factors: local and global gravitational instabilities, degree of
asymmetry in the host DM halo.

\end{itemize}
In the following, we shall discuss some of the above factors, but shall avoid
going into details about the internal dynamics of stellar bars (see review by
Lia Athanassoula, this volume) and general aspects of galactic dynamics (see
review by James Binney, this volume). This means that we shall emphasise the
external factors driving galaxy evolution, and only occasionally resort to
internal factors. Two such exceptions involve the physics of the central
kiloparsec and some aspects of the formation of SMBHs (Section~\ref{sec:kpc}).
We note, however, that this separation is largely artificial, and the action of
external triggers of galaxy evolution is frequently associated with `excitation'
of internal degrees of freedom. One such well-known example is the tidal
triggering of stellar bars. 

In order to grow, galaxies and their haloes must rely on the exterior reservoir
of baryons and DM, respectively. These can come in various degrees of
virialisation: from smooth to clumpy accretion (see
Section~\ref{sec:accretion}), from carrying nearly unvirialised to fully
virialised baryons and embedded in parent DM haloes. The latter lead to mergers
of various mass ratios, from major, to intermediate, to minor. The physics of
accretion of course differs profoundly depending on the smoothness of accreting
material. Mergers can also include stellar and DM components. In the following,
we are going to discuss the main processes which accompany merging: dynamical
friction, phase mixing and violent relaxation.

Mergers are a diverse phenomenon and their products are equally diverse. We
first discuss the criterion for merging, then merger demographics, i.e., their
role in forming and growing spheroids and gas-rich disks (wet/dry mergers). We
follow this up by reviewing the effect of major mergers on starbursts and AGN,
and the effect of minor mergers on disk heating. The relationship between
mergers and morphology evolution, stripping, and, briefly, the origin of
ellipticals are discussed next. Two questions will be emphasised: whether
mergers play an important role in creating spheroids, and whether dry mergers
move galaxies along the red sequence.

The process of merging can be defined as an encounter of two galaxies which
results in the formation of a single galaxy. Based on the mass ratios of the
merging objects, we distinguish between major mergers ($>$\,1:3), intermediate
mergers (from 1:3 to 1:10) and minor mergers ($<$\,1:10). Mergers in gas-rich
systems are highly dissipative and denoted as `wet' mergers, while mergers in
gas-poor systems are dissipationless and `dry'. Finally, mergers can be binary
or multiple when more than two galaxies are involved. 

Mergers are important because their rate grows rapidly with redshift,
$\propto$\,(1+$z$)$^m$, although the inferred rate exhibits a substantial
scatter. The differences come from matching the rate for rich clusters and for
field galaxies, but also because of using different methods, such as close pairs
and morphology. Extending the measure to fainter magnitudes generally shows an
increase in $m$. There is also a difference between observational and
theoretical estimates. Specifically, $m$\,=\,6\,$\pm$\,2 in rich clusters (e.g.,
van Dokkum {\it et al.} 1999), but it is much smaller in the field 
($m$\,=\,2.7\,$\pm$\,0.6; e.g., Le Fevre {\it et al.} 2000). A recent comparison
between optical and near-infrared (NIR) bands has revealed a substantial
difference in the major merger rate of $m$\,=\,3.43\,$\pm$\,0.49 and
2.18\,$\pm$\,0.18, respectively (Rawat {\it et al.} 2008). Diverse observations
have resulted in an overall range of $m$\,=\,2\,$\pm$\,2, while numerical
simulations result in a narrow range of $m$\,$\sim$\,3, although ignoring the
possibility of multiple galaxies per DM halo.  

The situation is more confusing with the redshift dependence of the cold gas
accretion rate as few observational constraints exist. On the other hand,
numerical estimates of accretion growth are possible. For $z$\,$\gtorder$\,2, 
accretion rates show a strong increase towards lower $z$, and a sharp decline
thereafter.

\begin{figure}[ht!!!]  
\begin{center}
\includegraphics[angle=0,scale=0.83]{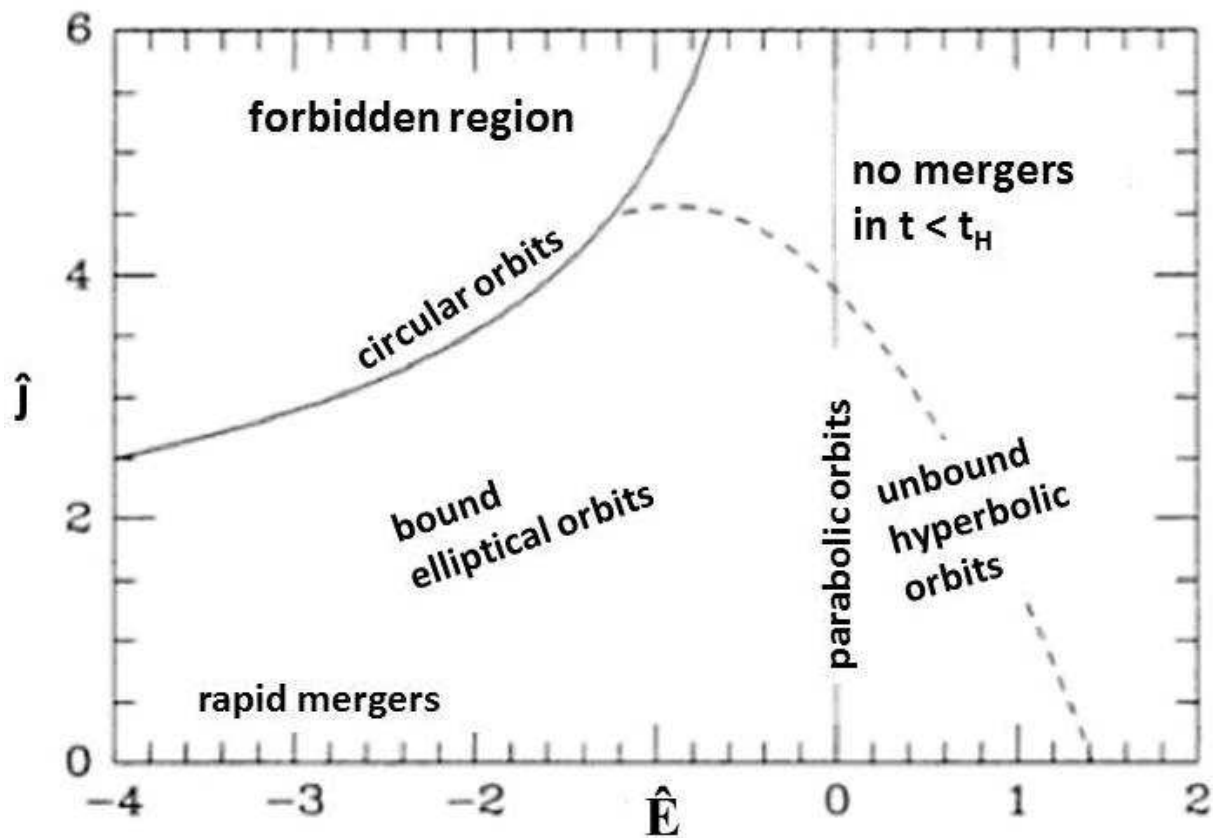}
\end{center}
\caption{\small Criterion for merging (modified Figure 7-9 from Binney \& Tremaine
1987), which is based on the time required for the merging of two galaxies  as a
function of the initial position of the binary orbit in the ${\hat j}-{\hat E}$
plane. Orbits are only possible outside the forbidden region delineated by the
solid lane (circular orbits). All elliptical (bound) and some hyperbolic
(unbound) orbits below the dashed lane will eventually lead to a merger.
However, merging time increases rapidly toward the upper-right corner, and above
the dashed lane the merging time exceeds the Hubble time.}
\label{fig:fig03}
\end{figure}

In order to delineate the relevant parameter space for mergers, we define two
dimensionless parameters, the energy ${\hat E}$\,$\equiv$\,$2E/\sigma^2$ and the
specific angular momentum ${\hat j}$\,$\equiv$\,$j/\sigma r_{\rm m}$, where
$\sigma$ is the inner dispersion velocity and $r_{\rm m}$ is the median
(half-mass) radius of an idealised galaxy. With a gross simplification of such
spherical, nonrotating equal-mass galaxies it is possible to delineate the
${\hat j}-{\hat E}$ parameter space of mergers with a characteristic timescale
of less than the Hubble time (Fig.~\ref{fig:fig03}). Note that merging can occur
also from unbound orbits. As a rule of thumb, only galaxies with relative
velocities less than their internal dispersion velocities, $v_{\rm
rel}$\,$\ltorder$\,$\sigma$, will merge. Dispersion velocities inside large
galaxies are $\sigma$\,$\sim$\,200--300\,km\,s$^{-1}$, while that of clusters of
galaxies $\sim$500--1000\,km\,s$^{-1}$. The merger orbital angular momentum
depends on $v_{\rm rel}$ and contributes to the spin. A simple recipe to
increase the parameter space for mergers is the dynamical friction mechanism.

\subsection{\textit{\textbf{Mergers and associated processes}}}
\label{sec:associate}

\subsubsection{Minor mergers and dynamical friction}
\label{sec:minorm}

Minor mergers are those with $m$\,$\ll$\,$M$. When the massive galaxy with mass $M$ moves among 
the background of low-mass $m$ galaxies (or else) with velocity ${\bf v}$, dynamical 
friction is a good approximation. The drag per unit mass resulting from a 
gravitational wake is given by (Chandrasekhar 1943)
\begin{equation}
\frac{d{\bf v}}{dt} = - \frac{4\pi {\rm ln}\,\Lambda G^2 \rho M}{v^2} 
    F\biggl(\frac{v}{\sqrt{2\sigma}}\biggr)\biggl(\frac{\bf v}{v}\biggr) ,
\label{eq:drag}
\end{equation}
where $\Lambda$\,$\approx$\,$b_{\rm max}/\max[r_{\rm h}; GM/v^2]$ is the Coulomb
logarithm, $b_{\rm max}$ is the maximum impact parameter, $\rho$ is the
background density of $m$ galaxies, and $F$ is the error function. In the
low-$v$ limit, the drag is $\sim$\,$v$, and in the high-$v$ limit, it is
$\sim$\,$v^{-2}$. Also, the dynamical friction force is $\sim$\,$M^2$, and the
DM appears as an important contributor to the friction during the merging
process. Assuming a circular orbit with a radius $r$ for $M$ moving around a
bound accumulation of $m$ masses, one can arrive at the analytical solution for
$r(t)$ which is a linear function of time. However, in a more realistic case,
$M$ plunges in a very elongated orbit, with a small pericentre, leading to a
substantial tidal disruption and an associated mass loss very early. This of
course complicates the analytical solution (e.g., Diemand {\it et al.} 2007;
Romano-D\'iaz {\it et al.} 2010) and tends to decrease the characteristic
timescale for the friction (e.g., Boylan-Kolchin {\it et al.} 2008).

The specifics of the dynamical friction are that it is a local and not a global
force in the Chandrasekhar approximation, which assumes that the $m$ particles
interact only with $M$. The situation is more complicated if self-gravitational
effects are taken into account, which is most relevant for galactic disks. The
most interesting corollary is the introduction of resonances between the orbital
motion of $M$ and motion of $m$ in the disk, both azimuthal, radial and vertical
(e.g., Lynden-Bell \& Kalnajs 1972). Other intricacies appear as well, but are
not discussed here (e.g., Mo {\it et al.} 2010).

The effect of dynamical friction on multiple mergers has been shown explicitly
in numerical simulations of Nipoti {\it et al.} (2003), who ran them with and
without friction with the goal of modelling the formation of cD galaxies in the
centres of galaxy clusters. The characteristic timescale for merging has been
substantially shortened when the Chandrasekhar drag has acted, compared to
merging in a fixed background potential. 

\subsubsection{Mergers: phase mixing and violent relaxation}
\label{sec:phasem}

The two-body relaxation timescale is too long to have an effect on mergers,
where the relaxation is dominated by collisionless processes, such as
phase mixing and violent relaxation -- relaxation in the mean field. Both
mechanisms have been introduced by Lynden-Bell (1967). Behind the idea of
phase mixing lies the time evolution of a coarse-grained distribution function
in a collisionless system. As the classical entropy is conserved in the absence
of collisions, the fine-grained function is time-independent, while the
coarse-grained function evolves to uniformly cover the available phase space for
the system, thus maximising the corresponding coarse-grained entropy. Phase
mixing, therefore, tends to destroy coherent phase-space structure.  

Violent relaxation is a relaxation in the time-dependent potential of the
system, when the specific energy of a particle is not conserved. This process
leads to a Maxwellian distribution of velocities in which the temperature is
proportional to the mass of the particle. So particle dispersion velocities
become independent of the particle mass. Violent relaxation is most relevant at
the time of virialisation of the system, so the characteristic time is the
crossing time of the system. This leads to a more complete relaxation in the
central regions compared to the outskirts, because the dynamical timescale
becomes prohibitively long there. On the other hand, galaxy interactions would
drive violent relaxation mostly in the outer regions. It is still unclear how
efficient this process is overall.

\subsection{\textit{\textbf{Dry and wet disk mergers: spheroids or rebuilding?}}}
\label{sec:drywet}

Mergers between systems that include stars and DM only are called dry mergers.
Examples: a merger of two elliptical galaxies, or one between an elliptical and a
lenticular galaxy. Only limited observational data exists on dry mergers, mostly
in clusters of galaxies (e.g., van Dokkum \textit{et al}. 1999). The endproduct
of this process is predicted to be an ellipsoidal system (e.g., Toomre \& Toomre
1972; Barnes 1992), as inferred from the Sersic law. The mixing appears
incomplete, and the metallicity gradient is not fully erased. The emergence of
the red sequence of massive galaxies has been linked to the increasing
importance of dry mergers after $z$\,$\sim$\,1, because of the seemingly
insufficient amount of massive blue galaxies that can serve as their precursors
(e.g., Khochfar \& Burkert 2003; Faber {\it et al.} 2007). How important the
contribution is of dry mergers in forming the red sequence, however, is
unclear. 

Dry mergers have been studied using the GEMS (Galaxy Evolution from Morphology
and SEDs [Spectral Energy Distributions]) survey in tandem with the COMBO-17
photometric redshift survey in order to constrain their frequency between
$z$\,$\sim$\,0.2--0.7 (e.g., Bell {\it et al.} 2006). Accompanying $N$-body
simulations have been used to explore the morphological signatures of such
mergers. An estimated rate of $\sim$0.5--2 major mergers between spheroids
over the $z$\,$\sim$\,0.2--0.7 period has been claimed to be consistent with the
limit of $\ltorder$\,1 such event in recent times, estimated using an
alternative semi-analytic method. This indicates that dry mergers can indeed be
an important factor in driving the evolution towards massive red galaxies at
present times, but more work is clearly required.

Scaling relations, such as Faber-Jackson (1976, FJ), Kormendy (1977), the
fundamental plane (Djorgovski \& Davis 1987), and others can provide, in
principle, important information about the formation of massive ellipticals, and
constrain it. $N$-body simulations have shown that multiple dry mergers preserve
the FJ relation, but produce lower central dispersion velocities and increase
the effective radius, although the fundamental plane of ellipticals remains thin
(Nipoti {\it et al.} 2003). Ciotti {\it et al.} (2007) confirmed that the FJ,
Kormendy and fundamental plane relations are robust against dry merging,
although caveats exist. An important question is when are these scaling
relations established? 

Wet mergers involve gas and are, therefore, dissipative. Here we focus on
mergers which involve galactic disks with various gas fractions. When disks are
involved, the outcome depends on many more parameters, including the disk plane
orientation with respect to the orbital plane, and the alignment of internal
spins (rotation) with the orbital spin, i.e., prograde versus retrograde 
encounters.

The first simulations of disk interactions were performed by Holmberg (1941)
using a light-bulb `supercomputer', with the important result that tidal forces
lead to the formation of spiral structure in galaxies. Modern numerical
simulations have revealed a rich library of processes involving disk
interactions and mergers: stretching, harassment, stripping, strangulation,
squelching, threshing, splashback and cannibalism. Additional effects include
enhanced star formation rates and quenching the star formation (e.g., recall
that spheroids are associated with quenched star formation). The outcome of disk
galaxy mergers can be either spheroidal or disk systems. 

The early arguments about disk merger remnants were based on $N$-body
simulations, without or with low-resolution gas. The merging proceeded via
dynamical friction against the DM component.  The DM haloes have been `soaking
up' the internal and orbital angular momenta of merging galaxies, and the
collision appeared sticky (e.g., Hernquist 1992). Overall, the simulations have
been successful in fitting the properties of the elliptical products, although
the remnants appeared too diffuse compared to observed massive ellipticals. The
morphology of interacting and merging galaxies has been closely matched (e.g.,
Hibbard \& Mihos 1995). Dubinski {\it et al.} (1996) have studied the merging of
pure stellar disks in live DM haloes, focussing on the shapes of tidal tails,
when disks are being stretched, imparting kinetic energy to the stars. The
length and mass of tidal tails have been found to be sensitive to the gradient
of the gravitational potential. Hence, tails can successfully map the DM
potential well and constrain the overall DM mass distribution. The problem was
in reproducing the high phase density observed in the centres of  elliptical
galaxies. This situation changed with the inclusion of a dissipative component
in simulations.

Numerical simulations have indicated that the dynamical role of gas is well in
excess of its mass fraction. This has been shown for isolated galactic models
(e.g., Shlosman \& Noguchi 1993; Heller \& Shlosman 1994) as well as for fully
cosmological models (e.g., Barnes \& Hernquist 1996). Due to its dissipative
nature, gas is always losing its energy and angular momentum, which leads to a
central accumulation, where the gas successfully competes with stellar and DM
contributions to the gravitational potential. The deepening of the potential
well by the gas resolves one of the outstanding issues we have mentioned above
-- high phase density in the centres of ellipticals. Barnes \& Hernquist (1996)
have found that the gas presence shortens the merging timescale and drives a
large fraction of gas to the very centre of the remnant. This evolution is
relatively insensitive to the detailed physics, given that the gas is able to
cool. Simulations of mergers involving disks with $\sim$10\% or less gas mass
fraction lead to the formation of a spheroidal stellar component with a surface
brightness of the de Vaucouleurs 1/4 law, and a central stellar cusp which is
not observed in such galaxies. These simulations have shown that wet mergers can
be responsible for the formation of some ellipticals. The question is what
fraction?

\begin{figure}[ht!!!] 
\begin{center}
\includegraphics[angle=0,scale=0.59]{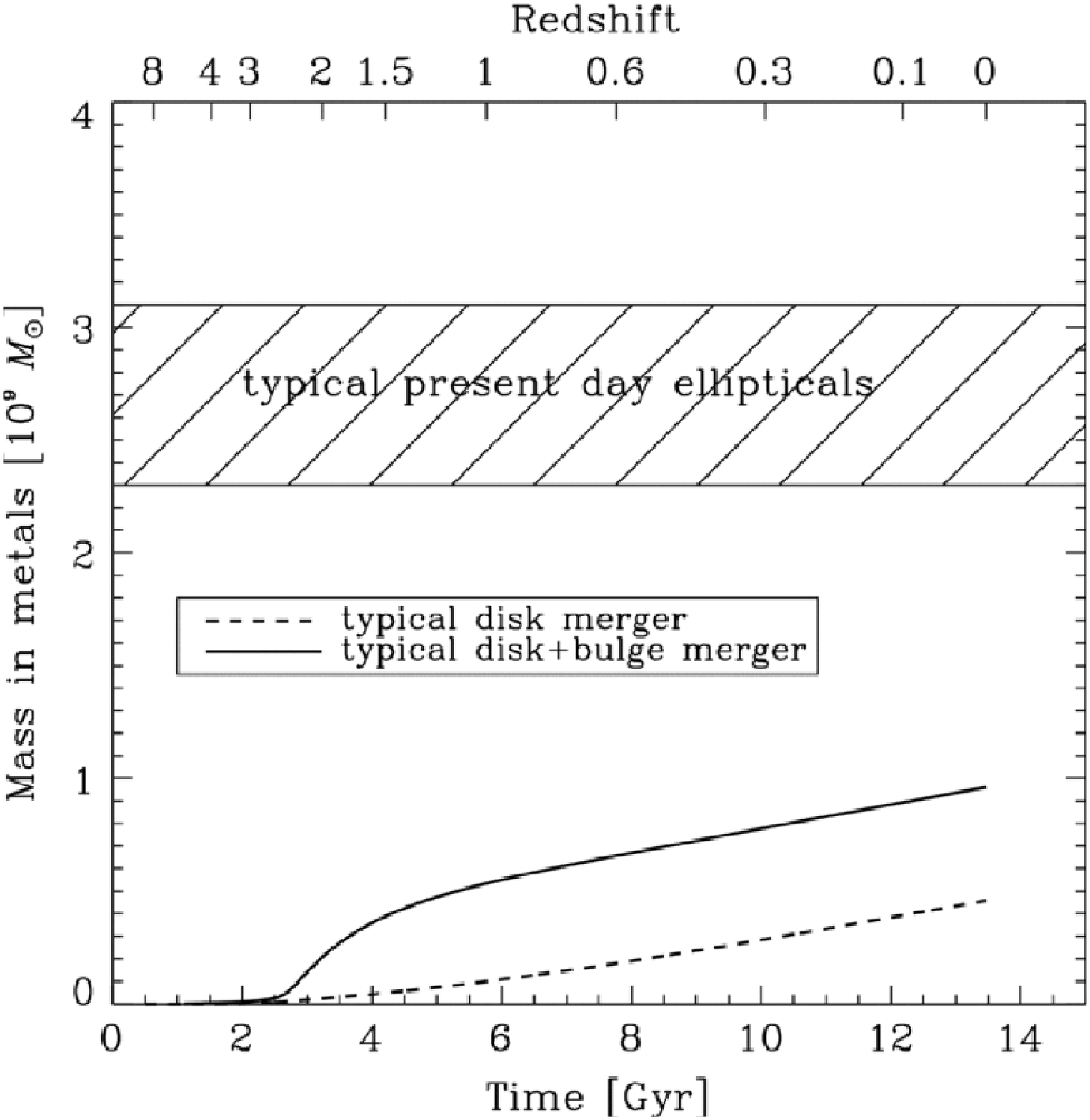}
\end{center}
\caption{\small Evolution of metals (in mass) for the merger of two $M_*$ disks with
(solid) and without (dashed) bulge component. The total mass of the progenitor
disks used was 2.9\,$\times$\,10$^{10}\,M_\odot$. The metal mass for $M_*$
ellipticals is indicated by the shaded area. Present-day ellipticals have (at
least) a factor of two more metals (from Naab \& Ostriker 2009). }
\label{fig:fig04}
\end{figure}

Difficulties with the scenario of ellipticals forming in binary major mergers of
disk galaxies include the following: typical ellipticals are more metal-rich
than typical present-day disks (see Fig.~\ref{fig:fig04}), ellipticals have
older stellar populations that seem to form on shorter timescales, and massive
ellipticals could not typically have formed from binary mergers of present day
disks (while they might form from high-$z$ disks whose descendants no longer
exist (e.g., Naab \& Ostriker 2009). In addition, binary mergers of any kind are
not isotropic, whereas massive ellipticals are.

Probably the most intriguing issue of disk mergers is whether disks can survive
mergers. Simulations of disk mergers without gas point to a clear trend of
mergers thickening and destroying the disks (e.g., Kazantzidis {\it et al.}
2008; Robertson \& Bullock 2008). The corollary is that the disks must grow
rather quiescently. But what about the more relevant situation where disks
contain gas, and there are plenty of `leftovers' from mergers? What is the
outcome when the disks not only contain gas but are gas-rich?  Is there a
critical gas fraction, $f_{\rm gas}$, for disk survival?

\begin{figure}[ht!!]  
\begin{center}
\includegraphics[angle=0,scale=0.84]{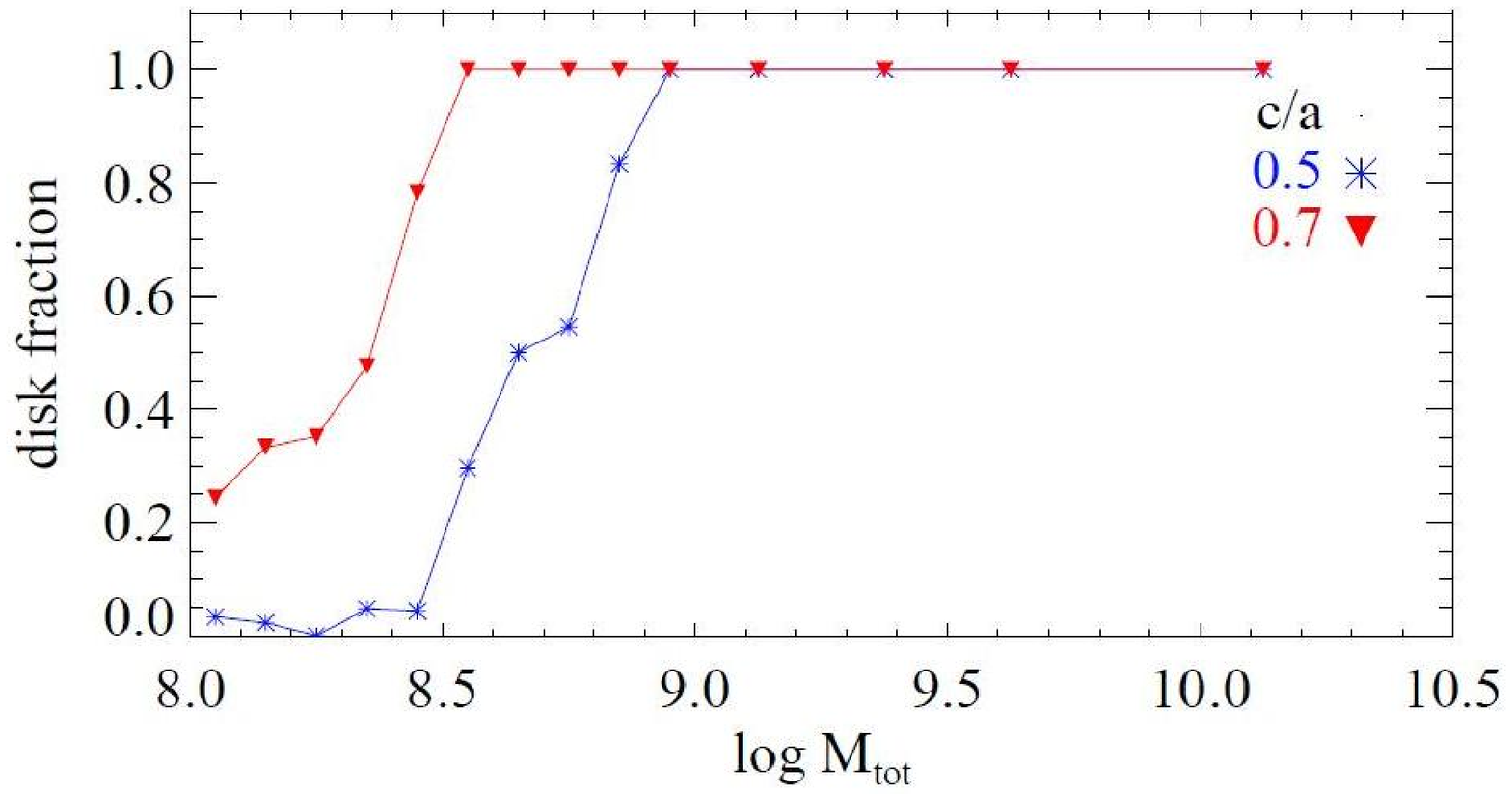}
\end{center}
\caption{\small Disk fraction as a function of its total (baryon$+$DM) mass at
$z$\,$\sim$10.2 (in $M_\odot$) within a high-resolution region of 7\,$h^{-1}$\,Mpc with
binning of 0.25 in total mass. Shown are two defining criteria for the gas disk,
$c/a$\,$\ltorder$\,0.5 (blue stars) and $c/a$\,$\ltorder$\,0.7 (red triangles),
where $a$, $b$ and $c$ are the major, intermediate and minor axes of gaseous
disks. Haloes in the process of merging have been omitted, overall four objects
within this mass range -- all of them had more than one disk per halo (from
Romano-D\'iaz {\it et al.} 2011b).}
\label{fig:fig05}
\end{figure}

Indications that disks can reform after {\it some} major mergers, if sufficient
amounts of gas can be maintained, come from simulations of pure-gas, bulgeless
disks on prograde parabolic orbits, in the presence of star formation (e.g.,
Springel \& Hernquist 2005). Moreover, gas-rich disks with $f_{\rm
gas}$\,$\gtorder$\,0.5 (e.g., Robertson {\it et al.} 2006; Robertson \& Bullock
2008), or continuous accretion of the cold gas following a destructive merger
(e.g., Steinmetz \& Navarro 2002) show a similar trend. Arguments that disk
heating has been overestimated in minor mergers have been advanced as well
(e.g., Hopkins {\it et al.} 2008; Romano-D\'iaz {\it et al.} 2008b). Recent
high-resolution simulations of over-dense regions in the Universe have shown a
resilient, disk-dominated population of galaxies (Fig.~\ref{fig:fig05}) at
$z$\,$\sim$\,8--10 (Romano-D\'iaz {\it et al.} 2011b). Subsequent analysis of disk
growth in such regions reveals that the dominant growth mode is not via major
mergers but rather through accretion of cold gas (Romano-D\'iaz {\it et al.}
2012), as we discuss in Sections~\ref{sec:accretion} and \ref{sec:highz}.

Attempts to understand the conditions for disk survival on cosmological
timescales have delineated the main contributing factors which can increase the
disk endurance: an existing reservoir of cold gas which is able to resupply it
on a short dynamical timescale; delayed star formation, in order to avoid a
destructive starburst which would quench the growth of the stellar disk; and
continued ability of the shocked gas to cool radiatively on a short timescale
(more about this in the next section). In other words, the disk rebuilding
processes should be more efficient than the destructive processes during the
merger event. No `universal' solution to this problem exists, although with a
sufficient amount of fine tuning, progress has been made.

Among the few successful examples of an efficient rebuilding of the disk
component is a numerical study of disk galaxy evolution following a 1.6:1 wet
merger at $z$\,$\sim$\,0.8 (Governato {\it et al.} 2009). The environment chosen
for this experiment was typical of field haloes and Milky Way parameters for the
re-simulated galaxy. SN feedback has been responsible for the delayed star
formation. For this purpose, the blastwave approximation has been used (see
Section~\ref{sec:type}), where the cooling shuts off over a Sedov crossing time
of 3\,$\times$\,10$^7$\,yr over $\sim$0.2--0.4\,kpc regions (corresponding
also to the resolution limit of the model). During the phase of $z$\,$<$\,3,
$f_{\rm gas}$ in the progenitor disks was below 0.25. Over the period of disk
rebuilding, $\sim$ few Gyr, the old stellar population, found in the {\it thick}
disk, has faded considerably. Thus, the formed thin disk dominated the light in
the $I$-band, while the thick disk contributed $\sim$70\% of the stellar mass,
and the stellar halo component faded by $z$\,=\,0.  

A number of corollaries follow attempts to rebuild and sustain disks over
cosmological times. First, it apparently requires the existence of a thick stellar
disk component, which represents the population of a pre-merger disk. A
beautiful example is that of NGC\,4762, which exhibits both thin and thick disks
(e.g., Burstein 1979). If the thickening has been abrupt, the radial extent of
the thick disk provides the size of the original disk at the merger event. The
absence of thick disks in some late-type galaxies, e.g., NGC\,4244, can be
interpreted as a challenge to numerical simulations. It is, therefore,
encouraging that Comer\'on \textit{et al.} (2011) claim to have identified a
sign of the thick disk in this object. On the other hand, it is plausible that
in some mergers the stellar disks are destroyed completely. 

The other issue lies in the prohibitively long disk-rebuilding timescale at low
$z$, a few Gyr. While this timescale severely limits the number of destructive
mergers a galaxy can have at low redshift, $z$\,$\ltorder$\,1, it is
unacceptable at high redshift $z$\,$\gtorder$\,6. Simulations show, however,
that characteristic timescales for similar processes are substantially shorter
at high $z$, by about a factor of ten, which maintains a robust population of
disks even in highly over-dense regions (e.g., Romano-D\'iaz {\it et al.} 2011b;
2012). The morphology-density relation during the epoch of reionisation,
therefore, does not follow the trend it exhibits at low $z$. At what redshifts
does this relation take the form of the observed one? 

Among the numerous corollaries of disk mergers is their plausible contribution
to classical bulges. The observed frequency and mass fraction of classical
bulges in disk galaxies are debatable at present, specifically with respect to
other bulge types. For example, the origin of so-called disky bulges is
unrelated to galaxy interactions, and they result mostly from stellar bar
instabilities (e.g., Combes {\it et al.} 1990; Pfenniger \& Friedli 1991; Raha
{\it et al.} 1991; Berentzen {\it et al.} 1998; Patsis {\it et al.} 2002;
Mart\'inez-Valpuesta {\it et al.} 2006; see also review by Kormendy \& Kennicutt
2004). This makes it even more difficult to obtain a quantitative estimate of
their link to mergers. On top of this, the merger outcome is sensitive to
various associated physical processes and kinematical parameters. As numerical
simulations themselves depend on subgrid (and sometimes unknown) physics,
attempts have focused on semi-analytical models, although their predictive power
has not been verified. 

Using observational constraints on disk masses and gas fractions, $f_{\rm gas}$,
in galaxies, Hopkins {\it et al.} (2010) have attempted to predict the
properties of merger remnants, and, specifically, to quantify the contribution
to classical bulge formation from mergers of various mass ratios. The main
result was that major mergers dominate the assembly of $L_*$ bulges, while minor
mergers dominate the formation of bulges in low-mass systems. The bulge-to-total
mass ratio, $B/T$, was found to trace the merger mass ratio, $\mu_{\rm gal}$. A
simple correlation, $B/T$\,$\sim$\,$\mu_{\rm gal}(1-f_{\rm gas})$, has been
identified. The straightforward corollary of this correlation is that increasing
the gas fraction tends to suppress the bulge formation, which has been
interpreted in terms of a reduced efficiency of gas angular momentum loss with
increasing $f_{\rm gas}$. The by-product of this conclusion is that
collisionless systems lose angular momentum more efficiently than dissipative
ones, something which is difficult to accept.

As a next logical step, one can ask whether gas-rich high-$z$ disks themselves
form in mergers. A sample of such massive, $\sim$10$^{11}\,M_\odot$, disk
galaxies at  $z$\,$\sim$\,2 has been analysed with SINFONI/VLT integral-field
spectroscopy (Genzel {\it et al.} 2008). Large random motions have been detected
and interpreted in terms of rapid inflows from cold accretion flows which
originated in cosmic web filaments (Section~\ref{sec:accretion}). Such
turbulence can reduce the viscous accretion timescale to below $\sim$1\,Gyr.
The detailed example of the galaxy BzK\,15504 at $z$\,$\sim$\,2.38 has been
studied in sufficient detail by Genzel {\it et al.} (2006). This object is
characterised by a high star formation rate of $\sim$140\,$M_\odot\,{\rm
yr^{-1}}$, a gas mass of $\sim$4$\times$\,10$^{10}\,M_\odot$, and a
gas-to-star mass ratio of $\sim$0.5. The ratio of circular to dispersion
velocity, $v/\sigma$\,$\sim$\,3, points to a geometrically thick disk, and has
been interpreted as the formation stage of the thick galactic disk.
Interestingly, BzK\,15504 shows no obvious signs of a recent or ongoing merger,
e.g., no obvious line-of-sight velocity asymmetry. So, is this object a
proto-disk caught in the stage of a rapid but secular evolution? Unfortunately,
there is no simple answer to this question. The problem lies in that the same
kinematic parameters can also characterise a merger remnant, as shown by
Robertson \& Bullock (2008).

%
%

\section{Mergers and their secondary by-products}
\label{sec:byprod}

\subsection{\textit{\textbf{Tidal dwarfs}}}
\label{sec:dwarfs}

The definition of a merger process as one which decreases the number of galaxies
(see Section~\ref{sec:mergers}) has its drawbacks. The deformation and the
possible destruction of interacting disks manifests itself also in the creation
of clumps of stars and molecular gas -- so-called tidal dwarfs, the by-products
of mergers. One can argue that tidal dwarfs are not {\it bona fide} galaxies as
they are not expected to contain a significant amount of DM. We shall stay away
from this dispute. There is an additional difference between these objects and
`normal' galaxies -- they are expected to be made of recycled material with
metals and dust, and not have the primordial composition of the first galaxies
or of low-metallicity dwarfs. Tidal dwarfs are usually associated with
antenna-type tidal tails of their massive parent galaxies, are gas rich, have
both old and young stellar populations, and contain both H{\sc i} and H$_2$, as
noted by Braine {\it et al.} (2001) in their survey. Tidal dwarfs are
characterised by a much larger ($\sim$100$\times$) of CO luminosity compared
to other dwarf galaxies of comparable optical luminosity. Because of the
relative proximity of these objects, they can serve as testing labs for our
understanding of the galaxy formation process, albeit different from that in the
early Universe.

\subsection{\textit{\textbf{Polar ring galaxies and ring galaxies}}}
\label{sec:rings}

While the tidal dwarfs represent a transient phenomenon during galaxy mergers,
polar rings are expected to describe a rather steady-state situation when the
externally acquired material finds stable orbits in the plane orthogonal to the
equatorial plane of the galaxy. A number of preferentially early-type disks or
ellipticals show such rings lying in their polar planes, and, therefore,
kinematically distinct from their parent galaxies. The rings appear younger than
their host galaxies, which seem to be depleted of cold gas. Polar rings include
young stellar populations, apparently formed after the capture, and are gas-rich
(a few times 10$^9\,M_\odot$) and dusty (e.g., van Driel {\it et al.} 2002).  

Two main alternative explanations, based on merger kinematics, include the
accretion or capture of satellites from a nearly circular orbit, or the
collisional destruction and subsequent capture of a donor from a rather radial
orbit. Under special conditions, the accretion of cold gas by the host galaxy
can also result in the formation of polar rings. In the accretion scenario
(e.g., Schweizer {\it et al.} 1983; Reshetnikov \& Sotnikova 1997), about 10\%
of the donor disk gas is captured in a polar ring, in less than 1\,Gyr. The
collision scenario of galactic disks (e.g., Bekki 1998) involves orthogonally
oriented disks in a head-on, low-velocity collision.  Bournaud \& Combes (2003)
have tested both alternatives in numerical simulations and conclude that the
accretion scenario is more supported by observations, although one cannot
exclude either possibility. We note that numerical simulations of galaxy
formation at higher $z$ have demonstrated routinely the formation of polar-ring
galaxies as a product of merging and interaction within the computational box
(e.g., Romano-D\'iaz {\it et al.} 2009; Roskar {\it et al.} 2010).

An interesting aspect of polar rings is that they can provide information about
the DM halo shapes (e.g., Sackett \& Sparke 1990). This is possible because the
rings are long-lived and, therefore, have sufficient time to settle on regular
orbits in the polar plane of the host galaxy, which is determined by the
extended DM halo. The measured flat rotation curves of the rings point to the
existence of such haloes around parent galaxies. The self-gravity of the gas
settling in the rings is also a stabilising factor in their dynamics, otherwise,
differential precession would destroy them in a short orbital time. This
conclusion is a clear outcome of orbital analysis in a triaxial potential (e.g.,
Sparke 1986; Arnaboldi \& Sparke 1994) and numerical simulations of ring
formation and evolution (e.g., Bournaud \& Combes 2003). Depending on the mass
and orientation of polar rings, a number of stable and unstable equilibria are
possible. If the flattening of the DM halo can be constrained independently,
e.g., from lensing, one can obtain bounds on the halo (or overall mass)
triaxiality, merely assuming that the observed rings are stable. 

The Cartwheel galaxy represents another class of rings, most probably
originating from head-on collisions involving at least one gas-rich disk. Unlike
polar rings, these rings are not stationary, are frequently off-centred, and
represent an expanding density wave which triggers star formation (Lynds \&
Toomre 1976). The relative velocity of these collisions appears to be much
higher than those leading to polar rings (e.g., Horellou \& Combes 2001). It is
characteristic of a galaxy cluster environment, as we discuss below. 

\subsection{\textit{\textbf{`Mergers' in clusters: galaxy harassment}}}
\label{sec:haras}

In a galaxy cluster environment, high-velocity encounters between galaxies have
relative velocities $v_{\rm gal}$\,$\sim$\,$\sigma_{\rm
cl}$\,$\gg$\,$\sigma_{\rm gal}$, where $v_{\rm gal}$\,$\sim$\,10$^3\,{\rm
km\,s^{-1}}$ is the typical relative velocity of galaxies in clusters,
$\sigma_{\rm gal}$\,$\sim$\,100--200\,km\,s$^{-1}$ the intrinsic velocity
dispersion in galaxies, and $\sigma_{\rm cl}$ the velocity dispersion in
clusters. The corollary is that high-velocity encounters dominate in clusters
and are more frequent than in the field environment, and that direct collisions
are rare. Hence, one should expect that galaxies are more morphologically
disturbed in clusters, which makes them vulnerable to future encounters, as well
as to the effects of cluster tides. The encounter dynamics can be approximated
by the impulse approximation. The cumulative effect of the above processes is
called galaxy harassment. One of the main questions is whether galaxies form
differently in clusters or whether environmental processes, as described above,
make them different.

Observationally, in the local Universe, cluster galaxies appear redder than in
the field and are more spheroid-dominated. Tidal disturbances are common from
close passages. Local clusters possess no spiral disks, as reflected by their
morphology-density relation (e.g., Dressler 1980). In comparison, already at
$z$\,$\sim$\,0.4, clusters contain many small disturbed spirals, which are
replaced by spheroidals at the faint end of the galaxy luminosity function (LF) at
$z$\,=\,0. They contain a substantial population of blue, star-forming and
starbursting galaxies -- a reflection of the Butcher-Oemler (1978) effect.
Star-forming rings are much more frequent in clusters than two-armed spirals
(e.g., Oemler {\it et al.} 1997). Furthermore, in hierarchical models of
structure formation, the field galaxy influx into clusters peaks around
$z$\,$\sim$\,0.4 (e.g., Kauffmann 1995), and the star formation declines
abruptly around $z$\,$\sim$\,0.5, resulting in a large population of passive,
post-starburst galaxies (e.g., Dressler {\it et al.} 1999; Poggianti {\it et
al.} 1999). This quenching of star formation, measured, for example, by a
decline in the H$\alpha$ emission, in the cluster environment is a reflection of
the overall trend in galaxy evolution which ultimately leads to the formation of
the so-called red sequence. In other words, the star formation and morphological
evolution in clusters appear to decouple at lower redshifts (e.g., Couch {\it et
al.} 2001).

One way to understand the environmental effects on galaxy evolution is to
compare and contrast the evolution of disk galaxies in rich clusters with that
of field galaxies. Numerical simulations have revealed the details of galaxy
harassment in such over-dense fields (e.g., Moore {\it et al.} 1998), where disks
are subject to interactions with brighter and more massive neighbours -- a
process that injects energy and makes them vulnerable to the cluster tidal
field. Moreover, the gas stripping process acts efficiently in the central regions
of clusters (e.g., Tonnesen {\it et al.} 2007). These processes affect disk
galaxies almost exclusively -- dense ellipticals are basically immune to the
effects of harassment. Furthermore, the $r^{1/4}$ de Vaucouleurs surface
brightness profile appears robust and invariant to harassment, even when a
galaxy loses $\sim$40\% of its mass during an interaction (e.g., Aguilar \&
White 1986).

\subsection{\textit{\textbf{Mergers and star formation rates}}}
\label{sec:sfrates}

Below, we discuss how mergers influence the star formation rates (SFRs). Here we
emphasise the morphological evolution these disks experience, which results in
loss of the gaseous component, partly ablated and partly falling to the centre,
and in a dramatic conversion of disks into spheroidals. Late-type Sc--Sd disks
appear to be more affected by this process. In addition, the ram pressure by the
intracluster hot gas is $\sim$\,$\rho_{\rm hot}\sigma_{\rm cl}^2$, while the
restoring force is $\propto$\,$2\pi\Sigma_{\rm tot}$, where $\Sigma_{\rm tot}$
is the disk total surface density (gas\,+\,stars). The stripping occurs when the
ram pressure exceeds the restoring force, leading to the transformation to
lenticular galaxies. When the outer hot gas, which is only loosely bound to the
DM halo, is stripped, this is called strangulation.

Simulations have also demonstrated an agreement with observations both in
accounting for intermediate-age stellar population in these spheroidals and in
their shapes -- which appear prolate and flattened by velocity dispersion
anisotropy. 

We now turn to the issue of merger-induced star formation. There is no doubt
that mergers are responsible for the largest starbursts, e.g., the
ultra-luminous infrared galaxies (ULIRGS). The most intensely star-forming galaxies
are in the advanced stage of merging. But is the reverse true? In other words,
does the merger rate drive the SFR?

\begin{figure}[ht!!!]  
\begin{center}
\includegraphics[angle=0,scale=0.41]{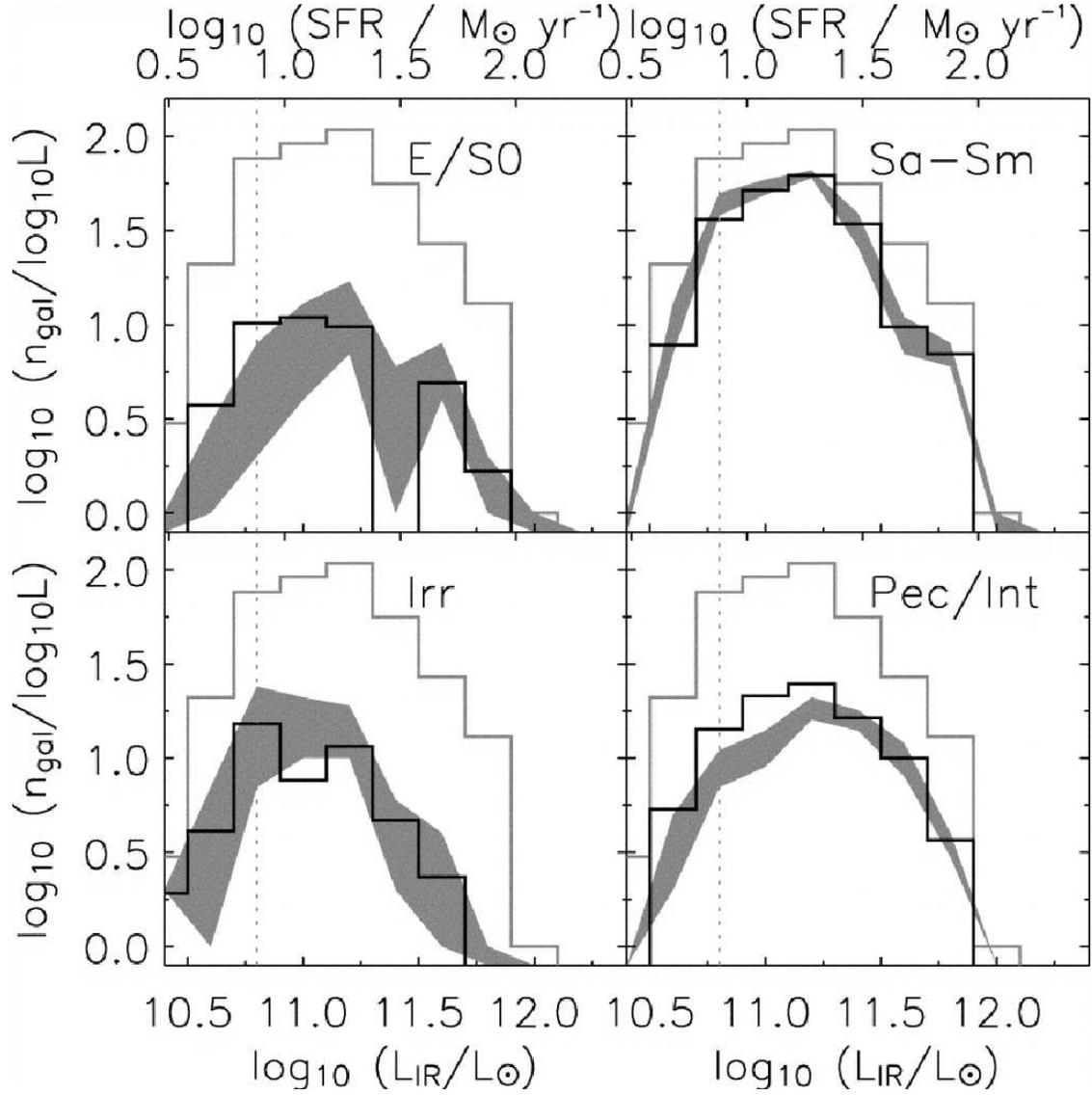}
\end{center}
\caption{\small Estimated 8--1000\,$\mu$m LF, split by morphological type for 397
galaxies at $z$\,$\sim$\,0.65--0.75 (see text for details). Only galaxies
detected at 24\,$\mu$m are shown, and no attempt to extrapolate to lower IR
luminosities has been made; the sample is grossly incomplete below
6\,$\times$\,10$^{10}\,L_\odot$ as denoted by the grey dotted line. In each
panel, the grey solid histogram shows the total IR LF. The shaded area shows the
IR LF split by galaxy type using GEMS-derived galaxy classifications, where the
extent of the shaded area explicitly shows the differences in IR LF given by the
three different classifiers. The black histogram shows the IR LF, averaged over
the three different classifiers and corrected to reproduce the increased
fraction of clearly interacting galaxies seen in GOODS-depth data (from Bell
{\it et al.} 2005).}
\label{fig:fig06}
\end{figure}

Since $z$\,$\sim$\,1, the cosmic SFR per unit comoving volume appears to
decrease by a factor of ten (e.g., Madau {\it et al.} 1998). Because over this
time period most of the star formation is associated with disk galaxies, the
inevitable conclusion is that the disks are shutting down their star-forming
activity. While in principle a number of processes can contribute to this
evolutionary trend, here we focus on the contribution from major mergers to
triggering the star formation. Because the LF of galaxies
at these redshifts is dominated by `normal' galaxies (Fig.~\ref{fig:fig06}), the
effect of mergers cannot be the principal one, as noted by Bell {\it et al.}
(2005). This conclusion is based on the analysis of the deep 24\,$\mu$m survey
made by the MIPS (Multiband Imaging Photometer for \textit{Spitzer}) Team (Rieke
{\it et al.} 2004), combined with the COMBO-17 redshift and SED survey (Wolf
{\it et al.} 2004) and the GEMS survey (Rix {\it et al.} 2004). The covered
redshift interval is 0.65--0.75 and includes about 1500 galaxies in the CDFS
(\textit{Chandra} Deep Field South). About 40\% of galaxies with stellar masses
$\gtorder$\,2\,$\times$\,10$^{10}\,M_\odot$ have been found to be undergoing a
period of elevated, intense star formation at $z$\,$\sim$\,0.7, while only
$\ltorder$\,1\% of similarly massive galaxies exhibit such star-forming activity
at $z$\,=\,0. Moreover, the IR LF and the SFR densities at
$z$\,$\sim$\,0.7 are dominated by morphologically undisturbed galaxies. More
than 50\% of the starbursting galaxies are spirals, but $\ltorder$\,30\% appear
to be strongly interacting. Hence the decline in the SFR is not `driven' by the
decline in the major merger rate. Rather, factors that do not strongly alter the
galaxy morphology are at play here, e.g., weak interactions and gas depletion.
Bell {\it et al.} (2005) argue that the selection procedure used should not
introduce any special bias against obscured starburst galaxies. 

To quantify the average enhancement in the SFR of major mergers between massive
$\gtorder$\,10$^{10}\,M_\odot$ galaxies, including pre- and post-mergers,
Robaina {\it et al.} (2009) used COMBO-17 and 24\,$\mu$m SFR from {\it Spitzer},
in tandem with the GEMS and STAGES {\it HST} surveys, for $z$\,$\sim$\,0.4--0.8.
Major interactions have been defined here as being resolved in {\it HST}
imaging, having a mass ratio of $\ge$\,1:4 based on the luminosity ratio, and
exhibiting clear signs of interaction. Major merger remnants have been
identified using a highly disturbed `train wreck' morphology, double nuclei, and
tidal tails of similar length, or spheroidal remnants with large-scale tidal
debris. Prominent disks with signs of merging, i.e., highly asymmetric spirals
or a single tidal tail, have been assumed to be minor mergers.

In addition, the enhancement in the SFR has been evaluated as a function of the
projected galaxy separation. While confirming that most starbursting galaxies
are in the process of merging, Robaina {\it et al.} (2009) have found that SFRs
in major-merger systems are only elevated by a factor of $\sim$1.8 compared to
those in non-interacting ones, when averaged over all interactions and all
stages of interaction (see also Li {\it et al.} 2008; Sommerville {\it et al.}
2008). The main enhancement is visible for close pairs with projected
separations of $\ltorder$\,40\,kpc. Overall, about 8\%\,$\pm$\,3\% of the total
star formation has been estimated to be directly triggered by major
interactions. This indeed confirms the conclusion of Bell {\it et al.} (2005),
mentioned above, that major mergers are not the dominant factor in building the
stellar mass at $z$\,$\ltorder$\,1, and, therefore, they are not responsible for
the decline in the SFR over this time.

We now discuss the input from numerical simulations which test the above
observational results on the relation between major mergers and SFRs. Di Matteo
{\it et al.} (2007) have focussed on this issue by modelling galaxy collisions
of all Hubble types while varying both bulge-to-disk mass ratios and $f_{\rm
gas}$. Direct and retrograde orbits have been used, and star formation in
interacting and merging galaxies has been compared to that in isolated galaxies.
The main outcome is that the retrograde orbits seem to produce more starbursts,
and the star formation efficiency is higher (in the sense of star formation per
unit mass). Moreover, these starbursts are essentially nuclear starbursts, from
the gas inflow triggered by the gravitational torques from asymmetries induced
by the tides. 

In a comprehensive study of unequal-mass mergers, Cox {\it et al.} (2008) have
quantified the effect of tidal forces on the star formation. Specifically, they
have focused on the effect of mass ratio, merging orbits and galaxy structure on
merger-driven starbursts. These kinematical and morphological parameters are of
prime importance for the main issue -- the relation between mergers and SFRs. It
was found that merger-induced star formation is a strong function of the merger
mass ratio, which spans over a factor of $\sim$23, being negligible for small
mass-ratio mergers -- a straightforward dependence on the tide's strength. An
additional parameter that is helpful to measure the induced star formation is
starburst efficiency -- the fraction of gas that is converted into stars over
the interaction time. The starburst efficiency was found to be insensitive to
the details of the feedback parameterisation from stellar evolution -- this is
very helpful because the feedback physics is sufficiently uncertain. Overall,
the burst efficiencies have been reduced compared to previous studies. 

However, while the burst efficiency for equal-mass mergers does not depend on
the merger orbital parameters, disk orientation, or the primary galaxy
properties, this differs for unequal-mass mergers.  Direct coplanar-orbit
mergers produce the most significant bursts at close passages. The other
important factor appears to be the mass distribution in the primary galaxy. For
example, a massive concentrated stellar bulge stabilises the disk and suppresses
the induced star formation. More gas driven above the threshold density for star
formation reduces the burst efficiency, and so is a more efficient feedback.  

In summary, recent studies of mergers agree that the evolution of the merger
rates after $z$\,$\ltorder$\,1 is not responsible for the overall decrease in
SFR in the Universe. The obvious direction for improving the numerical
simulations of merger-induced star formation requires incorporating them into
the cosmological context -- so far simulations deal with isolated pairs of
interacting and merging galaxies. This will allow accounting for the effect of
cold-gas accretion which is expected to compete with merger-induced galaxy
growth, especially at higher redshifts. Lastly, increases in the number of
particles, both collisional and collisionless, are important for modelling the
disk response during the interaction and merging periods. Small numbers of the
particles are known to destabilise the disks owing to increased noise.

%
%

\section{Disk growth: cold accretion}
\label{sec:accretion}

Figure~\ref{fig:fig02} displays two plausible modes of galaxy growth: galaxy
mergers and gas accretion. While we know that the frequency of interactions and
mergers increases steeply with redshift, the availability of unvirialised gas
increases as well. In Sections~\ref{sec:mergers} and \ref{sec:byprod} we
reviewed galaxy growth driven by mergers, mainly the growth of galactic disks.
Here we focus on the alternative process of galaxy growth via the accretion of
unvirialised baryons and DM.

\subsection{\textit{\textbf{The standard view and the new paradigm}}}
\label{sec:standard}

The ability of galaxies to grow by means of accretion has been known for some
time. The standard view has been that gas falling into a DM halo shocks to a
virial temperature, $T_{\rm vir}$, at around the halo virial radius, $R_{\rm
vir}$, and fills it up, remaining in a quasi-hydrostatic equilibrium with
$T_{\rm vir}$\,$\sim$\,10$^6\,(v_{\rm circ}/167\,{\rm km\,s^{-1}})^2$\,K. Hot,
virialised gas cools from the inside out, loses its pressure support and settles
into a centrifugally-supported disk (Rees \& Ostriker 1977; White \& Rees 1978;
Fall \& Efstathiou 1980).  

\begin{figure}[ht!!!]  
\begin{center}
\includegraphics[angle=0,scale=0.86]{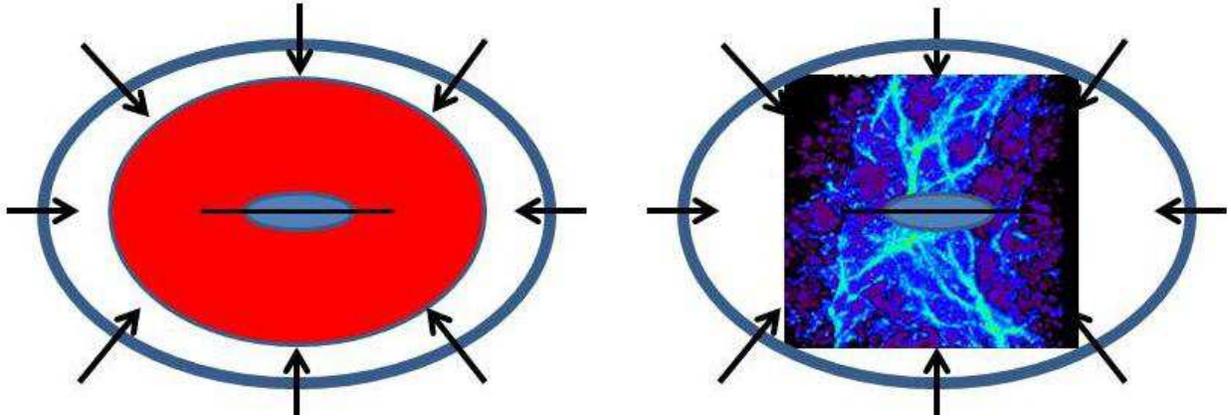}
\end{center}
\caption{\small Left: The standard view -- gas shocks at $R_{\rm vir}$, becomes
pressure-supported, then cools down and settles in a disk; Right: The new
paradigm -- some of the gas shocks, but the majority enters the DM halo in the
cold phase along the filaments feeding the disk growth.}
\label{fig:fig07}
\end{figure}

This view has been substantially modified now in that not all the gas is shocked
when it enters the halo. Instead, much of the gas is capable of entering the
halo along denser filaments and penetrating deeply -- this radical shift in
understanding has led to a new paradigm (Fig.~\ref{fig:fig07}).

\subsection{\textit{\textbf{Accretion shock?}}}
\label{sec:shock}

Birnboim \& Dekel (2003) have performed an idealised analytical study of gas
accretion on a spherical DM halo, assuming two alternatives: an adiabatic
equation of state and radiative cooling. The solution has been tested with a 1-D
hydrodynamic code. The incoming gas is not virialised and therefore its motion
is supersonic, creating favourable conditions for the virial shock --  its
existence and stability have been analysed. The crucial support for this shock
comes from the post-shock gas. If the virialised gas is adiabatic or its cooling
is inefficient, the shock-heated gas becomes subsonic (with respect to the
shock) and its support for the shock remains stable, with the shock positioned
at $\sim$\,$R_{\rm vir}$.  This is always the case for the adiabatic gas, which
is also stable against gravitational collapse (i.e., Jeans instability) if the
adiabatic index $\gamma$\,$>$\,4/3. Gravitationally unstable gas will collapse
to the centre, thus removing support from the shock, which will rapidly move
inwards. The gas can be treated as adiabatic when the radiative cooling
timescale is longer than the collapse timescale. The gravitational stability
condition is slightly modified for gas with radiative cooling to an effective
adiabatic index which includes the time derivatives, $\gamma_{\rm
eff}$\,$\equiv$\,$(d\,{\rm ln}\,P/dt)/(d\,{\rm ln}\,\rho/dt)$. Its critical
value, $\gamma_{\rm eff}$\,$>$\,$\gamma_{\rm
crit}$\,$\equiv$\,$2\gamma/(\gamma+2/3)$\,=\,10/7, is close to the adiabatic
case. Here $P$ and $\rho$ are thermal pressure and density in the gas. For a
monatomic gas with $\gamma$\,=\,5/3, this stability condition can be rewritten
as 
\begin{equation}
\frac{\rho_0 R_{\rm s}\Lambda(T_1)}{|u_0|^3} < 0.0126,
\label{eq:stable}
\end{equation} 
where $\rho_0$ is the preshock gas density, $u_0$ and $u_1$ are the pre- and
post-shock gas velocities, $T_1$\,$\propto$\,$u_0^2$ is the post-shock
temperature, $R_{\rm s}$ is the shock radius, and $\Lambda(T)$ is the cooling
function.

While the 1-D hydrodynamics is an obvious simplification of both the halo and
gas properties, its simplicity has a certain advantage in that it allows one to
follow the analytical solution closely. It shows that the adiabatic shock exists
always and gradually propagates outwards, coinciding with the virial radius,
$R_{\rm vir}$. On the other hand, for radiative cooling in the gas with
primordial composition, the shock exists only where the inflow encounters the
disk, initially. With the halo growth this shock also moves outwards and
stabilises around  $R_{\rm vir}$. In the following, we shall argue that the cold
inflow can join the disk smoothly, without being shock-heated -- i.e., that 1-D 
hydrodynamics cannot capture this solution.

This simple 1-D model predicts a critical value for the DM halo mass above which
the shock is supported at $R_{\rm vir}$. A weak dependence on the redshift of
halo virialisation exists, and a stronger dependence on the gas metallicity as
well (because it affects the cooling significantly). The mass range for the
critical halo mass appears to be $\sim$10$^{11}\,M_\odot$ for a primordial gas
composition, and $\sim$5\,$\times$\,10$^{11}\,M_\odot$ for about 0.05 of the
Solar metallicity. For this metallicity, Press-Schechter $M_*$ haloes will
generate stable shocks only by $z$\,$\sim$\,1.6. The corollary: virial shocks
will form only in the massive haloes mentioned above, at low redshifts. The
general condition for shock stability is that the cooling timescale of the
shocked gas should be longer than the compression timescale. 

A number of issues can complicate these conclusions: arbitrary triaxial halo
shapes, the interaction between the supersonic gas infall and the forming disk,
and the possible trapping of Ly$\alpha$ photons within the halo gas. The first
two issues can be resolved in terms of numerical simulations (see below).
The trapping of Ly$\alpha$ photons during gravitational collapse and its effects
on the fragmentation and related issues of proto-disk formation are under
investigation (e.g., Spaans \& Silk 2006; Latif {\it et al.} 2011).

Following the analytical/1-D hydro approach discussed above, numerical
simulations have been performed addressing a number of issues, e.g., what is the
maximum temperature of the gas entering the DM haloes? The standard view has
been that $T_{\rm max}$\,$\sim$\,$T_{\rm vir}$, but as we have already
discussed, not all of the gas is shocked to  $T_{\rm vir}$. It is helpful to
define two modes of accretion -- first, a cold mode with a  maximum temperature
of $T_{\rm max}$\,$<$\,$T_{\rm vir}$\,$\sim$\,10$^5$\,K, which was not
shock-heated, is distributed anisotropically, and follows filaments into the DM
halo. Second, a hot mode with  $T_{\rm max}$\,$>$\,$T_{\rm
vir}$\,$\sim$\,10$^5$\,K, which has been shock-heated at $\sim$\,$R_{\rm vir}$,
cools down while in a quasi-static gas halo and is accreted quasi-isotropically.
The filamentary inflow of the cold gas exhibits much lower entropy,
$\propto$\,$T/\rho^{2/3}$, compared to the shocked halo gas (e.g., Nagai \&
Kravtsov 2003).

Simulations reveal a more complex picture when some of the gas is accreted via
filaments, and some from cooling of the hot halo gas (Keres {\it et al.} 2005;
Dekel \& Birnboim 2006). Only about half of the gas follows the expected path of
accretion, which is heated to $T_{\rm vir}$, then cools down and participates in
the star formation. The rest of the gas stays much cooler at all times. The
overall emerging picture is that of a bi-modal evolution of accreting gas.
Specifically, the cold mode dominates in low-mass galaxies and DM haloes,
$M_{\rm gal}$\,$\ltorder$\,2\,$\times$\,10$^{10}\,M_\odot$ and $M_{\rm
h}$\,$\ltorder$\,2.5\,$\times$\,10$^{11}\,M_\odot$, respectively, and the hot
mode dominates in the higher-mass objects. As a result the cold mode is expected
to dominate at high redshift and in the low-density environment at low
redshifts. The hot mode will dominate in the high-density environment at low
$z$, such as in galaxy clusters.  

The quoted critical (baryonic) mass, $M_{\rm
gal}$\,$\sim$\,2\,$\times$\,10$^{10}\,M_\odot$, obtained from numerical
simulations is close to the observed characteristic mass for a shift in galaxy
properties, \\ $M_{\rm gal}$\,$\sim$\,3\,$\times$\,10$^{10}\,M_\odot$ (e.g.,
Kauffmann {\it et al.} 2004; Kannappan 2004), based on a complete sample of SDSS
galaxies. These studies focussed on the environmental dependence of various
parameters which describe the galaxies, such as morphology, stellar mass, SFR,
etc., quantifying the distribution of these properties with respect to galaxy
mass. For stellar masses above the critical $M_{\rm gal}$ no dependence on 
environmental factors has been found for the distribution of sizes and
concentrations at fixed stellar masses, whereas for less massive galaxies, the
trend has been detected for galaxies to be somewhat more concentrated and more
compact in denser regions. The star formation history has been found to be much
more sensitive to the environment: e.g., the relation between the
$\lambda$\,=\,4000\,\AA\ break, specific SFR (per unit stellar mass) and $M_{\rm
gal}$ -- with the same separator of $\sim$3\,$\times$\,10$^{10}\,M_\odot$. The
drop in the specific SFR for less massive galaxies was about a factor of ten
over the density interval used in the study, much stronger in comparison with
the more massive galaxies.  

An interesting corollary is the apparent similarity between the redshift
evolution of galaxy properties and their change as a function of a (local)
density. In retrospect, this result is almost a common wisdom, reflecting the
`sped up' evolution of galaxies in over-dense regions in the Universe.

Hence, compelling observational evidence exists that galaxies below the critical
mass of $M_{\rm gal}$\,$\sim$\,3\,$\times$\,10$^{10}\,M_\odot$ are much more
active in forming stars, have larger gas fractions, lower surface densities, and
exhibit late-type morphologies. More massive galaxies have old stellar
populations, supplemented with low gas fraction, higher surface densities, and
early Hubble types. This bimodal behaviour can have its origin in the
fundamental way the galaxies grow, or rather in the way their growth is limited.
If indeed a large fraction of the accreted gas is not heated to the virial
temperatures, it can join the disk and be converted into stars. We discuss the
associated processes in Section~\ref{sec:disk}. The shock-heated gas, on the
other hand, can also contribute to the star formation if its cooling time is
sufficiently short. So, some of the hot-mode gas in haloes somewhat smaller than
the critical one to sustain the shock (e.g., in low-density regions and/or at
higher $z$) will cool down if not subjected to feedback. This gas can
contribute to disk or spheroidal buildup over time.

However, as pointed out by Dekel \& Birnboim (2006), above the halo mass of
$M_{\rm h}$\,$\sim$\,10$^{12}\,M_\odot$, the cooling timescale for
$\sim$10$^{6-7}$\,K hot- and low-density gas becomes longer than the Hubble
time, and the gas, once heated to the virial temperature, will never cool down
and hence will not contribute to the disk growth in any direct way. This hot gas
which fills up the halo can also be subject to additional heating by AGN
feedback, both mechanical (throughout the halo) and radiative (at the base),
because of its high covering factor. For massive haloes, we expect that the only
real effect of this feedback can be in generating an outflow of the overheated
gas. 

While this is only a circumstantial observational argument in favour of the
existence of cold filamentary flows, it is nevertheless very intriguing by
bringing up the same bimodality in galaxy properties. The prime observational
issue of course remains the detection of these flows. Hot gaseous haloes have
been detected in X-rays around individual galaxies, groups and clusters of
galaxies (e.g., Crain {\it et al.} 2010a,b; Anderson \& Bregman 2011). There is
no direct observational evidence in favour of cold accretion flows, although
accretion of cold patchy gas has been observed (e.g., Rauch {\it et al.} 2011).
For higher-redshift galaxies, contradictory claims exist regarding the
possibility that diffuse Ly$\alpha$ emission around them comes from cold
accretion flows (e.g., Dijkstra \& Loeb 2009) and represents the cooling
radiation (Fardal \textit{et al.} 2001), or, alternatively, is the scattered light coming
from the H{\sc ii} regions (e.g., Furlanetto {\it et al.} 2005; Rauch {\it et
al.} 2011). Because of various reasons, including the low emissivity, absorption
against bright sources like quasi-stellar objects (QSOs) is the most promising
way to detect the cold accreting gas, especially in Ly$\alpha$. Van de Voort
{\it et al.} (2012) have argued that the high column density H{\sc i} absorption
detected at $z$\,$\sim$\,3 originates mostly in accreting gas with
$T$\,$\ltorder$\,3\,$\times$\,10$^5$\,K, based on numerical simulations. Rakic
{\it et al.} (2012) have interpreted results of the Keck Baryonic Structure
Survey of H{\sc i} Ly$\alpha$ absorption in the vicinity of star-forming
galaxies at $z$\,$\sim$\,2--2.8 as due to large-scale infall. It is not clear
whether the individual H{\sc i} absorbers can be attributed to cold accretion,
based only on their proximity to the galaxy and a low metallicity (e.g.,
Giavalisco {\it et al.} 2011). 

High-velocity clouds around the Milky Way galaxy (for a recent review see
Sancisi {\it et al.} 2008) can be closely related to the cold gas accretion
phenomenon discussed here. Sancisi {\it et al.} (2008) have detected accretion
rates of $\sim$0.2\,$M_\odot$ in H{\sc i} clouds, which is possibly a lower
limit for our Galaxy, that has a SFR\,$\sim$\,1\,$M_\odot\,{\rm yr^{-1}}$. There
are numerous ways in which cold gas clouds can form in the accreting matter
without being processed by the DM substructure. One such possibility involves
Rayleigh-Taylor instabilities in the halo-penetrating filaments (Keres \&
Hernquist 2009). But additional options exist as well. These clouds can 
subsequently be accreted by the galaxy and contribute to the ongoing steady star
formation there.

So why has cold accretion not been detected in a decisive manner so far? One can
bring up the similar situation and difficulties in detecting cold accreting gas
in AGN. At the same time, outflows are commonly detected both in AGN and in
starburst galaxies. The plausible explanation may be in small cross sections,
low emissivity and very high column densities along the line-of-sight due to the
gas accumulation in the `equatorial' plane.

\subsection{\textit{\textbf{Cold flows: redshift evolution}}}
\label{sec:redshift}

The coexistence of cold and hot modes of accretion can have interesting
implications for galaxy growth. These modes depend differently on the
environment, as well as on the feedback from stellar evolution and AGN.
Additional issues raised so far in the literature involve a plausible difference
of the associated initial mass functions (IMFs). 

Cosmic star formation exhibits a broad maximum at $z$\,$\gtorder$\,1 and a steep
decline below this redshift (e.g., Madau {\it et al.} 1996). This decline can be
associated with the decay of the cold accretion flows (e.g., Keres {\it et al.}
2005; Dekel \& Birnboim 2006). Below $z$\,$\sim$\,2, massive
$\sim$10$^{12}\,M_\odot$ haloes become typical, the shocks are stable around
$R_{\rm vir}$, and the cooling time of the shocked gas becomes too long,
effectively quenching the cold-mode accretion. This defines the critical
redshift, $z_{\rm crit}$\,$\sim$\,2--3. After $z_{\rm crit}$ the star formation
will be suppressed in massive haloes and especially in galaxy clusters. Under
these conditions, the observed bimodality in galaxy properties can find a
natural explanation. In terms of the prevailing colours of stellar populations,
which are determined by the stellar ages and SFRs, this shutdown of the cold
accretion mode will result in the relatively quick transformation of galaxies
with high SFRs.  This means that the origin of the red sequence can be traced
directly to the switching of the prevailing accretion mode, as noted by Dekel \&
Birnboim (2006). It would be a strong argument in favour of this picture if a
number of bi-modal correlations, such as the colour-magnitude, bulge-to-disk
ratio, or morphology-density ones, can be explained as corollaries of the cold
gas supply shutdown at various redshifts and environments. However, there is a
caveat: the bulge-to-disk ratio can be affected and even dominated strongly by
other processes (e.g., Combes {\it et al.} 1990; Raha {\it et al.} 1991;
Mart\'inez-Valpuesta {\it et al.} 2006). Other correlations may exhibit similar
trends.

Clearly, accretion flows that have been investigated for decades as the
mechanism to power AGN are capable of playing a substantial role in growing
galaxies embedded in DM haloes. Moreover, within the CDM framework, cold-mode
accretion forms naturally because of the low dispersion velocities in the gas
that has cooled down in the expanding Universe during the Dark Ages. The large
turnover radii, corresponding to the accretion radius, and the substantially
sub-Keplerian spin parameter $\lambda$, assure a strong dependence on the
accretor mass, i.e., DM halo mass. It also means that the cold-mode accretion
should dominate at high redshifts, and the hot mode should only pick up at low
redshifts, if at all. 

Keres {\it et al.} (2009a) have investigated the cosmological evolution of
smooth accretion flows using simulations with sufficient resolution to follow
up growth of galaxies in massive haloes only. Cold flows appear to dominate the 
global gas supply to galaxies basically at all times, especially in small
galaxies residing in $\ltorder$\,10$^{12}\,M_\odot$ haloes for $z$\,$\gtorder$\,1. At
these redshifts, the galaxy growth was found to be only a function of its mass.
At $z$\,$\ltorder$\,1, the cold accretion on smaller galaxies has decreased sharply.
These results have been confirmed by Brooks {\it et al.} (2009) -- for galaxies
up to $L^*$ the cold accretion fuels the star formation. Romano-D\'iaz {\it et
al.} (2008b) argued that late minor mergers with DM substructure ablate the
cold disk gas and quench the star formation there. Cold accretion dominated
growth has also been inferred in high-resolution simulations of galaxies at redshifts 
$z$\,$\gtorder$\,6 (Romano-D\'iaz {\it et al.} 2012).

Moreover, the total gas accretion rate has been found to peak at $z$\,$\sim$\,3
and to exhibit a broad maximum between $z$\,$\sim$\,2--4, the same as the cold
accretion. Hot accretion which consists of a shocked virialised gas that is able
to cool down over relative short time has been found to {\it contribute little}
over time, except lately, for $z$\,$\ltorder$\,1, after peaking at
$z$\,$\sim$\,1.5. Mergers become globally important only after $z$\,$\sim$\,1.
Finally, the SFR has been estimated to correlate with the smooth accretion rate
and to be about a factor two of the Madau diagram.

So, according to Keres \textit{et al.} (2009a), galaxies grow via the accretion
of cold and never-shocked gas, while the contribution of the hot mode of cooling
shock-heated virialised gas is not important at all. This is a dramatic
turnaround and a {\it paradigm shift} with respect to the standard picture
described in Section~\ref{sec:standard}. If verified, the implications of this
are broad: it is the cold mode of accretion that drives the star formation in
galaxies. However, taken at face value, this star formation will lead to fast
conversion of gas into stars -- an overproduction of the stellar mass already at
an early time, if the feedback from stellar evolution and AGN is not efficient
enough. In short, a mechanism to suppress the star formation is necessary. (This
is discussed in Section~\ref{sec:feedback}). Another corollary is the possible
shock at the inflow-disk interface. Is it avoidable? (See
Section~\ref{sec:phase} for more options.) Is it observable? Lastly, the
dominant cold mode of accretion must be incorporated into the prescriptions for
semi-analytic models.

\subsection{\textit{\textbf{Cold flows: between the virial radius and the disk}}}
\label{sec:disk}

Understanding the kinematics and dynamics of the cold flow which penetrates the
DM halo and is not shocked to virial temperatures is crucial in order to
estimate the flow's contribution to disk growth. As the filaments penetrate deep
into the halo, their temperature stays approximately constant, because of the
efficient cooling, and they are compressed by the surrounding hot halo gas, if
it exists. The efficiency of this inflow contribution to the disk growth process
is unclear at present. In principle, it can be expected to depend on at least
two parameters: the angular momentum in the cold gas and the shape of the
background gravitational potential. These will determine the prevailing
trajectories within the DM haloes and to some degree the amount of dissipation
in the infalling gas. As a result, we shall be able to estimate the infall
timescale (which will be longer than the free-fall time within the DM halo) and
the way this gas joins the growing disk, by smoothly merging or experiencing a
strong shock. In the former case, the infall energy of the gas is transformed
into rotational energy. In the latter, part of the infall energy will be
radiated away sufficiently close to the shocked interface.

In smaller haloes and especially at higher redshifts, the virial shock is not
sustainable at $\sim$\,$R_{\rm vir}$, and the forming galactic disk can be
directly affected by the deposition of matter, linear and angular momenta, and
energy by the inflow of the cold gas. How much dissipation is involved in this
process? Is the local, i.e., inflow-disk interface, shock-unavoidable?  

The 1-D case discussed above is not representative here, as the shock is
unavoidable (if the cold inflow exists) and the angular momentum plays no role.
Based on 3-D numerical simulations, Dekel \& Birnboim (2006) argue that cold
streams intersecting among themselves and with the forming disk will trigger
starbursts, characterised by the most common mode of star formation in the
Universe. The strength of the starburst will determine whether the disk will
continue to grow relatively quiescently or whether the process will contribute
to the spheroidal component. 

\begin{figure}[ht!!!!!!!!!!]  
\begin{center}
\includegraphics[angle=0,scale=0.86]{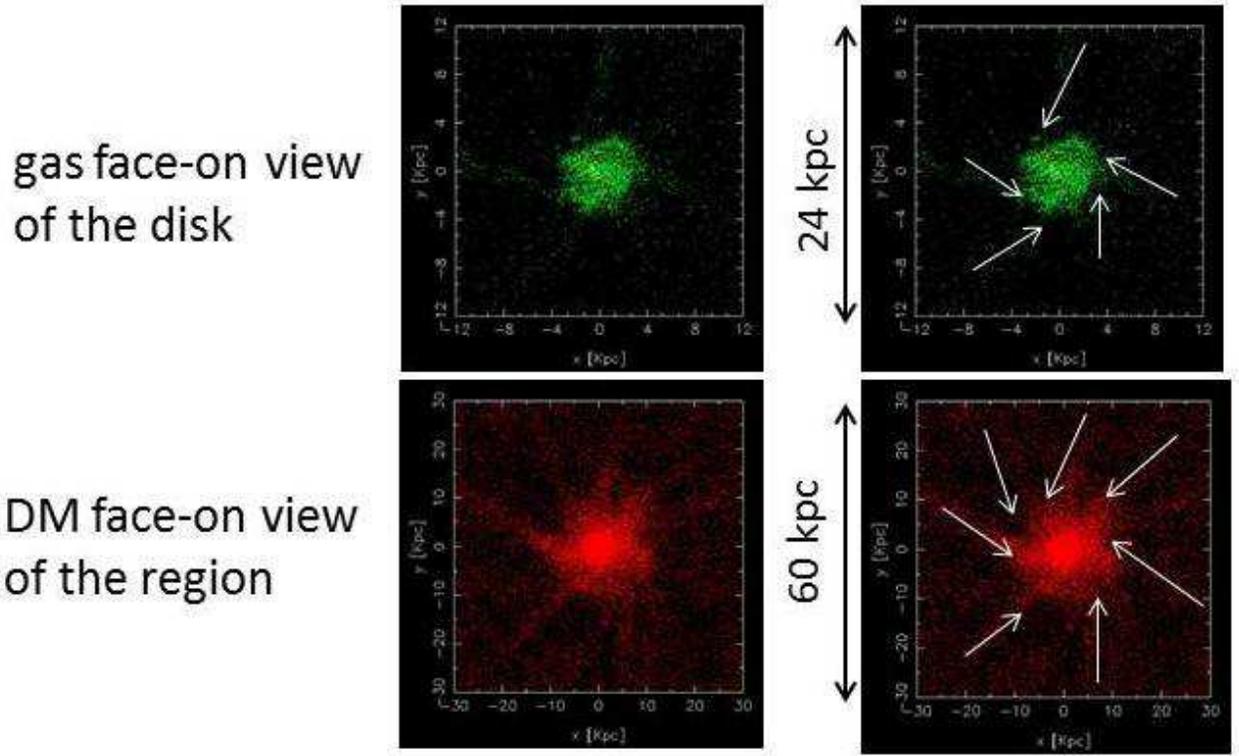}
\end{center}
\caption{\small `Cat's cradle' morphology: face-on view of the gas disk (upper frames)
and the extended DM regions (lower frames) showing the cold gas inflow joining
the disk. The white arrows (right frames) underline the DM filaments and the
associated gas inflow. Note that the gas streamlines join the disk at tangent
angles which preclude strong shocks from forming and rather assure a smooth
unshocked transition flow (from Heller \textit{et al.} 2007b).}
\label{fig:fig08}
\end{figure}

Most cosmological simulations lack the necessary resolution to investigate the
inflow-disk interface. The simplest way to circumvent this is to reproduce the
thermal histories of the gas particles. Brooks \textit{et al.} (2009) found that most of
the gas joins the disk unshocked in a smooth accreting component, opposite to
the gas accreted with the substructure, i.e., clumpy gas. The only exception is
the disk evolution in the most massive halo, well above $L^*$. For this halo,
the SFR is not exactly balanced by the accretion rate onto the halo, as a
substantial delay in star formation occurs due to the prolonged cooling time of
the shocked gas. 

Heller {\it et al.} (2007b) have shown that the cold gas filaments can smoothly
join the outer disk, being deflected from the disk rotation axis by the
centrifugal barrier -- no standing shock has been detected there. In this case
the infall kinetic energy is converted into rotational energy. Interestingly,
the gas filaments are supported by the DM filaments in a configuration which
resembles the `cat's cradle' -- a small amorphous disk fuelled via nearly radial
string patterns (e.g., Fig.~\ref{fig:fig08}).

If the inflow-disk interface shock does not exist or is sufficiently mild, what
additional signatures of a recent accretion can be expected deep inside the host
haloes?  For a number of reasons discussed above and in
Section~\ref{sec:angmom}, most probably the gas has a non-negligible amount of
angular momentum and will settle in some `equatorial' plane outside the growing
stellar disk. However, the orientation of this plane can differ substantially
with respect to the stellar disk plane. This will lead to the formation of
inclined and polar rings, warps, etc. Indeed, numerical simulations of such
disks in a cosmological setting have demonstrated the formation of rings and
warps, as a rule rather than an exception (e.g., Romano-D\'iaz {\it et al.}
2009; Roskar \textit{et al.} 2010; Stewart {\it et al.} 2011). Specifically,
Romano-D\'iaz {\it et al.} (2009) have demonstrated that the mutual orientation of the
rotation axis of the disk, DM halo, and the accreting gas fluctuate dramatically
over time, even during the quiescent periods of evolution (see their Fig.~19). 

Dekel {\it et al.} (2009b) have shown that galaxies of
$\sim$10$^{11}\,{M_\odot}$ at $z$\,$\sim$\,2--3 -- at the peak of SFR in the
Universe -- have been fed by cold accretion streams, rather than by mergers.
About 1/3 of the stream gas mass has been found in clumps, leading to mergers of
$\gtorder$\,1/10 mass ratio; the rest in the smooth streams. The deep
penetration of cold streams happened even in DM haloes of 
$>$\,10$^{12}\,{M_\odot}$ which are above the critical mass for virial shock
heating (Section~\ref{sec:shock}). Dekel \& Birnboim (2006) have noted that the
cold gas streams are supported by DM streams whose characteristic width is
smaller than $R_{\rm vir}$, and which are denser than the diffuse halo material.
They cross the shock basically staying isothermal because of the short cooling
distance. We return to the issue of penetrating streams in
Section~\ref{sec:highz} on high-$z$ galaxies.

There is a possibility that the extended XUV disks detected by \textit{GALEX}
(\textit{Galaxy Evolution Explorer}), whose population can reach $\sim$20\%
locally, have their origin in accretion flows (e.g., Lemonias \textit{et al.}
2011; Stewart \textit{et al.} 2011). A strong argument in favour of such a
scenario comes from the observations of such disks around massive early-type
galaxies. Moreover, there is no indication that XUV disks prefer
tidally-distorted disk galaxies, so they cannot originate as a result of a close
passage or a merger event. 

In the presence of a disk, one would expect that the gas accretion will join its
outer parts, at least the majority of the inflow, as discussed above. The
low-$j$ material that can come closer to the rotation axis would be accreted
earlier and such orbits would be depopulated quickly.

\subsection{\textit{\textbf{Cold accretion flows in the phase space}}}
\label{sec:phase}

The phase space provides the maximum information about filamentary cold flows.
Even 2-D phase space allows for a clear display of the accretion flows. It is
especially suitable in order to follow up the phase mixing and violent
relaxation processes discussed in Section~\ref{sec:phasem}. Various
complementary presentation options exist here, such as using $R-v_{\rm R}$,
$r-v_{\rm circ}$ and/or $r-\sigma$, where $R$ and $r$ correspond to the
spherical and cylindrical radii, and $v_{\rm R}$, $v_\phi$ and $\sigma$ to the
radial and azimuthal velocities and to the dispersion velocity, respectively.

\begin{figure}[ht!!!]  
\centerline{
\includegraphics[angle=0,scale=1.0]{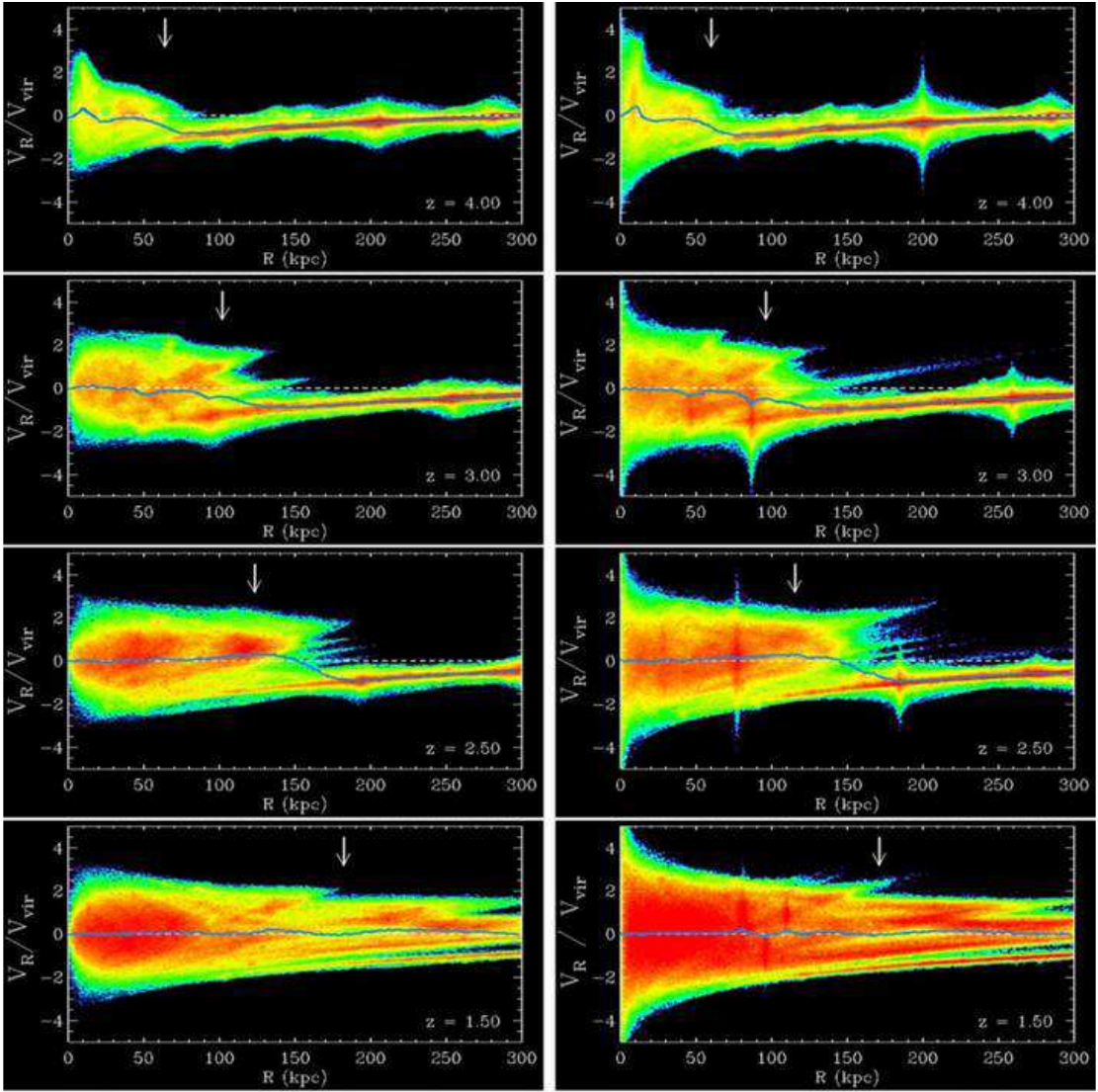}}
\caption{\small Phase space evolution of a DM halo without (left) and with (right)
baryons in the $R-v_{\rm R}$ plane, run from identical initial conditions. The
halo has been projected to collapse by $z$\,$\sim$\,1.3 with a mass of
$\sim$10$^{12}\,h^{-1}M_\odot$, based on the top-hat model. The epoch shown
here corresponds to intensive merger activity and to the cold accretion growth
in these simulations. The corresponding redshifts,
$z$\,$\sim$\,4\,$\rightarrow$\,1.5, are shown in the lower-right corners of each
frame. Note the appearance of `fingers' and `shell' structure inside and outside
of the halo -- much more pronounced in the presence of baryons. The inflow
containing both substructure and a smooth component is clearly visible as a
stream penetrating deeply inside the DM halo, especially in the right-hand
frames. The shape of the denser region is `smashed' against the $v_{\rm R}$-axis
in the presence of baryons and has a convex shape in the pure DM case. The
colours correspond to the DM volume density. The vertical arrows display
instantaneous value of $R_{\rm vir}$, the dashed white line shows $v_{\rm
R}$\,=\,0, and the solid blue line --  mass-averaged $v_{\rm R}$ at each $R$.
The velocity axis is normalised by $v_{\rm circ}$ -- the circular velocity at
$R_{\rm vir}$ (from Romano-D\'iaz {\it et al.} 2009).}
\label{fig:fig09}
\vspace{-1.7cm}
\end{figure}

Comparison of the evolution of pure DM and DM$+$baryon models in the $R-v_{\rm
R}$ plane reveals the effect of baryon inflow on the kinematics of the DM halo
(Fig.~\ref{fig:fig09} and Romano-D\'iaz {\it et al.} 2009). The mass-averaged
radial velocities are negative outside  $R_{\rm vir}$ and lie below the $v_{\rm
R}$\,=\,0 line, and change to positive inside the halo, initially. At later
times, the mass-averaged velocity is zero, as the inner halo reaches its virial
equilibrium.      Both major and minor mergers (substructure) can easily be
distinguished by the vertical spikes,  and are much more prominent, by a factor
of $\sim$2, for models with baryons, before the tidal disruption. Moreover, the
smooth accretion can be well separated from the substructure. The width of the
inflowing stream, which represents the mass accretion flux, declines with time,
while that of the rebounding material increases. In the process of tidal
disruption, inclined `fingers' form, again more prominent in the presence of
baryons. The subsequently forming `shell' structure reveals the insufficient
mixing of merger remnants in the form of a $R-v_{\rm R}$ correlation --
`streamers', which appear to be long-lived. Streamers formed after
$z$\,$\sim$\,1 largely survive to $z$\,=\,0. The phase space also delineates the
kinematical differences of the inner DM haloes in these models: note the outline
of $v_{\rm R}(R)$ at small radii. This effect can be explained in terms of the
gravitational potential shape there, which represents a more
centrally-concentrated object. 

So the phase-space analysis shows that DM haloes, while reaching virial
equilibrium, are far from relaxed in other aspects. Streamers are probably the
best example of this inefficient relaxation, and are strengthened by the
presence of baryons. The degree of relaxation in DM haloes can be further
quantified using the smoothing kernel technique (Romano-D\'iaz {\it et al.}
2009). This procedure allows us to estimate the contribution of the {\it excess}
DM mass fraction associated with density enhancements (i.e., subhaloes, tidal
tails, and streamers) above some smoothed reference density which is
time-adjusted. This excess mass fraction in the substructure becomes more
prominent with time.

A complementary option to study the buildup of DM haloes is in the $r-v_{\rm
circ}$ plane. In this plane, the halo kinematics is much more symmetric with
respect to the $v_{\rm circ} = 0$ line. Mergers disrupt this symmetry which is
quickly restored. Both major and minor mergers are easily traced in such a
diagram. The high degree of symmetry between the number of prograde- and
retrograde-circulating particles is very important in order to understand the
dynamical state of DM haloes and growing stellar disks, especially the disk-halo
resonant and non-resonant interactions which ultimately affect the disk ability
to channel baryons toward the centre (e.g., Shlosman 2011). One note of caution:
at higher redshifts, the haloes appear substantially triaxial in the range of
radii, and hence $v_{\rm circ}$ provides a bad approximation to the mass
enclosed within $r$. The overall symmetry in the $r-v_{\rm circ}$ diagram (e.g.,
Romano-D\'iaz {\it et al.} 2009) confirms that there is very little {\it net}
circulation of the DM within the halo. The tumbling of the halo figure is also
found to be negligible -- the halo appears to be orientated along the main
filament which feeds its growth.

%
%

\section{Disk growth: feedback}
\label{sec:feedback}

Why is it that such a small fraction of baryons has been converted into stars
over the Hubble time? Strong arguments exist, as we have discussed in the
previous sections, that an additional process which lowers the efficiency of
conversion of gas into stars must be taken into account when considering the 
cosmological evolution of galaxies. High-resolution numerical simulations have
demonstrated the necessity for this process.  Energy, momentum and mass
deposition from stellar and AGN evolution can have a profound effect on the
state of the gas within $R_{\rm vir}$, when one considers a simple spherical
geometry and their isotropic deposition. Whether this indeed happens when the
geometry becomes more complex must be verified. This should be performed by
including the by-products of any mechanical feedback on various spatial scales:
stellar, AGN and galactic winds, as well as radiative feedback from these
objects and from the cosmological background radiation.

A compelling example can be found in the comparison study of disk evolution in a
cosmological setting, with and without feedback, (Section~\ref{sec:catast}; see
also Robertson {\it et al.} 2004), where in the absence of feedback the gas
quickly violates the Toomre criterion and fragments. Moreover, without feedback
there is an overproduction of metals, especially in small galaxies. Even more
revealing is the overcooling problem (Section~\ref{sec:catast}) which leads
directly to the angular momentum catastrophe and involves gas which is bound to
the DM substructures, cools down and falls to the centre of the substructure.
The latter spirals in to the inner parts of the parent DM halo losing its $J$
via dynamical friction (e.g., Maller \& Dekel 2002, and Fig.~\ref{fig:fig10}).
The gas, therefore, hitchhikes to the bottom of the potential well without being
ablated and is deposited there when the substructure is dissolved by the tidal
forces, and possibly contributes to the growth of a (classical) bulge.

\begin{figure}[ht!!!]  
\begin{center}
\includegraphics[angle=0,scale=0.82]{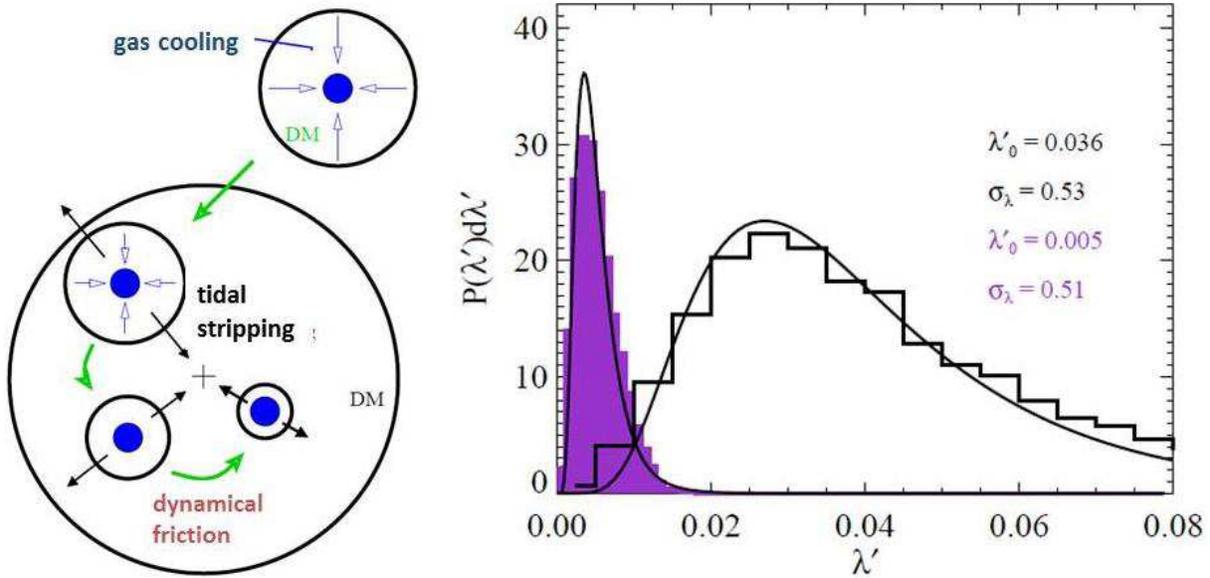}
\end{center}
\caption{Sketch of the overcooling problem (from Maller \& Dekel 2002). Left:
gas cools down and falls toward the centre of a satellite galaxy which is
merging with the main DM halo. Dynamical friction accelerates the process. The
cold gas hitchhikes with the sinking satellite and is losing its orbital $J$ but
is immune to stripping until the satellite falls to the centre of the parent
halo and dissolves there. So gas overcooling leads to the $J$ catastrophe
(Section~\ref{sec:angmom}). Right: The effect of overcooling on the spin
distribution of baryons compared to that of the DM. Baryons (filled curve)
being tightly bound at the satellite centre spiral in with the satellite toward
the inner parent halo and lose most of $J$. The figure shows that the $J$
deficiency in baryons is almost a factor of ten.}
\label{fig:fig10}
\end{figure}

Weak feedback has been tested, confirming that such models overproduce the
observed baryonic mass function for galaxies, especially for the most and least
massive objects (e.g., Keres {\it et al.} 2009b). Various `preventive' feedbacks
have been used to remove this discrepancy, such as a highly efficient AGN `radio
mode' -- this did not improve the fit at the high-mass end. Furthermore, it has
been impossible to recover the population of massive quiescent galaxies. The
overall conclusion is that the solution should come from a more `sophisticated'
feedback mechanism which substantially suppresses the star formation in a
fraction of galaxies, which increases with mass, while leaving the star
formation in the remaining galaxies unchanged. 

\subsection{\textit{\textbf{Bulgeless disks}}}
\label{sec:bulge}

\begin{figure}[ht!!!]  
\begin{center}
\includegraphics[angle=0,scale=0.8]{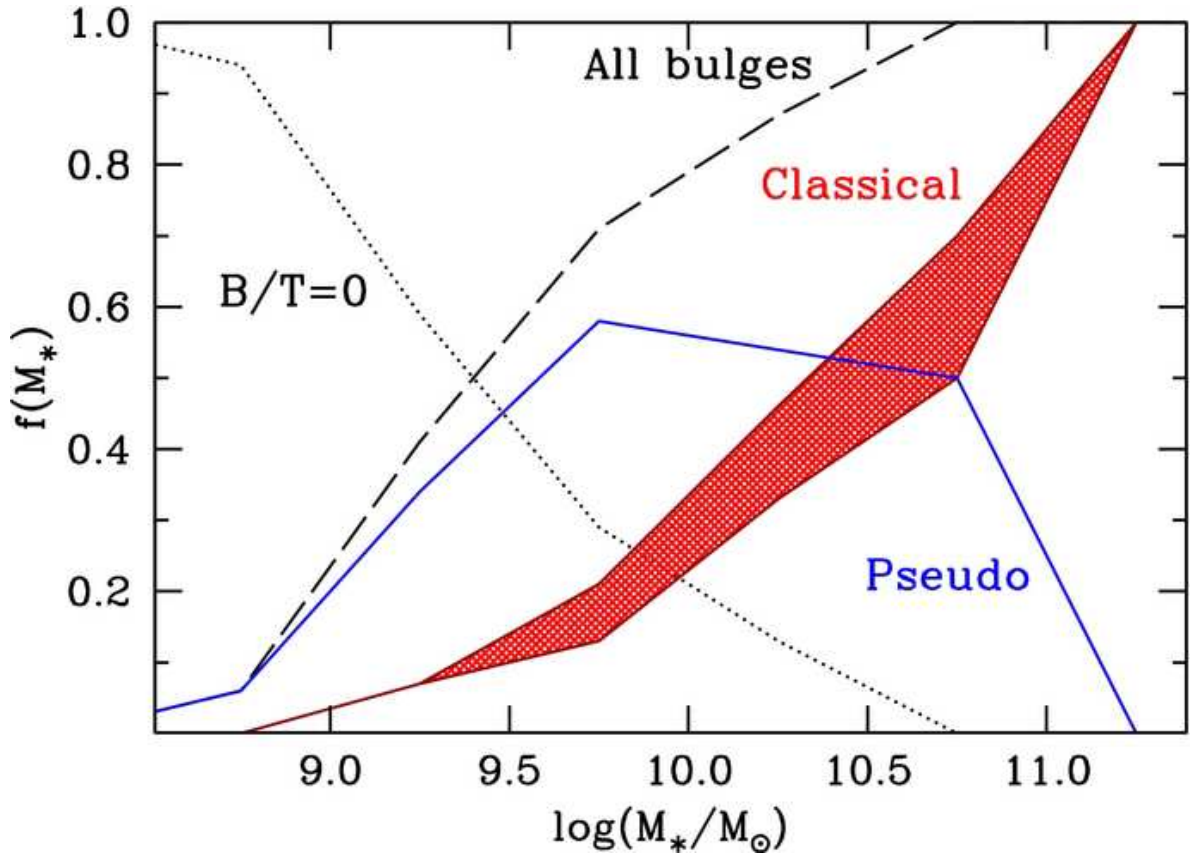}
\end{center}
\caption{Relative numbers of galaxies with classical bulges and elliptical
galaxies (red lines), galaxies with disky bulges (i.e., pseudobulges) (blue
line), and all disky bulges (black dotted line) as a function of galaxy stellar
mass (from Fisher \& Drory 2011).}
\label{fig:fig11}
\end{figure}

Galactic bulges may provide a testing ground for our understanding of angular
momentum redistribution in a forming disk and for various mechanisms that
regulate the star formation. An unexpectedly large fraction, $\sim$76\%, of
massive, $\gtorder$\,10$^{10}\,M_\odot$ galactic disks can be fit with a Sersic
index $n$\,$\ltorder$\,2 and $\sim$69\% have a $B/T$ mass ratio of
$\ltorder$\,0.2, both in barred and unbarred galaxies (Weinzirl {\it et al.}
2009). This result has been obtained from ground-based imaging of a local sample
of 182 galaxies that covers all Hubble types, S0/a--Sm, in the optical and the
NIR. Disks inclined by $i$\,$\geq$\,70$^\circ$ have been excluded.
Two-dimensional bulge-disk-bar decomposition has been performed on $H$-band
images. This result is in stark contrast with predictions based on the numerical
modelling of such objects -- only disks that did not experience a major merger
event since $z$\,$\sim$\,2 are expected to have such low $B/T$ ratios. This
number is more than an order of magnitude smaller than the observed fraction of
low $B/T$ disks. Taken at face value, this result points to a serious
contradiction between the observationally detected trend and our theoretical
understanding, and specifically to the role of major mergers in bulge formation.
  
This contradiction becomes even stronger in the recent study of the bulge
population within a sphere of 11\,Mpc radius using {\it Spitzer} 3.6\,$\mu$m and
\textit{HST} data (Fisher \& Drory 2011). The dominant galaxy type in the local
Universe has been found to possess pure disk properties, i.e., having a disky
bulge (Section~\ref{sec:drywet}) or being bulgeless. The fraction of these
galaxies exceeds 80\% of the number of galaxies with a stellar mass of
$\gtorder$\,10$^9\,M_\odot$. Classical bulges and elliptical galaxies account
for about 25\% and disks for about 75\% of the total stellar mass within
11\,Mpc. Moreover, $\sim$2/3 of the star formation in these galaxies occurs in
galaxies with disky bulges. Figure~\ref{fig:fig11} delineates the fractions of
various bulge types found as a function of stellar mass. Below
$\sim$3\,$\times$\,10$^9\,M_\odot$, galaxies are most likely to be bulgeless.
If disky bulges do not originate in mergers (as we have discussed in
Section~\ref{sec:drywet}), the results of this analysis reinforce those of
Weinzirl {\it et al.} (2009) in the most dramatic way.

These results reinforce the opinion that additional physical processes are
needed to explain much less massive spheroidal components in disk galaxies.

\subsection{\textit{\textbf{What type of feedback?}}}
\label{sec:type}

Feedback, or `feed back', is typically defined as a cause-and-response chain of
events that forms a (causal) loop. The loop can be stable or unstable. In simple
terms, in the former case, the process repeats itself and we define it as a
self-regulated one, and the feedback is negative. In the latter case, the
initial conditions for the process are modified and the system must find a new
stable loop or disintegrate. In this case the feedback is positive.

In the context of disk evolution, positive or negative feedback can increase or
decrease the SFR in a disk. The list of relevant mechanisms is not short and
involves radiation (stellar and AGN), winds (stellar, AGN, galactic), AGN jets
and backflow cocoons of large-scale relativistic jets, turbulence, SN, bubbles
and superbubbles, spiral density waves, stellar and gaseous bars, and cold
accretion along cosmological filaments. All the above mechanisms can induce both
positive and negative feedback on the SFR. The disk response points to a
self-regulation achieved with respect to the star formation process.  

As we have discussed in Section~\ref{sec:angmom}, the absence of any feedback in
disk evolution leads to an overestimate of the spheroidal stellar component.
Hence, additional processes must regulate the fragmentation and star formation
in disk galaxies. The cause for this can lie beyond the disk itself but fully
understanding the disk response is crucial for predicting the feedback loop.

The physics of the various types of feedback is very complex and in many cases,
e.g., AGN feedback, not yet fully understood. In some cases, e.g.,
turbulence, our understanding is mostly empirical. Moreover, many relevant
processes, like star formation and turbulence, are well below the resolution
limit of current numerical models of galaxy evolution in the cosmological
context, and are therefore treated on the subgrid level, i.e., purely
phenomenologically. We first discuss the main feedback mechanisms, and then, in
Section~\ref{sec:feedsf}, survey the star formation and feedback algorithms
used in numerical simulations.

\subsubsection{Supernova feedback}
\label{sec:supern}

The simplest to model is {\it gravitational feedback}, which involves
collisional heating of the gas. It occurs during gas-rich mergers. The
collision-induced shocks help to virialise gas -- a process whose efficiency
depends on the orbits of the merger components. Most affected appears to be the
low-density gas in the outer parts of the galaxy. The fate of the gas, however,
is to remain largely bound to the merger product. 

The SN feedback or yield is probably the most analysed in the literature. We
follow estimates by Dekel \& Woo (2003). Kinetic energies are similar for
type\,Ia and type\,II SNe, although Ia are irrelevant here as they deposit their
energy away from the star formation sites. A number of SN explosions resulting
from a steady SFR of ${\dot M}_*$ from gas mass $M_{\rm gas}$
embedded in the DM halo of $M_{\rm vir}$ can be estimated as
\begin{equation}
E_{\rm SN}\sim \epsilon\nu {\dot M}_*t_{\rm rad},
\label{eq:esn}
\end{equation} 
where $\epsilon$\,$\sim$\,10$^{51}$\,erg is the initial energy released by a
typical SN, $\nu$\,$\sim$\,5\,$\times$\,10$^{-36}\,{\rm g^{-1}}$ is the number
of SNe per unit mass of stars (for the Salpeter IMF, $\nu$\,$\sim$\,1 per
100\,$M_\odot$ of stars), and $t_{\rm rad}$ marks the end of the adiabatic phase
and the onset of radiative phase in the SN expansion. The SFR 
can be written in terms of the free-fall time as ${\dot
M}_*$\,$\sim$\,$M_*/t_{\rm ff}$. So $E_{\rm SN}$\,$\sim$\,$\epsilon\nu
M_*(t_{\rm rad}/t_{\rm ff})$. In the relevant range of gas temperatures, a few
$\times$\,10$^4$ to a few $\times$\,10$^5$\,K, the ratio $t_{\rm rad}/t_{\rm
ff}$\,$\sim$\,10$^{-2}$ is constant. In order to expel (unbind) the amount of
gas, $M_{\rm gas}$, from a galaxy which has virial velocity $v_{\rm vir}$, the
energy released by the SNe should be $E_{\rm SN}$\,$\gtorder$\,0.5\,$M_{\rm
gas}v_{\rm vir}^2$ -- the binding energy of this gas. For this to happen, the
velocity of the gas must exceed the critical velocity of $v_{\rm
crit}$\,$\sim$\,0.14\,$(\epsilon\nu M_*/M_{\rm
gas})^{1/2}$\,$\sim$\,100\,$(M_*/M_{\rm gas})^{1/2}$\,km\,s$^{-1}$. In other
words, the gas can be removed from DM haloes with virial velocities $v_{\rm
vir}$\,$\ltorder$\,$v_{\rm crit}$\,$\sim$\,100\,km\,s$^{-1}$, if a large
fraction, say $M_*$\,$\sim$\,$M_{\rm gas}$, of the gas is turned into stars.

The associated DM virial mass which supports the formation of $M_*$ of stellar
mass for such $v_{\rm vir}$ is $M_{\rm
vir}$\,$\sim$\,2\,$\times$\,10$^{11}\,M_\odot$, and the corresponding stellar
mass  $M_*$\,$\sim$\,3--4\,$\times$\,10$^{10}\,M_\odot$, in apparent agreement
with the bifurcation mass for bimodal evolution discussed in
Section~\ref{sec:shock}. So the SN feedback can reheat the ISM, push it into the
halo and even expel a large fraction of the gas from smaller galaxies. One SN
can in principle unbind a giant molecular cloud. The above estimate makes it
plausible that the gas can be driven out of a galaxy by the SN feedback, {\it if
a large fraction of the original gas has been converted into stars.} But SNe are
not very efficient in converting the stellar rest mass into kinetic energy
because $f_{\rm SN}$\,$\equiv$\,$\epsilon/100\,M_\odot
c^2$\,$\sim$\,2.8\,$\times$\,10$^{-6}$. While progress has been made in
high-resolution hydrodynamical simulations of SN shells sweeping up the ISM, the
details are not clear yet, especially how efficiently the energy is distributed
among the baryons.

\subsubsection{OB stellar winds and AGN feedback}
\label{sec:agn}

Stellar winds from OB stars are driven by radiation pressure in UV resonance
lines of CNO elements and inject about the same amount of kinetic energy over
their lifetime as SNe. The asymptotic velocities of these winds are
$\sim$2000--3000\,km\,s$^{-1}$. On the subgrid level, these winds from
individual stars have been incorporated into numerical simulations (Heller \&
Shlosman 1994). Understanding the formation of galactic winds under a combined
action of OB stars and SNe is more complicated. 

AGN winds originate within the central pc -- kpc region from the SMBH and are
energy- and/or momentum-driven. We limit our discussion here to sub-relativistic
winds in radio-quiet QSOs only and exclude the jets. The mechanisms responsible
for these winds are based on radiation power, or are hydromagnetically (MHD)
driven from the underlying accretion disk (e.g., Blandford \& Payne 1982;
Shlosman {\it et al.} 1985; Emmering {\it et al.} 1992; K\"onigl \& Kartje
1994). The former can be driven by absorption (scattering) in the resonance
lines  similar to those in OB stars, and by the dust opacity. The MHD winds can
feed on the rotational energy of the disk. Note that the MHD winds are much more
efficient in extracting the angular momentum from the underlying accretion
disks. Unlike the radiation-driven winds, they can reduce the mass accretion
rate substantially -- the mass outflow rate in an MHD wind can exceed the inflow
rate for some period of time. At the same time, magnetic torques can extract the
angular momentum from the disk {\it without} much of the outflow at all. This
cannot be achieved with the radiatively-driven winds -- in order to extract
angular momentum from the disk, they must maintain a very high mass flux in the
wind. Because this is beyond the scope of our review, we avoid discussing here
the magnetorotational instability (MRI) which can generate
turbulence and transport $J$ within the disk.

Outflows with clear signatures in UV and optical emission and absorption lines,
and in the radio, have been known for a long time. But reliable measurements of
a mass outflow rate has been obtained only recently, e.g., for a broad
absorption line QSO, SDSS\,J0318-0600, at $\sim$120\,${M_\odot\,{\rm
yr^{-1}}}$ (Dunn \textit{et al}. 2010). The kinetic luminosity for this object
has been estimated at $\sim$0.001 of its bolometric luminosity, at the lower
end of assessments available in the literature. Powerful molecular outflows from
AGN have been detected with a mechanical luminosity estimated from a few to 100\%
of the AGN bolometric luminosity (e.g., Reeves {\it et al.} 2009). This means
that they cannot be driven by SNe but require a higher efficiency associated
with radiatively-driven winds, or most probably with MHD winds. It also means
that such outflows will deplete the {\it in situ} reservoir of molecular gas in
a relatively short time. The molecular tori observed in type~2 AGN in the NIR
can represent the base of such molecular outflows (Elitzur \& Shlosman 2006). 

What is the AGN analogue of the SN feedback efficiency or yield, $f_{\rm SN}$,
estimated above? The total energy, $E_{\rm AGN}$, produced by an AGN depends on
the conversion factor, $\epsilon_{\rm AGN}$\,$\sim$\,0.1, of its accretion
energy, i.e., $E_{\rm AGN}$\,$\sim$\,$\epsilon_{\rm AGN}\eta M_\bullet c^2$,
where $\eta$\,$\sim$\,10$^{-3}-1$ is the conversion factor of the bolometric
luminosity, $L_{\rm bol}$, of the AGN into mechanical luminosity, and
$M_\bullet$ is the mass of the central SMBH. The exact value of $\eta$ is not
known, and it is expected to depend on the nature of the major contributor to
$L_{\rm bol}$, on the ratio of $L_{\rm bol}$ to the Eddington luminosity, and
possibly on additional parameters of the flow. We therefore leave $\eta$
unconstrained and note that in the epoch of galaxy formation at high redshift,
which can be characterised by plentiful fuel supply and high Eddington ratios
for the AGN, the value of $\eta$ can be as high as $\sim$1. Moreover, if we
assume for brevity that the $M_\bullet-\sigma$ relation is maintained at these
redshifts (quite unlikely!), $M_\bullet/M_*$\,$\sim$\,10$^{-3}$ and 
\begin{equation}
E_{\rm AGN}\sim 10^{-4}\left(\frac{\epsilon_{\rm AGN}}{0.1}\right)\eta M_* c^2,
\label{eq:eagn}
\end{equation} 
which means that the efficiency of the high-$z$ AGN is $f_{\rm
AGN}$\,$\sim$\,10$^{-4}\eta$. This estimate can be about 10--50 times higher
than $f_{\rm SN}$, but remains speculative.

The first serious attempt to account for AGN feedback in high-resolution
numerical simulations used purely thermal feedback, and no delivery or coupling
mechanisms were specified (Springel 2005). A fraction of the isotropic
bolometric luminosity of the AGN has been assumed to be deposited locally.
Momentum transfer has been ignored. Application of this feedback resulted in
limiting the growth of the SMBH and expulsion of the ISM from the galaxy.  Less
realistic but more detailed simulations have shown that the momentum transfer
dominates because of the very short cooling timescales and the inability to
retain thermal energy in the dense gas found in galactic centres (e.g., Ostriker
{\it et al.} 2010). This is also symptomatic of OB stellar winds and winds from
accretion disks with effective temperatures in the UV. 

So AGN can have a potentially dramatic effect on the ISM and IGM based on their
energy output. If this energy, momentum and mass around the SMBHs were
distributed in a highly symmetric fashion, this could ensure an efficient
coupling with baryons. However, besides direct evidence in the form of
collimated jet-ISM/IGM interaction in radio galaxies (both in X-rays and radio
bands) and in clusters of galaxies, additional evidence is scarce. Furthermore,
the relevant physics of AGN output deposition in baryons is still poorly
understood. {\it How exactly is the coupling to baryons achieved?} Even when the
feedback energy exceeds the gas binding energy, the gas can escape along the
preferential directions or via the fluidised bed-type phase transition,
dramatically reducing the feedback.

While models based on energy- and momentum-driven outflows have been proposed,
these are mostly phenomenological models. The detailed physics of energy and
momentum deposition is under investigation now, as well as attempts to translate
it into subgrid physics. The important issues involve the driving mechanisms for
the wind, their dependence on mass accretion flows, on Eddington ratio,
geometric beaming, and additional factors.

\subsubsection{Galactic winds}
\label{sec:galwind}

While at present there are still theoretical difficulties with almost any type
of wind from any object (e.g., star, accretion disk), these winds nevertheless
exist and there are clear incentives in working out the corollaries of this
existence. The presence of galactic outflows is supported by numerous
observations (e.g., Heckman 1994, and references therein). In most cases they
are driven by the SNe and winds from OB stars in galaxies that experience
starbursts. The contribution of  AGN is debatable at present. The driving by the
SNe and stellar winds occurs when ejecta of individual sources form a bubble of
hot gas, $\sim$10$^{7-8}$\,K, which expands due to the strong overpressure
down the steepest pressure gradient and enters the `blow-out' stage. Besides
the mass, energy and momentum injected by such winds, this is probably the main
way the highly-enriched material can be injected into the halo and further out
into the IGM. The winds appear inhomogeneous and carry embedded cold,
$\sim$10$^4$\,K clouds with them, so they represent a multiphase ISM. Their
velocities appear correlated with the galaxy stellar mass or its SFR (e.g.,
Martin 2005). 

The overall energetics of these bubbles and superbubbles points to the important
and even crucial role galactic winds play in galaxy evolution, from determining
the size of galactic disks to regulating the star formation process in galaxies,
and removing the overcooling problem discussed earlier. In this context, the
overcooling results from under-resolved ISM which will radiate away all the
thermal energy deposited by the feedback. The proper resolution must correctly
treat the merging of individual SN bubbles producing low-density superbubbles
which will remain hot for a prolonged period of time -- this resolution is not
yet achievable in cosmological or individual-galaxy numerical simulations. Only
when small volumes representing a region in the disk are modelled, these
processes can be followed. To summarise, any modelling of galaxy formation and
evolution must include the development of galactic outflows triggered either by
stellar or AGN feedback. 

To circumvent this lack of numerical resolution, the necessary physics can be
introduced on the subgrid level. We discuss four such algorithms: the constant
velocity, the delayed cooling, the blastwave, and the variable wind models.

\begin{itemize}

\item{\it \textbf{Constant wind model} (Springel \& Hernquist 2003; Springel 2005)}\\ 
This wind model is based on a multiphase ISM treatment. Two phases, cold and
hot, are not directly resolved but coexist in a single SPH particle. So
dynamically they cannot be separated. The two phases are followed above the
critical gas density which corresponds to the threshold, $\rho_{\rm SF}$, above
which the star formation is allowed to occur. The SNe directly heat the ambient
hot phase whose cooling timescale is long. This effectively modifies the
equation of state for the ISM. The cold phase is heated and evaporated via
thermal conduction from the hot phase. Mass-transfer equations between the two
ISM phases are solved analytically. The galactic wind is triggered by modifying
the behaviour of some gas particles into the `wind' particles. The wind
particles are not subject to hydrodynamical forces and experience the initial
kick from the SN. All wind particles have the same constant velocity and the
same mass-loading factor, $\beta_{\rm w}$\,$\equiv$\,$\dot M_{\rm w}/\dot M_{\rm
SF}$, where $\dot M_{\rm w}$ is the mass loss in the wind and $\dot M_{\rm SF}$
is the SFR. The observational constraints on the value of the loading factor
$\beta_{\rm w}$ are weak.

\item{\it\textbf{Delayed~cooling~wind~model} (Thacker \& Couchman 2000; Heller \textit{et al.} 2007b)}\\
The energy injection by the SN and OB stellar winds affects the fixed number of
neighbouring SPH particles. This process is resolved by at least five timesteps
and extends to $t_{\rm in}$\,$\sim$\,3\,$\times$\,10$^7$\,yr -- the feedback
timescale. Radiative cooling is disabled for these neighbours. The SN energy is
deposited in the form of thermal energy and converted into kinetic energy via an
equation of motion and using the energy thermalisation parameter. The affected
particles do not interact hydrodynamically over the time period, which depends
on the minimum of $t_{\rm in}$ and the time it takes for the wind particle to
move to an ambient density below some threshold density. 

\item{\it \textbf{Blastwave wind model} (Stinson \textit{et al.} 2006)}\\
This model is based on the adiabatic (Sedov-Taylor) and snowplough phases in the
SN expansion. The blastwave is generated by the collective explosion of many SNe
of type\,II. The maximum radius of the blastwave is given by the Chevalier
(1974) and McKee \& Ostriker (1977) formalism. Radiative cooling is disabled for
$R$\,$\leq$\,$R_{\rm blast}$. The timescale for the blastwave to reach this
radius is of the order of a timestep resolution. The blastwave wind model has
been efficient in moving the peak of the star formation to lower redshifts, well
beyond the last major merger, and has significantly reduced the SFR during
mergers due to the feedback in progenitors.

\item{\it \textbf{Variable wind model} (Choi \& Nagamine 2011)}\\
This model adopts the subgrid multiphase ISM. All the previous algorithms have
loaded galactic winds with star-forming particles and their close neighbours
only. In this model particles from low-density/high-temperature and
high-density/low-temperature are selected. The main parameters are chosen as the
wind load $\beta_{\rm w}$ defined above and the wind  velocity $v_{\rm w}$, and
both are constrained by observations.  Typically, observations express these two
parameters in terms of the host galaxy stellar mass, $M_*$, and the SFR. This
requires that simulations compute both parameters on the fly as the model is
running and not in the post-processing stage, which is challenging. In the
original version, the outer density of baryons in a galaxy is limited by
0.01\,$n_{\rm SF}$, where $n_{\rm SF}$\,$\sim$\,0.01--0.1\,cm$^{-3}$ (defined
above) is the threshold for star formation. The value of $n_{\rm SF}$ is based
on the translation of the threshold surface density in the Kennicutt-Schmidt
law, SFR\,$\sim$\,$\Sigma_{\rm gas}^\alpha$, where SFR is the disk surface
density of star formation, $\Sigma_{\rm gas}$ is the surface density of the
neutral gas, and $\alpha$\,$\sim$\,1--2, depending on the tracers used and on
the relevant linear scales. 

In the variable wind model, the wind velocity is calculated as a  fraction of
the escape speed from the host DM halo,  $v_{\rm w}$\,=\,$\zeta v_{\rm esc}$,
where $\zeta$\,=\,1.5 for momentum-driven and $\zeta$\,=\,1 for energy-driven
winds is a scaling factor. The SFR is determined from the empirical relation
with $v_{\rm esc}$, 
\begin{equation}
{\rm SFR} = 1.0\, \left(\frac{v_{\rm esc}}{130\,{\rm km\,s^{-1}}}\right)^3
\left(\frac{1+z}{4}\right)^{-3/2}\, M_\odot\,{\rm yr^{-1}},
\label{eq:sfr}
\end{equation}
which is consistent with observations (e.g., Martin 2005). Hence, the wind
velocity is an increasing  function of redshift and the SFR. The load factor,
$\beta_{\rm w}$, is assumed to represent the energy-driven wind in the low-density 
case, $n$\,$<$\,$n_{\rm SF}$, and the momentum-driven wind for 
$n$\,$>$\,$n_{\rm SF}$.

\end{itemize}

\subsection{\textit{\textbf{Feedback and star formation}}}
\label{sec:feedsf}

The physics of isolated stellar, gaseous and gas+stars disks have been analysed
extensively over the last few decades by means of analytic methods and numerical
simulations. The parallel study of galaxy evolution in a cosmological setting
placed emphasis on the initial conditions and on the fact that galaxies are open
systems that can exchange mass, momentum and energy with their environment.
Under these conditions, the rate of secular evolution can be dramatically
accelerated and the direction of this evolution altered significantly. The
approximation that galaxies are in dynamical equilibrium remains, except during
the merger events. 

Probably the most important corollary of studying galaxies in cosmology is
narrowing the range of initial conditions and requiring that the evolution
complies with them. Understanding that early disks have been much more gas-rich
than disk galaxies in the local Universe brings about the natural question of
what prevented the full conversion of this gas into stars over the Hubble time.
This can be achieved either by keeping the gas at low densities or high
temperatures, to prevent the Jeans instability from developing. However, it is
difficult to maintain low densities when the gas assembles in disks, unless
strong deposition of momentum, energy or both expels the gas from the disk.

Most of the prescriptions for star formation can be followed from Katz (1992)
with some modifications and involve the Jeans instability. An additional
constraint introduces the critical volume density for star formation, $n_{\rm
SF}$, which corresponds to the total H{\sc i}\,+\,H$_2$ (atomic and molecular
hydrogen) surface density threshold in the Kennicutt-Schmidt (K-S) law,
$\Sigma_{\rm SF}$\,$\sim$\,3--10\,$M_\odot\,{\rm pc^{-2}}$ (e.g., Schaye \&
Dalla Vecchia 2008). However, new observational evidence points to a
considerable dispersion in the values of the threshold surface density,
$\Sigma_{\rm SF}$ and the slope $\alpha$, as well as to a substantial steepening of
the K-S law above $z\sim 3$ (e.g., Gnedin \& Kravtsov 2011 and references
therein). A growing body of evidence points to a dependence of the SFR on the
surface density of the molecular hydrogen, rather than on the total surface
density of the neutral hydrogen. 

An intriguing question is whether the Jeans instability in the neutral gas
triggers the formation of H$_2$, or whether it is the formation of H$_2$ that
triggers the Jeans instability and subsequent gravitational collapse. Current
efforts focus on understanding the various factors which regulate the formation
and destruction of H$_2$, such as dust abundance and metallicity, UV background,
gravitational instabilities facilitating the gas cooling, etc. A fully
self-consistent model of the molecular gas balance in the ISM will be developed
in the next few years.

We have discussed the mechanisms that are responsible for the feedback from
stellar and AGN evolution and pointed out the importance of understanding the
ways in which energy and momentum are distributed among the ISM and the IGM.
While the sources of energy and momentum operate on very small scales,
$<$\,1\,pc, compared to characteristic galactic scalelengths and scaleheights,
$>$\,1\,kpc, they are deposited on much larger scales and can have a global
effect. In addition, there is broad agreement between observations, theory and
numerical simulations that the multiphase ISM is required to treat the feedback
properly. 

\begin{figure}[ht!!!]  
\begin{center}
\includegraphics[angle=0,scale=0.8]{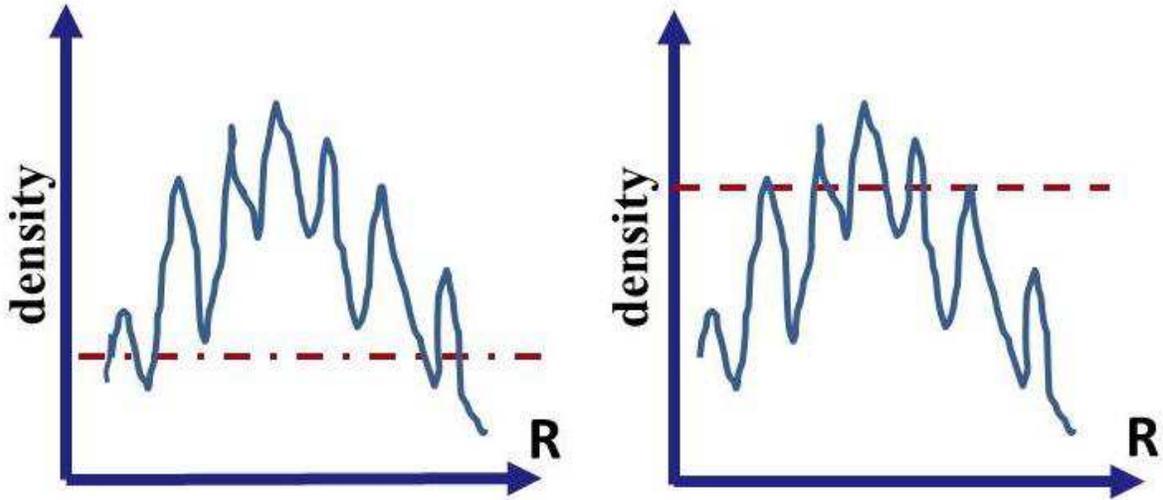}
\end{center}
\caption{Effect of density threshold on star formation. Left: a low-density
threshold -- a single massive star formation region exists; Right: a
high-density threshold. In the latter case, a much smaller region becomes
Jeans-unstable and a number of smaller star formation sites exist, resulting in
lower star formation efficiencies when averaged over all phases of the ISM.}
\label{fig:fig12}
\end{figure}

One possibility for progress in this direction lies in a further increase of the
spatial resolution of numerical models, to an extent that at least part of the
subgrid physics is in fact simulated, e.g., turbulent cells in the ISM.
Sub-50\,pc resolution has shown the formation of hot SN bubbles, superbubbles
and chimneys, in tandem with turbulent flows there (e.g., Ceverino \& Klypin
2009). Galactic winds develop naturally under these conditions, without
additional ad hoc assumptions. Unfortunately, such resolution is difficult to
achieve in present-day fully self-consistent cosmological simulations. 

Another by-product of high-resolution modelling is the possibility of using a
much higher threshold for star formation than usual, corresponding to molecular
gas. The effect of going to higher critical densities for star formation is
shown in Fig.~\ref{fig:fig12}, and results in lower star formation efficiencies
when averaged over all phases of the ISM. A proper treatment of the cold gas
phase is also crucial in order to account for dissipation in supersonic flows.
In fact, in order to reproduce the observed log-normal probability density
function (PDF) in the ISM, it is necessary to resolve the broad density range,
and in particular to resolve the critical density of $10^5$ times that of the
average density in the ISM (e.g., Elmegreen 2002). This requirement in tandem
with the ISM heating by gravitational instabilities leads to self-regulation,
when turbulence limits the efficiency of the star formation process. 

Gas-rich disks are prone to fragmentation -- the so-called Toomre (1964)
instability. The idea of a marginally gravitationally stable gaseous disk was 
originally proposed by Paczynski (1978), who specifically considered the case of
 turbulence driven by gravitational instabilities in a disk that was able to
cool below $Q$\,$\sim$\,1. The resulting turbulent gravitational viscosity was
found to be responsible for the disk re-heating, angular momentum transfer and
inflow. Analysis and numerical simulations of isolated two-component gas$+$stars
disks indeed have demonstrated the formation of massive clumps that migrate to
the centre and heat up the stellar component as a result of dynamical friction,
in agreement with analytical estimates (Shlosman \& Noguchi 1993; Noguchi 1999).
 The characteristic timescale for spiraling in toward the central kpc is about a
couple of orbital periods. Noguchi (1999) further argued that these clumps
contribute to the buildup of galactic bulges.

The developing massive clumps will migrate to the centre over a
few dynamical times, while maintained in a marginally stable state with
Toomre's $Q$\,$\sim$\,1, and contribute there to the bulge growth (e.g., Dekel
{\it at al.} 2009a). In reality, it is not trivial to stabilise the
self-gravitating clumps against runaway star formation. Dekel {\it et al.}
propose that gravitational interactions between the clumps will also induce
turbulence inside the clumps, which will stabilise them against gravitational
collapse over the spiralling-in timescale  -- an interesting possibility so far
not verified. 

In the cosmological context, including star formation, these clumps, in
addition to forming in the disk itself, can be supplied by the incoming flows
from cosmological filaments (Heller {\it et al.} 2007b; Dekel {\it et al.}
2009a; Ceverino {\it et al.} 2010). The clumps did, however, show vigorous
star formation.  Rather than being supported by clump-clump interactions,
Heller {\it et al.} found that the feedback from stellar evolution has provided
support for the clumps, while some of them have been sheared and/or
collisionally destroyed before they entered the central kpc. Moreover, the
energy/momentum feedback parameter has been varied and the clumps have appeared
earlier and have been more numerous in models where this feedback has been
smaller. The final bulge-to-disk mass ratio has also shown a clear
anti-correlation with the amount of the feedback. Hence, as shown by Heller {\it
et al.} energy/momentum feedback inside the gas-rich disk delays its
fragmentation, and, at the same time, decreases the star-forming activity inside
these massive clumps. Internal turbulent pressure in massive self-gravitating
clumps can indeed play the role of delaying the star formation, but it is not
clear whether it can be driven by the outside turbulence in the disk. The
possible internal drivers of turbulence can be their gravitational collapse and
energy input from newly formed massive stars and SNe.

Dense molecular clouds in the ISM form via supersonic turbulence in the ISM. The
supersonic velocities decrease on smaller scales as $v$\,$\sim$\,scale$^{0.5}$
(Larson 1981), and may represent a turbulent field dominated by shocks. This
result is supported by numerical simulations of supersonic turbulence and by
analytical calculations (e.g., Padoan \& Nordlund 2002, and references therein),
with the possible extention to include the cloud column density,
$v$\,$\sim$\,$\Sigma^{0.5}$\,scale$^{0.5}$ (Ballesteros-Paredes {\it et al.}
2011). The velocity-scale relation may determine the characteristic scale for
gravitational instabilities and sites for star formation. The turbulent driving
comes from the larger scales and dissipation occurs within the clouds. Recent
results show that turbulent motions inside molecular clouds are driven by
gravitational collapse (e.g., Ballesteros-Paredes {\it et al.} 2011).

%
%

\section{High-redshift galaxies}
\label{sec:highz}

The epoch of galaxy formation follows the end of the Dark Ages when baryons
could start to accumulate within the DM haloes and star formation was 
triggered. The scope of this review does not allow us to go into the details of
this fascinating subject. Here we shall focus on galaxy evolution during the
reionisation epoch, at redshifts $z$\,$\sim$\,6--12. We shall not discuss the
formation and evolution of the Population\,III stars either, which has been
largely completed by the onset of the reionisation process, except maybe in
low-density regions. Section~\ref{sec:kpc} will touch upon some aspects of SMBH
formation in 10$^8\,M_\odot$ DM haloes. All the problems discussed in the
previous sections remain relevant at these high redshifts.

The rapidly increasing list of objects above $z$\,$\sim$\,6 makes it possible to
study the population of galaxies during reionisation. Deep imaging in multiband
surveys using the Wide Field Camera 3 (WFC3) on the \textit{HST}, as well as
some ground-based observations using 8\,m telescopes, have revealed galaxies via
absorption at wavelengths shorter than Ly$\alpha$ from the intervening neutral
hydrogen (e.g., Bouwens {\it et al.} 2010). In many cases, these photometric
redshifts could be verified spectroscopically, up to $z$\,$\sim$\,7 (e.g.,
Pentericci {\it et al.} 2011). The majority of reionisation-epoch galaxies are
faint, but much rarer brighter galaxies have also been identified at
$z$\,$\sim$\,8 by means of a large-area medium-deep \textit{HST} survey
(Brightest of Reionizing Galaxies, BoRG) along random lines of sight, including
the candidate for the most distant protocluster (Trenti {\it et al.} 2011). 
Even fainter galaxies have been found using gravitational lensing by massive
galaxy clusters. 

\subsection{\textit{\textbf{The high-redshift galaxy zoo}}}
\label{sec:zoo}

One of the most successful methods to search for reionisation-epoch galaxies is
the dropout method based on the absorption short of some characteristic
wavelength, the 912\,\AA\ Lyman break and a smaller break at Ly$\alpha$
1216\,\AA, which originate in the intervening neutral hydrogen (e.g., Steidel
{\it et al.} 1996). Using multiwavelength imaging and filters, objects
`disappear' (drop out) when a particular and progressively redder filter is
applied. The resulting break in the continuum spectrum allows us to determine
the photometric redshift of the object. For $z$\,$\sim$\,6, the break lies at
$\sim$8500\,\AA. This technique has been applied first to $U$-band dropouts --
galaxies that lack flux in the $U$-band ($z$\,$\sim$\,3), then to $g$-band
dropouts ($z$\,$\sim$\,4). The choice of the filter determines the targeted
redshift. Additional dropouts have been named according to the relevant bands,
$i_{775}$ ($z$\,$\sim$\,6), $z_{850}$ ($z$\,$\sim$\,7), $Y$ ($z$\,$\sim$\,8--9)
and $J$ ($z$\,$\sim$\,10). Existing data from NICMOS, GOODS/ACS and UDF can
reveal dropouts up to $z$\,$\sim$\,10. The population of detected galaxies has
already provided substantial constraints on the galaxy growth in the Universe at
that epoch. 

The expanding classification of high-$z$ galaxies has its origin in diverse
observational techniques used for their detection and study, resembling the
early stages of AGN classification, before unification. Galaxies that exhibit a
break in the Lyman continuum redshifted to the UV and other bands have been
called Lyman break galaxies (LBGs). Complementary to continuum-selected surveys,
the Ly$\alpha$ galaxies, or so-called Ly$\alpha$ emitters (LAEs), have been
mostly detected in narrow-band imaging surveys. Such surveys typically miss the
LBGs because of the faint continuum. Spectroscopic identification of
$z$\,$\gtorder$\,6 LBGs is only possible if they have strong Ly$\alpha$
emission, and are bright (e.g., Vanzella \textit{et al.} 2011). 

An important question is what is the relationship between various classes of
high-$z$ galaxy populations and what are their low-$z$ counterparts. Especially
interesting is their relationship to sub-mm galaxies, found at
$z$\,$\sim$\,1--5. These sub-mm galaxies have been detected in the
200\,$\mu$m\,--\,1\,mm band, via redshifted dust emission, using the Sub-mm
Common-User Bolometer Array (SCUBA) camera. These objects have a negative
$K$-correction\footnote{The $K$-correction is the dimming of a source due to the
$1+z$ shift of the wavelength band and its width.} because the Rayleigh-Jeans
(RJ) tail of the Planck blackbody distribution. Galaxies in the RJ tail become
brighter with redshift. They are generally not SBGs because of the weak UV
emission. The sub-mm galaxy population consists of very luminous objects with
bolometric luminosity $\sim$10$^{12-13}\,L_\odot$, emitted mostly in the IR.
Powered by intense starbursts, their estimated SFRs are
$\sim$10$^{2-3}\,M_\odot\,{\rm yr^{-1}}$.

\subsection{\textit{\textbf{Mass and luminosity functions}}}
\label{sec:mfunc}

\begin{figure}[ht!!!]  
\begin{center}
\includegraphics[angle=0,scale=0.8]{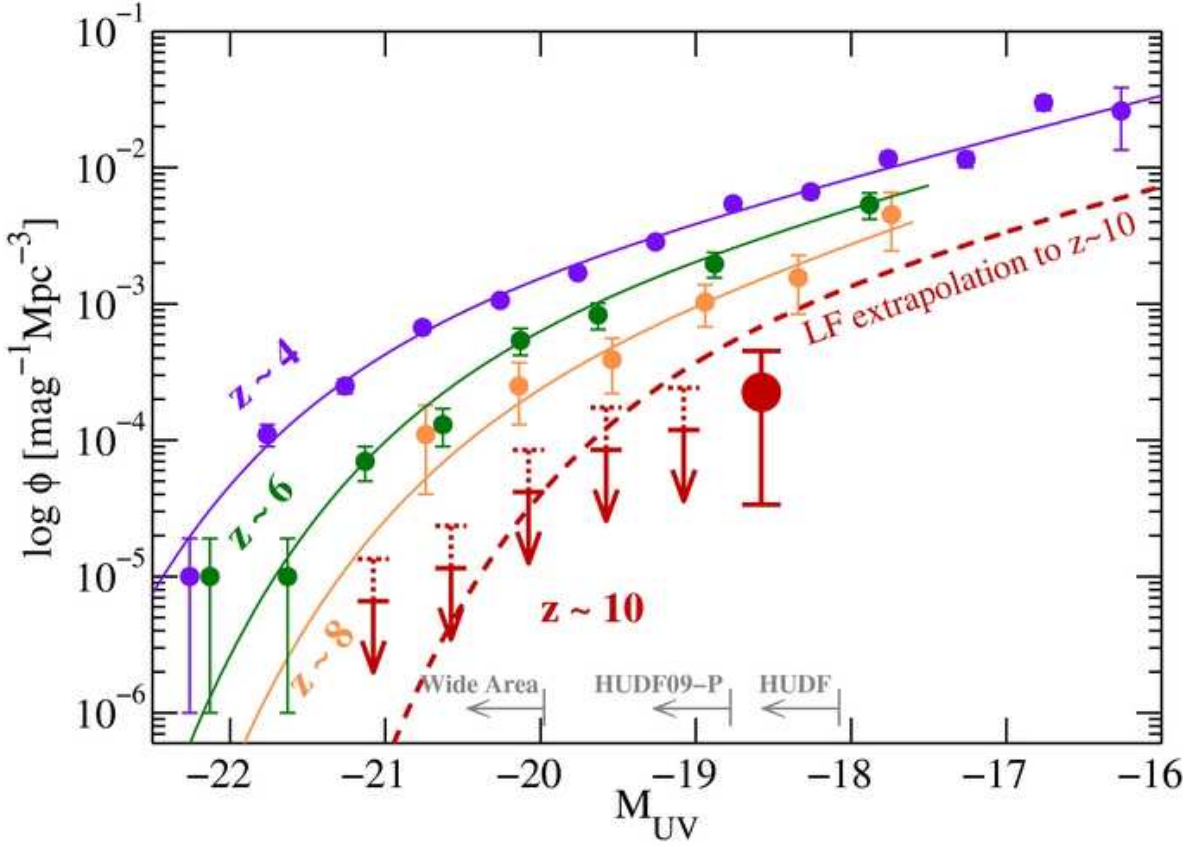}
\end{center}
\caption{\small UV LFs for $z$\,$\sim$\,4, 6, 8 and projected LF at $z$\,$\sim$\,10
(Oesch {\it et al.} 2012). The $z$\,$\sim$\,10 LF extrapolated from fits to
lower-redshift LBG LFs is shown as a dashed red line (see also the text). For
comparison the $z$\,$\sim$\,4 and $z$\,$\sim$\,6 LFs are plotted, showing the
dramatic buildup of UV luminosity across $\sim$1\,Gyr of cosmic time. The
light-grey vectors along the lower axis indicate the range of luminosities over
which the different data sets dominate the $z$\,$\sim$\,10 LF constraints.}
\label{fig:fig13}
\end{figure}

Observations of $z$\,$\gtorder$\,6 galaxies have shown a rapidly evolving
galactic LF which agrees with the predicted DM halo mass
function (e.g., Bouwens {\it et al.} 2011). The UV LF of LBGs has been
established with its faint end exhibiting a very steep slope (e.g., Bouwens {\it
et al.} 2007). Using the Schechter function fit,
$\phi(L)$\,=\,$(\phi^*/L^*)(L/L^*)^\alpha {\rm exp}(-L/L^*)$, the {\it faint}
end of this LF at $z$\,$\sim$\,7 has the slope of
$\alpha$\,=\,$-1.77$\,$\pm$\,0.20, and
$\phi^*$\,=\,1.4\,$\times$\,10$^{-3}\,{\rm Mpc^{-3}\,mag^{-1}}$, which is
consistent with no evolution over the time span of $z$\,$\sim$\,2--7 (e.g.,
Oesch {\it et al.} 2010). The bright end of the LF evolves significantly over
this time period. An even steeper faint end of the LF,
$\alpha$\,=\,$-1.98$\,$\pm$\,0.23, has been claimed recently (Trenti 2012). The
SFR appears to decline rapidly with increasing redshift. So by $z$\,$\sim$\,6,
the number of ionising photons is just enough to keep the Universe ionised, and
most of them come from objects fainter than the current detection limit of the
\textit{HST} (e.g., Oesch {\it et al.} 2010; Trenti {\it et al.} 2010; Trenti
2012). 

\begin{figure}[ht!!!]  
\begin{center}
\includegraphics[angle=0,scale=0.77]{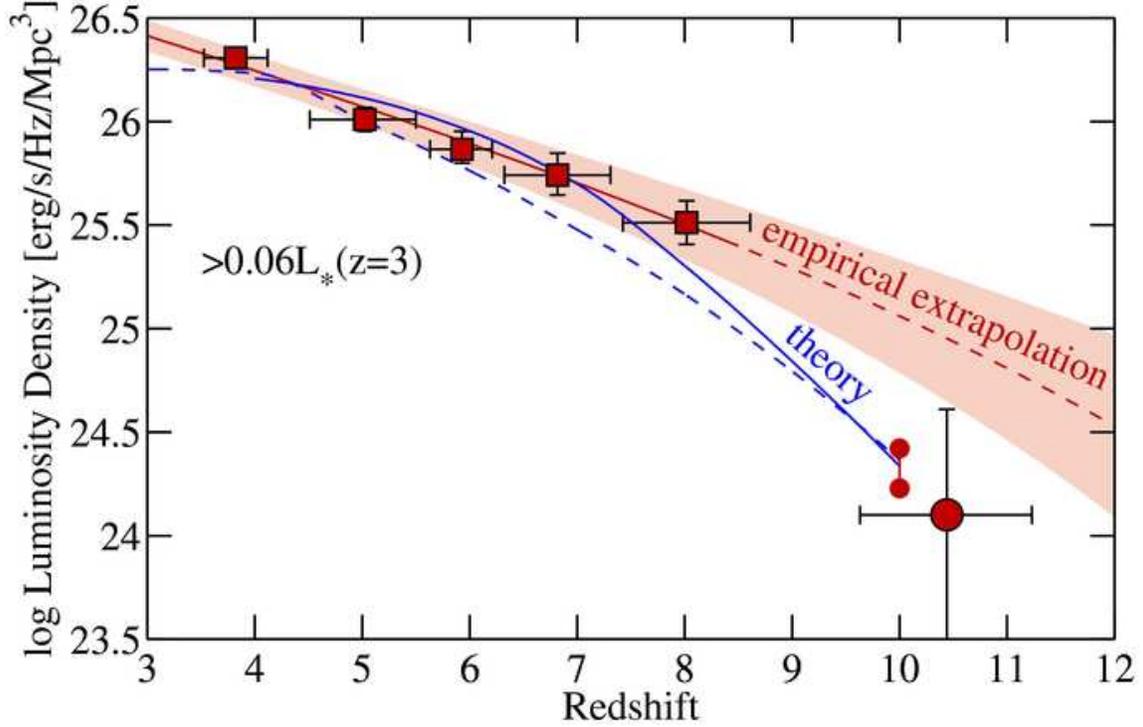}
\end{center}
\caption{\small Evolution of the UV LD above
$M_{1400}$\,=\,$-18$\,mag [$>$\,0.06\,L$^*(z=3)$] (Oesch {\it et al.} 2012). The
filled circle at $z$\,$\sim$\,10.4 is the LD directly measured for the galaxy
candidate. The red line corresponds to the empirical LF evolution. Its
extrapolation to $z$\,$>$\,8 is shown as a dashed red line.}
\label{fig:fig14}
\end{figure}

An accelerated evolution of galaxies during reionisation has been predicted and
observed (e.g., Bouwens \textit{et al.} 2007, 2010; Trenti {\it et al.} 2010;
Lacey {\it et al.} 2011; Oesch {\it et al.} 2010, 2012). Strong evolution is
expected for $z$\,$\sim$\,8--10, by about a factor of $\sim$2--5. The
estimated number of $z$\,$\sim$\,10 galaxies has been derived from the observed
LF at $z$\,$\sim$\,6 and 8 (Fig.~\ref{fig:fig13}). Using this LF, six objects
are expected to be present in the field at $z$\,$\sim$\,10, but only one has
been detected. Hence, the LF appears to drop even faster than expected from the
previous empirical lower-redshift extrapolation. The resulting accelerated LF
evolution in the range of $z$\,$\sim$\,8--10 has been estimated at
$\gtorder$\,94\% significance level (Oesch \textit{et al}. 2012). 

An important conclusion from the above studies has been the realisation that the
UV luminosity density (LD) originating in the high-$z$ galaxy population levels
off and gradually falls toward higher $z$, in the range $z$\,$\sim$\,3--8
(Fig.~\ref{fig:fig14}).  The LD data at $z$\,$\sim$\,4--8 are taken from Bouwens
{\it et al.} (2007) and Bouwens {\it et al.} (2011). As can be seen in
Fig.~\ref{fig:fig14}, the LD increases by more than an order of magnitude in
170\,Myr from $z$\,$\sim$\,10 to 8, indicating that the galaxy population at
this luminosity range evolves by a factor $\gtorder$\,4 more than expected from
low-redshift extrapolations. The predicted LD evolution of the semi-analytical
model of Lacey {\it et al.} (2011) is shown as a dashed blue line, and the
prediction from theoretical modelling (Trenti {\it et al.} 2010) is shown as a
blue solid line. These reproduce the expected LD at $z$\,$\sim$\,10 remarkably 
well. 

A strong decline in the LF beyond $z$\,$\sim$\,8 has corollaries for the
reionisation by the more luminous galaxies at this epoch, as the number of
luminous sources appears insufficient for this process. These data point clearly
to a strong evolution of the galaxy population, but what is the cause of this
evolution?

Analysis and modelling of the available data point to the underlying cause: the
accelerated evolution is driven by changes in the DM halo mass function (HMF),
as follows from theoretical considerations (e.g., Trenti {\it et al.} 2010) and
semi-analytical modelling (e.g., Lacey {\it et al.} 2011), and not by the star
formation processes in these galaxies. Interestingly, the rapid assembly of
haloes at $z$\,$\sim$\,8--10 alone can explain the LF evolution (Trenti {\it et
al.} 2010). However, this assumption has never been put to a self-consistent
test using high-resolution simulations with the relevant baryon physics. The
possible link between LF and the DM HMF has been studied by means of the
conditional LF method (e.g., Trenti {\it et al.} 2010 and references therein) to
understand the processes regulating star formation. The main conclusions can be
summarised as (1) a significant redshift evolution of galaxy luminosity vs halo
mass, $L_{\rm gal}(M_{\rm h})$, (2) only a fraction $\sim$20--30\% appear to
host LBGs, and (3) the LF for $z$\,$\gtorder$\,6 deviates from the Schechter
functional form, in particular, by missing the sharp drop in density of luminous
$M$\,$\ltorder$\,$-20$ galaxies with $L$. For example, due to the short
timescales -- $\Delta z$\,$\sim$\,1 corresponds to $\ltorder$\,170\,Mpc -- it
becomes difficult to rely on the fast evolution of $L_{\rm gal}(M_{\rm h})$,
while $M_{\rm h}$ evolves rapidly at these redshifts.

Due to the nature of the hierarchical growth of structure, high-$z$ galaxies
should appear and grow fastest in the highest overdensities, and therefore are
expected to be strongly clustered around the density peaks. For example, Trenti
{\it et al.} (2012) infer the properties of DM haloes in the BoRG\,58 field at
$z$\,$\sim$\,8 based on the found five $Y_{098}$-dropouts, using the Improved
Conditional Luminosity Function model. The brightest member of the associated
overdensity appears to reside in a halo of
$\sim$(4--7\,$\pm$\,2)\,$\times$\,10$^{11}\,M_\odot$ -- a 5\,$\sigma$ density
peak which corresponds to a comoving space density of
$\sim$(9--15)\,$\times$\,10$^{-7}\,{\rm Mpc^{-3}}$. It has $\sim$20--70\%
chance of being present within the volume probed by the BoRG survey. Using an
extended Press-Schechter function, about 4.8 haloes more massive than
10$^{11}\,M_\odot$ are expected in the associated region with the (comoving)
radius of 1.55\,Mpc, compared to less than $10^{-3}$ in the random region.  For
higher accuracy, a set of 10 cosmological simulations (Romano-D\'iaz {\it et
al.} 2011a) has been used, tailored to study high-$z$ galaxy formation in such
an over-dense environment. A DM mass resolution of 3\,$\times$\,10$^8\,M_\odot$
has been used, and, therefore, haloes with masses $\gtorder$\,10$^{11}\,M_\odot$
have been well resolved. The constrained realisation (CR) method (e.g.,
Bertschinger 1987; Hoffman \& Ribak 1991; Romano-D\'iaz {\it et al.} 2007, 2009,
2011a,b) has been instrumental in modelling these rare over-dense regions. We
describe this method below. 

The CR method consists of a series of linear constraints on the initial density
field used to design prescribed initial conditions. It is not an
approximation but an {\it exact} method. All the constraints are of the same
form -- the value of the initial density field at different locations, and are
evaluated with different Gaussian smoothing kernels, with their width fixed so
as to encompass the mass scale on which a constraint is imposed. The set of mass
scales and the location at which the constraints are imposed define the
numerical experiment. Assuming a cosmological model and power spectrum of the
primordial perturbation field, a random realisation of the field is constructed
from which a CR is generated. The additional use of the zoom-in technique
assures that the high-resolution region of simulations is subject to large-scale
gravitational torques. The CRs provide a unique tool to study high-$z$ galaxies
at an unprecedented resolution. It allows one to use much smaller cosmological
volumes, and, without any loss of generality, accounts for the cosmic variance.

The initial conditions for the test runs described above have been constrained
to have a halo of mass $\sim$10$^{12}\,M_\odot$ by $z$\,$\sim$\,6. This halo
has reached $\sim$5\,$\times$\,10$^{11}\,M_\odot$ by $z$\,$\sim$\,8 in
compliance with BoRG\,58-17871420. Within the field of view of $70''\times 70''$
and the redshift depth of $\Delta z$\,$\sim$\,19\,Mpc about 6.4 haloes more
massive than $\sim$10$^{11}\,M_\odot$ have been expected, and the highest
number found in the simulations was 10 (Fig.\,\ref{fig:fig15}). A random
(unconstrained) region of the same volume has been estimated to host
$\sim$0.013 such haloes. The probability of contamination in such a small area
is negligible, $\sim$2.5\,$\times$\,10$^{-4}$. In summary, if indeed the
brightest member of the BoRG\,58 field lives in a massive DM halo, the fainter
dropouts detected in this field are part of the overdensity that contributes to
the protocluster, depending of course on spectroscopic confirmation. 
Simulations provide some insight into the fate of this overdensity with a total
DM mass of $\sim$(1--2)\,$\times$\,10$^{13}\,M_\odot$ -- it has collapsed by
$z$\,$\sim$\,3, and is expected to grow to
$\sim$(1--2)\,$\times$\,10$^{14}\,M_\odot$ by $z$\,=\,0.

\begin{figure}[ht!!!]  
\begin{center}
\includegraphics[angle=0,scale=0.84]{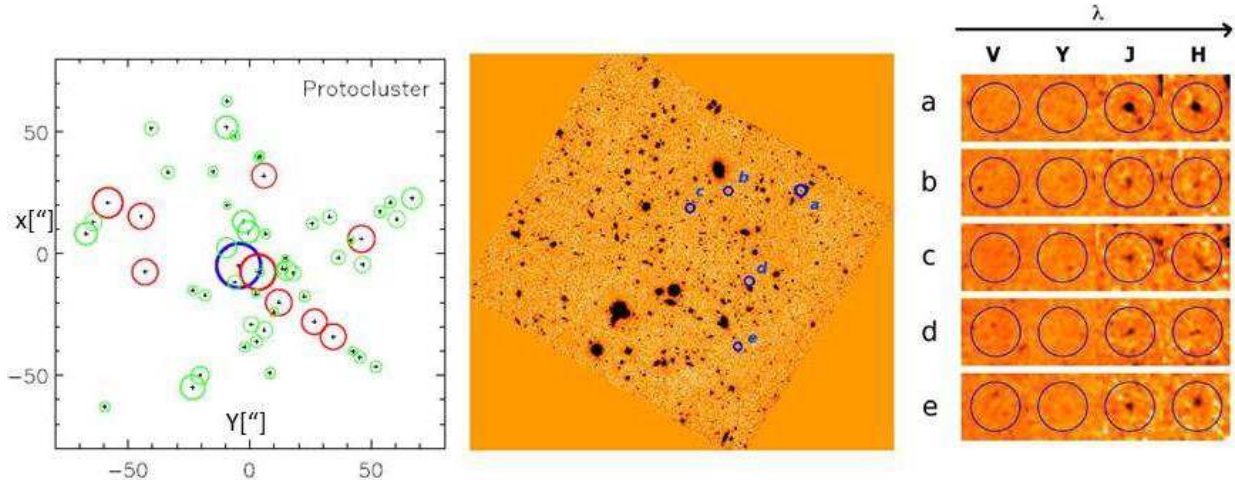}
\end{center}
\caption{\small The most distant candidate protocluster at $z$\,$\sim$\,8 (Trenti {\it
et al.} 2012). Left: DM halo distribution for a simulated protocluster in a
comoving volume of 11\,$\times$\,11\,$\times$\,19\,Mpc$^3$ from Romano-D\'iaz
{\it et al.} (2011a). The largest (blue) circle represents the most massive halo
in the simulation, $\sim$5\,$\times$\,10$^{11}\,M_\odot$; red circles, haloes
above $10^{11}\,M_\odot$; green circles, haloes of
$10^{10}$--$10^{11}\,M_\odot$. Middle: $J_{125}$ image of BoRG\,58 field,
with $Y_{098}$-dropouts indicated by (blue) circles. Right: Postage-stamp images
($3.2''\times 3.2''$) of sources: BoRG\,58-17871420, BoRG\,58-14061418,
BoRG\,58-12071332, BoRG\,58-15140953, and BoRG\,58-14550613 fields (top to bottom).}
\label{fig:fig15}
\end{figure}

The evolution of the HMF is very sensitive to the assumed cosmology, because the
halo growth rate depends on the average matter density in the Universe. As the
DM is not observable directly, numerical simulations are indispensable in
studying the halo growth, and analytic techniques provide an additional tool.
The process of DM halo formation quickly becomes non-linear which makes an analytical follow-up difficult. Analytically, one relies on modelling the
spherical or ellipsoidal collapses, but only $N$-body simulations reveal the
complexity of the process which is hierarchical in Nature. Numerically, the halo
growth depends on the force resolution used and on the size of the computational
box. The $N$-body simulations of halo evolution are very accurate, $\sim$1\%,
and the analytical methods are $\sim$10--20\% (e.g., Press \& Schechter 1974,
Bond {\it et al.} 1991). Nevertheless, the analytical HMF can reproduce the
numerical results at least qualitatively, and can be defined{\footnote{A variety
of definitions of the HMF exist in the literature. We use the differential HMF.}
as $dn/dM$, where $n(M)$ is the number density of haloes in the range $dM$
around mass $M$ at redshift $z$ (e.g., Jenkins {\it et al.} 2001),
\begin{equation}
\frac{dn(M)}{dM} = f(\sigma)\frac{<\rho>}{M^2}\frac{d\,{\rm ln}\, 
            \sigma^{-1}(M)}{d\,{\rm ln}\,M},
\label{eq:mhalos}
\end{equation}
where $\sigma^2$ is the variance of the (linear) density field smoothed on the
scale corresponding to $M$, and $<\rho>$ is the average density in the Universe.
In the spherical collapse approximation developed by Press \& Schechter (1974),
$f(\sigma)$\,=\,$(2/\pi)^{1/2}(\delta_{\rm c}/\sigma)\,{\rm exp}({-\delta_{\rm
c}^2/2\sigma^2})$, where $\delta_{\rm c}$\,$\approx$\,1.686. Press \& Schechter
assumed that all the mass is within the DM haloes, i.e.,
$\int_{-\infty}^{+\infty} f(\sigma)d\,{\rm ln}\,\sigma^{-1}$\,=\,1. An extension
for arbitrary redshift is achieved by taking $\delta_{\rm c}$\,=\,$\delta_{\rm
c}(z=0)/D(z)$, $D(z)$ being the linear growth factor.

\begin{figure}[ht!!!]  
\begin{center}
\includegraphics[angle=0,scale=0.83]{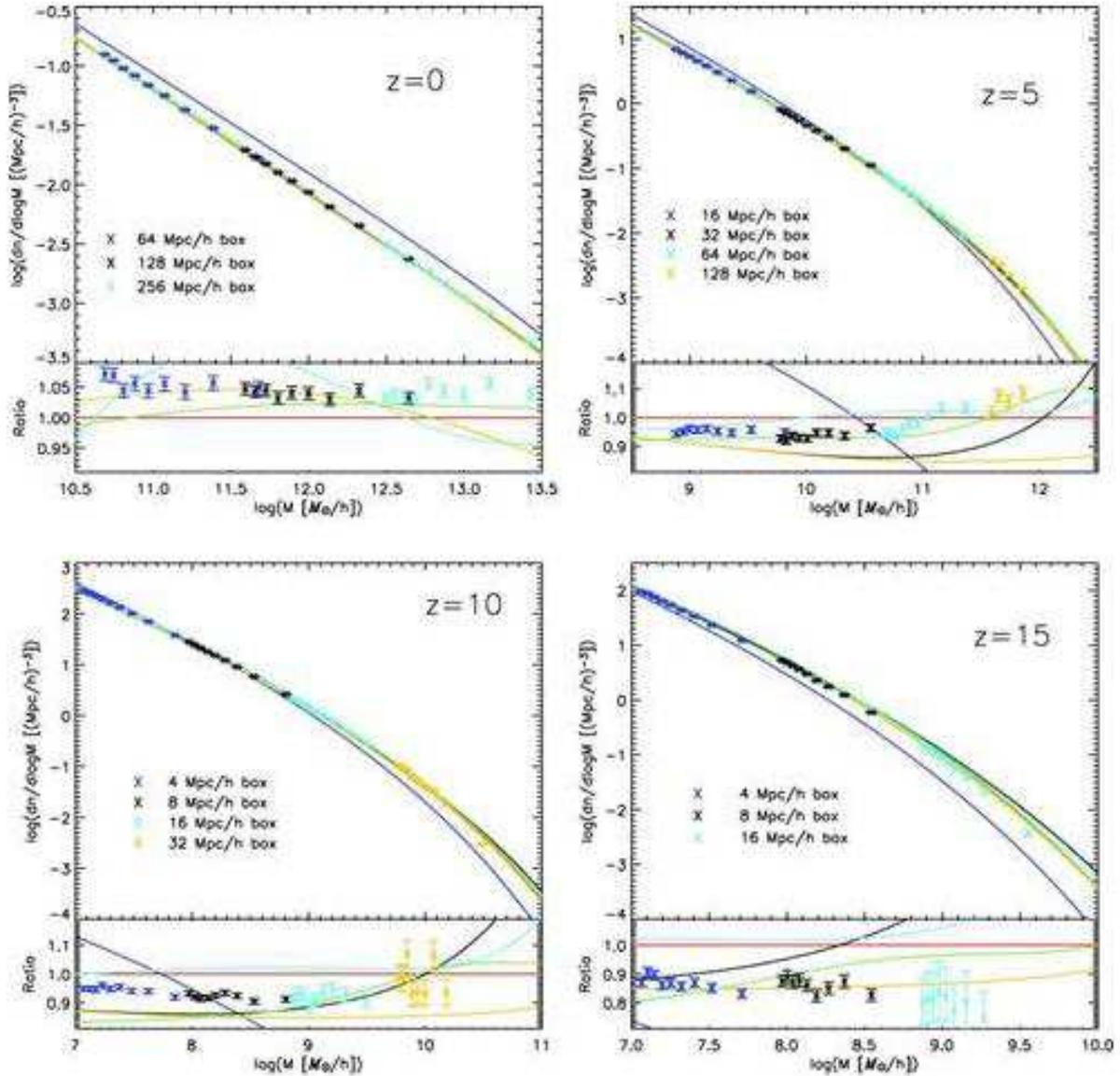}
\end{center}
\caption{\small The HMFs at four redshifts ($z$\,=\,0, 5, 10 and 15) compared to
different fitting formulae, analytic and numerical (coloured curves). Note that
the mass ranges are different at different redshifts. The bottom panels show the
ratio with respect to the Warren {\it et al.} (2006) fit, agreeing at the 10\%
level for $z$\,$\ltorder$\,10, and with a systematic offset of 5\% at $z$\,=\,0.
At higher redshifts, agreement is at the 20\% level. Agreement becomes very
close once finite-volume corrections are applied. Press-Schechter is a bad
fit at all redshifts, especially at high redshifts, $z$\,$\gtorder$\,10, where
the difference is an order of magnitude. From Lukic {\it et al.} (2007).}
\label{fig:fig16}
\end{figure}

Discrepancies between the analytically derived and numerically obtained HMFs can
be sufficient to affect our understanding of galaxy growth during the
reionisation epoch, as shown in Fig.~\ref{fig:fig16} (Lukic {\it et al.} 2007).
It is, therefore, important that the shape of the HMF  can have a universal
character, independent of epoch, cosmological parameters and the initial power
spectrum, in particular representations (Jenkins {\it et al.} 2001), although
this must be taken with caution. Violations of universality have been found both
at low ($z$\,$\ltorder$\,5 at $\sim$20\% level, Fig.~\ref{fig:fig16}), and
high ($z$\,$\sim$\,10--30) redshifts, but the issue is still unsettled due a
number of numerical concerns (e.g., Lukic {\it et al.} 2007; Reed {\it et al.}
2007).

%
%

\section{Disk evolution: the central kpc and the SMBHs}
\label{sec:kpc}

A wealth of issues dominate our understanding of the central regions in galaxies
and their role in the overall galaxy evolution on cosmological timescales -- the
secular evolution. But is there a {\it dynamically} distinct central kiloparsec
region in galaxies?  The answer appears to be positive, as a major resonance
between the bar and/or spiral arm pattern speed, $\Omega_{\rm p}$, on the one
hand and the linear combination of the epicyclic frequency, $\kappa$, and the
angular velocity, $\Omega$, on the other is positioned in this area, i.e., 
\begin{equation}
\Omega_{\rm p} = \Omega - \frac{\kappa}{2}.
\label{eq:omegap}
\end{equation}
Incidentally, the right-hand side of this relation represents the precession
frequency of stellar orbits. The resonance between the orbit precession
frequency and the pattern speed is called the inner Lindblad resonance (ILR). A
multiple number of ILRs can exist in the neighbourhood but typically their
number does not exceed two. The ILR(s), if they are not saturated, dampen the
propagation of waves in the stellar `fluid'. This resonance can trigger various
processes in the region, e.g., gas accumulation in the form of nuclear ring(s),
nuclear starbursts, nuclear bars, etc. (Shlosman 1999, and references therein).
The ILR(s) can pump the kinetic orbital energy into vertical stellar motions. 
So while there are naturally strong interactions between the inner and outer
disks, different processes dominate both regions. 

The next question is whether there is a {\it morphologically} distinct central
kiloparsec region in galaxies. The answer is positive again -- the ILR(s) act(s) as
separators between the inner and outer disk, resulting in detached bars and
spiral patterns. The inner region is generally dominated by the bulge and hosts
the SMBH. 

A number of important issues, which also include the inner kpc directly or
indirectly, are discussed by Lia Athanassoula and James Binney (this volume). We
shall attempt to avoid unnecessary overlap, although some overlap is actually
welcomed. In the discussion below, we shall focus, therefore, on various
asymmetries in the mass distribution that drive the evolution, such as disk and
halo asymmetries, large-scale stellar bars (briefly), and the dynamics of nested
bars. In Section~\ref{sec:origin} we shall touch on the issues related to the
formation and evolution of SMBHs at high redshifts. We have already reviewed, to
some extent, the feedback from AGN in Section~\ref{sec:feedback}. The immediate
environment of the SMBHs, e.g., the role of molecular tori, is beyond the scope
of this discussion. 

Two types of torques can have a dramatic effect on the dynamics within the
central kpc, namely magnetic and gravitational torques. The former can dominate
the central 1--10\,pc from the SMBHs, while the latter can have an effect
outside this region, on scales of $\gtorder$~few tens of parsecs. Viscous
torques can be important near the major resonances and can be neglected in other
regions, in comparison with magnetic and gravitational torques. 

\subsection{Bars and the morphology of the central kpc}
\label{sec:bars}

Stellar bars can be formed either as a result of a (spontaneous) break of axial
symmetry -- the so-called bar instability (e.g., Hohl 1971), or via tidal
interaction between galaxies (e.g., Noguchi 1988) or between galaxies and DM
subhaloes (e.g., Romano-D\'iaz {\it et al.} 2008b). Stellar bars themselves are
subject to dynamical instabilities and secular evolution which affect the disk
as well. Of these, we shall single out the vertical buckling instability (e.g.,
Combes {\it et al.} 1990). This instability has both dynamical and secular
aspects. Dynamically, this instability exhibits a spontaneous break of the
equatorial symmetry in the $rz$ plane (e.g., Pfenniger \& Friedli 1991; Raha
{\it et al.} 1991). The action of the vertical ILR effectively converts the
rotational kinetic energy of the star in the disk into vertical oscillations.
This results in a vertical thickening of the stellar disk at radii smaller than
the position of the vertical ILR, and in the appearance of a characteristic
peanut/boxy-shaped bulge. The symmetry is always restored on the dynamical
timescale (e.g., Fig.~\ref{fig:fig17}, note the flip-flow at
$\sim$2.3--2.4\,Gyr). Moreover, if the equatorial symmetry is (artificially)
imposed, this bulge nevertheless appears, although on a longer timescale and
driven by the same resonance.

\begin{figure}[ht!!!]  
\begin{center}
\includegraphics[angle=0,scale=0.82]{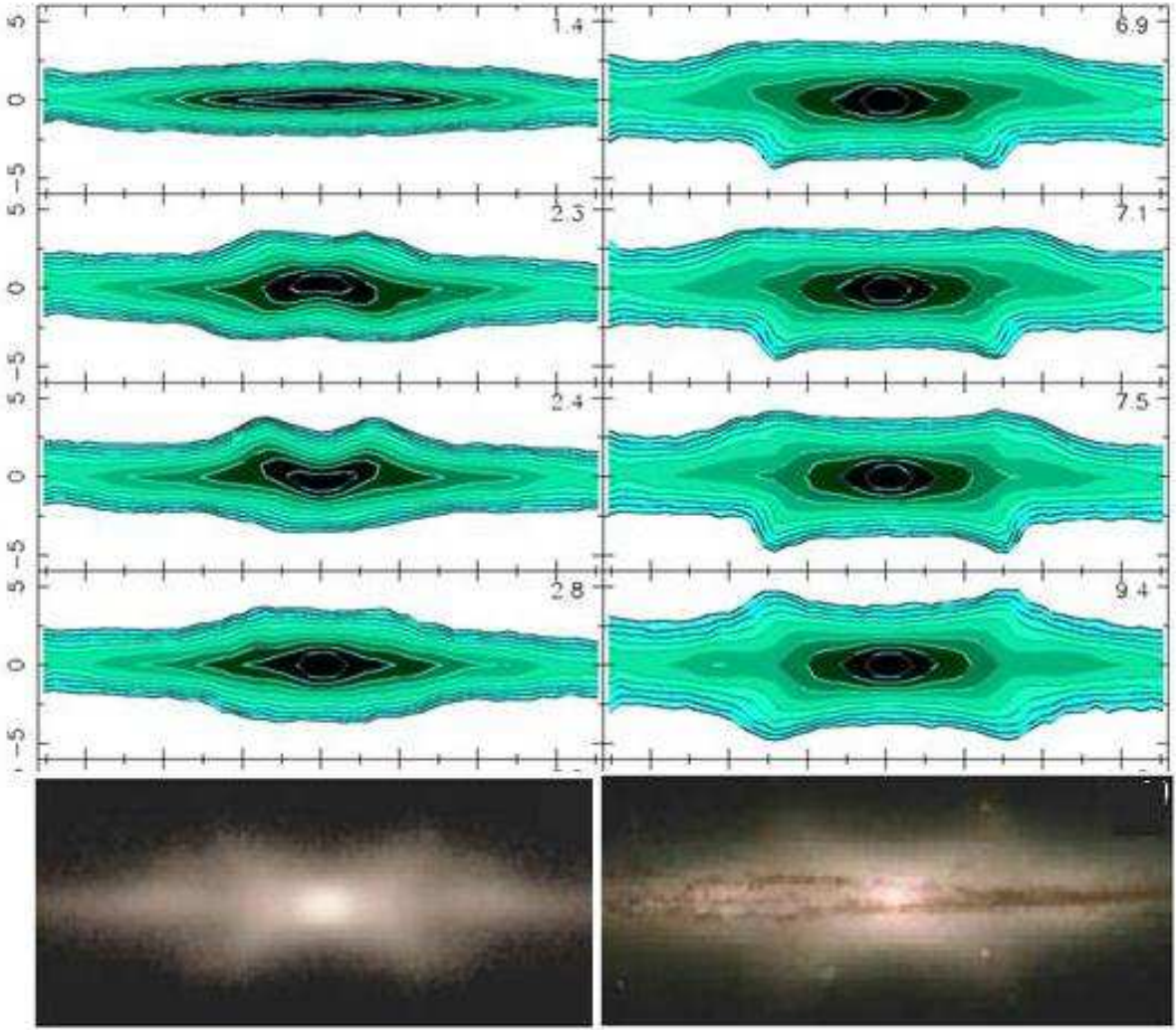}
\end{center}
\caption{\small The {\it recurrent} buckling instability. Upper frames (from
Mart\'inez-Valpuesta {\it et al.} 2006): Evolution of the vertical structure in
the bar: edge-on view along the bar's minor axis. The length is given in kpc and
the values of the projected isodensity contours are kept unchanged. The time in
Gyr is given in the upper-right corners. Note the bar flip-flop at 2.3--2.4\,Gyr
and the persistent vertical asymmetry at 5.2--7.5\,Gyr. Lower frames: smoothed
version of the above figure at 9.4\,Gyr (left), and a matching galaxy from HCG\,87
group of galaxies  (Hubble Heritage Team), courtesy of J.~H.\,Knapen (right).}
\label{fig:fig17}
\end{figure}

What the low-resolution simulations have failed to capture, and what has been
obtained by Mart\'inez-Valpuesta {\it et al.} (2006) for the first time, is the
{\it recurrent} break in the equatorial symmetry occurring on a much slower
timescale of a few Gyr, around 5.2--7.5\,Gyr. This slow buildup of the bar
asymmetry long after the first vertical buckling occurred is rooted in the secular
evolution of stellar orbits driven by the low-order vertical resonances. Unlike
the first buckling, the second phase displays persistent asymmetry. The Fourier
amplitude of the symmetry breaking decreases with the next stage of the
instability.

About 50\% of edge-on disks show peanut/boxy bulges (Mart\'inez-Valpuesta {\it
et al.} 2006 and references therein), which appear to be a clear signature of
stellar bars. While we do understand the reasons for the dynamical stage of the
buckling instability,  we cannot predict the onset of the (second) secular stage
of this instability. Models with a gas component show that the amplitude of this
instability decreases with increasing gas fraction (Berentzen {\it et al.} 1998;
2007).  

\subsubsection{Nested bars: observational perspective}
\label{sec:nested}

Getting rid of the angular momentum is a major issue for astrophysical systems
(e.g., Sections~\ref{sec:angmom} and \ref{sec:mergers}). Given that a
substantial mass is involved, gravitational torques appear as the most efficient
mechanism for redistribution of angular momentum on various spatial scales. The
formation of disks, therefore, is a reflection of the inefficiency of this
process. Gravitational torques are triggered by a non-axisymmetric distribution
of matter. The most frequent asymmetry in the disk is in the form of spiral
arms, which lead to torques with an amplitude of $\sim$1\% -- a quasilinear
perturbation, if defined as a ratio of tangential to radial acceleration. On the
other hand, bars are strongly non-linear perturbations on the level of
$\sim$10--100\%. The importance of bars, at least after $z$\,$\sim$\,1, is
reflected in the existence of the branch of barred galaxies in the Hubble fork,
although the exact statistics is still being debated (e.g., Jogee {\it et al.}
2004; Sheth {\it et al.} 2008).

But the efficiency of bars in extracting angular momentum, for example from
the gas, is limited by a decade in radius, due to a strong decay in the
gravitational multipole interactions (Shlosman {\it et al.} 1989, 1990). So it
is only natural that bars `repeat' themselves on progressively smaller scales.
In retrospect, it is not surprising that such `bizarre' systems of {\it
nested} bars exist in Nature. 

The first known observation of a system with nested bars has been performed on
NGC\,1291, classified as an SB0/a galaxy (Evans 1951). An inner twist of the
optical isophote has barely been detected, and explained as an inner spiral. A
much later observation of this object resulted in a large-scale bar of
$\sim$9.9\,kpc and a nuclear bar of $\sim$1.8\,kpc, misaligned at
$\sim$30$^\circ$, with the inner bar leading the outer bar (de Vaucouleurs
1975). A question has been asked on whether the presence of a second bar in
NGC\,1291 is an `oddity of Nature' or is a fairly common, perhaps typical,
structural feature. De Vaucouleurs concluded that `the lens-bar-nucleus
structure on two different scales in barred lenticular galaxies is probably not
rare, and raises an interesting problem in the dynamics of stellar systems' but
did not follow up on this issue. A morphological survey of 121 barred galaxies
has revealed additional objects with inner structure misaligned with the outer
bar, but this has been interpreted as a bulge distorted by the large-scale
bar (Kormendy 1979). In a subsequent study, Kormendy (1982) has analysed bulge
rotation in barred galaxies and summed up that the kinematics of these triaxial
bulges is similar to those of bars while the light distribution is as in
elliptical galaxies. All these bars have been stellar in origin. Their mutual
interactions have not been discussed and they have not been considered as a
dynamically coupled system. Shlosman {\it et al.} (1989, 1990) have suggested
that nuclear bars can be of multiple types, and that nested bars form a new
dynamically coupled system which redistributes angular momentum in galaxies.

In principle, nested bars can be of a few types: stellar/stellar,
stellar/gaseous and gaseous/gaseous. The first two types have been observed now,
while the third type has not yet been observed\footnote{One can envisage also
the existence of gaseous/stellar bars, where the large-scale bar is gaseous and
the inner bar is stellar.}. In addition, we do not count as a separate class
stellar nested bars which are gas-rich or vice versa.

The first catalogues of double-barred galaxies have been published recently.
Laine {\it et al.} (2002) used an \textit{HST} sample of 112 galaxies in the
$H$-band, Erwin \& Sparke (2002) analysed a sample of 38 galaxies in the
optical, and Erwin (2004) considered 67 galaxies, mostly from Laine {\it et
al.}. The main conclusion of these studies has been that about 25\%\,$\pm$\,5\%
of all disk galaxies host nested bars, and about 1/3 of barred disks possess
secondary bars. These numbers show decisively that galaxies with nested stellar
bars are not a marginal phenomenon, but form an important class of dynamical
systems.

Molecular gas is abundant in galactic centres, but it is not clear what fraction
of this gas is in a `barred' state. While there have been surveys of molecular
gas within the central $\sim $\,kpc, the available resolution was insufficient
for the detection of nuclear gaseous bars. Some nearby galaxies host gaseous
bars, e.g., IC\,342, NGC\,2273, NGC\,2782, Circinus, etc., but their statistical
significance is not clear. Surveys with ALMA (the Atacama Large
Millimetre/submillimetre Array) will probably resolve this issue.

Detecting gas-dominated nuclear bars requires surveys of gas morphology and 
kinematics in the central few hundred parsec of disk galaxies at a resolution of 
$\sim$10\,pc. An additional detection problem can arise from gaseous bars being 
very short-lived -- an issue related to the stability of these objects, which we 
discuss below.

An analysis of the nested bars in the Laine {\it et al.} (2002)
sample has clarified some of the basic properties of nested bar systems.
Firstly, it has shown that the size distributions of large-scale and nuclear
bars differ profoundly and exhibit a bimodal behaviour. Whether bar sizes are
taken as physical or normalised by the galaxy size, $r_{25}$, there is little
overlap between their distributions. In physical units, this division lies at
$r$\,$\sim$\,1.6\,kpc, in normalised -- around $r/r_{25}$\,$\sim$\,0.12. This
bimodality can be explained in terms of a disk resonance, the ILR, and the above
radii can be identified with the position of this resonance. The ILR acts,
therefore, as a dynamical separator. The ILRs are expected to form where the
gravitational potential of the inner galaxy switches from three-dimensional to
two-dimensional. This normally happens at the bulge-disk interface, or
alternatively, where the disk thickness becomes comparable to its radial
extension. To summarise, while the sizes of large-scale bars in nested bars, as
well as those of single bars, exhibit a linear correlation with disk size,
nuclear bars do not show the same behaviour. 

Secondly, nuclear bars almost always come in conjunction with large-scale bars
-- a clear signature that this is a prerequisite for their existence, although
one can envisage a scenario where they form separately. Thirdly, the size
distribution of nuclear rings in these galaxies peaks at the same radii of
$r/r_{25}$\,$\sim$\,0.12 (see above), which signals again at the crucial role
the ILR plays in the dynamics of these systems. Finally, nuclear bars in nested
systems have smaller ellipticities than their large-scale
counterparts.  

The search for stellar nuclear bars is limited to optical and NIR surface
photometry -- an insufficient method based on the isophote fitting which can be
affected by the presence of nuclear clusters, starbursts and dust. A suggestion
to look for the characteristic offset dust lanes delineating shocks in nuclear
bars, similar to those observed in large-scale bars, did not work. The reason
for this is a different gas-flow response in these systems (see below). Nuclear
bars are not scaled-down versions of large-scale bars. 

Probably the mostly intriguing aspect of nested bars is that their pattern
speeds are different, at least during some particular stages of evolution, as
predicted (Shlosman {\it et al.} 1989, 1990), supported by the detection of
random mutual orientations of bars (Friedli {\it et al.} 1996), and confirmed in
a direct measurement of their pattern speeds (Corsini {\it et al.} 2003). 

\subsubsection{Theoretical perspective: bars-in-bars mechanism}
\label{sec:theory}

From a theoretical point of view, nested bar systems provide a great laboratory
to study non-linear dynamics in physical systems. How do such systems form? Are
they long- or short-lived? What is their role in driving the secular evolution
of disk galaxies? As in the case of single bars, numerical simulations of such
systems are indispensable. 

The first attempts to simulate pure stellar nested bar systems have succeeded in
forming both bars via the bar instability, but the lifetime of such a system was
short (e.g., Friedli \& Martinet 1993). The problem with the lifetime has been
purely numerical -- when the resolution of $N$-body simulations has been
increased, the system of two stellar bars has lived indefinitely long (e.g.,
Pfenniger 2001). The issue of the initial conditions for such systems, however,
is much more fundamental. It is quite revealing that in pure stellar
systems one can create nested bars only by {\it assembling} the stellar disk as
bar-unstable on both large and small spatial scales -- naturally, because of the
shorter timescale of the inner bar, it will develop first. But why would Nature
create such a strongly bar-unstable stellar disk? No process known to us can
create such a disk by means of a dissipationless `fluid'. Therefore, strong
arguments appear in favour of a dissipational process which naturally involves
the gas (Shlosman 2005, and references therein). The initial conditions
necessary to create a nested bar system become nearly a trivial matter when
dissipation is involved.

The role of stellar bars in angular momentum redistribution is most important
when the gaseous component is involved. Bars are very efficient in extracting of
the angular momentum from the gas (and during the bar instability also from the
stars) and depositing it in the outer disk, beyond the corotation radius. The
ability of the DM halo to absorb the angular momentum has been noticed long ago
(e.g., Sellwood 1980; Debattista \& Sellwood 1998) and quantified in terms of
the lower resonances recently (Athanassoula 2002, 2003; Mart\'inez-Valpuesta
{\it et al.} 2006). 

The outward flow of the angular momentum in barred galaxies is associated with
the inward flow of the gas. Skipping the details, this gas accumulates within
the central kpc, where the gravitational torques from the large-scale bar are
minimised (e.g., Shlosman 2005). The star formation in the bar is largely
dampened because of the substantial shear behind the large-scale shocks. So one
should not be concerned with gas depletion before it crosses the ILR. Hence, the
action of the large bar leads to a radial inflow -- the rate of this inflow
depends on a number of factors. In the cosmological setting which is the subject
of this review, the outer bar will interact with and can in fact be triggered by
the asymmetry (e.g., triaxiality) of the DM halo (Heller {\it et al.} 2007a), or
by interaction with DM substructure (Romano-D\'iaz {\it et al.} 2008b). 

\begin{figure}[ht!!!]  
\begin{center}
\includegraphics[angle=0,scale=0.84]{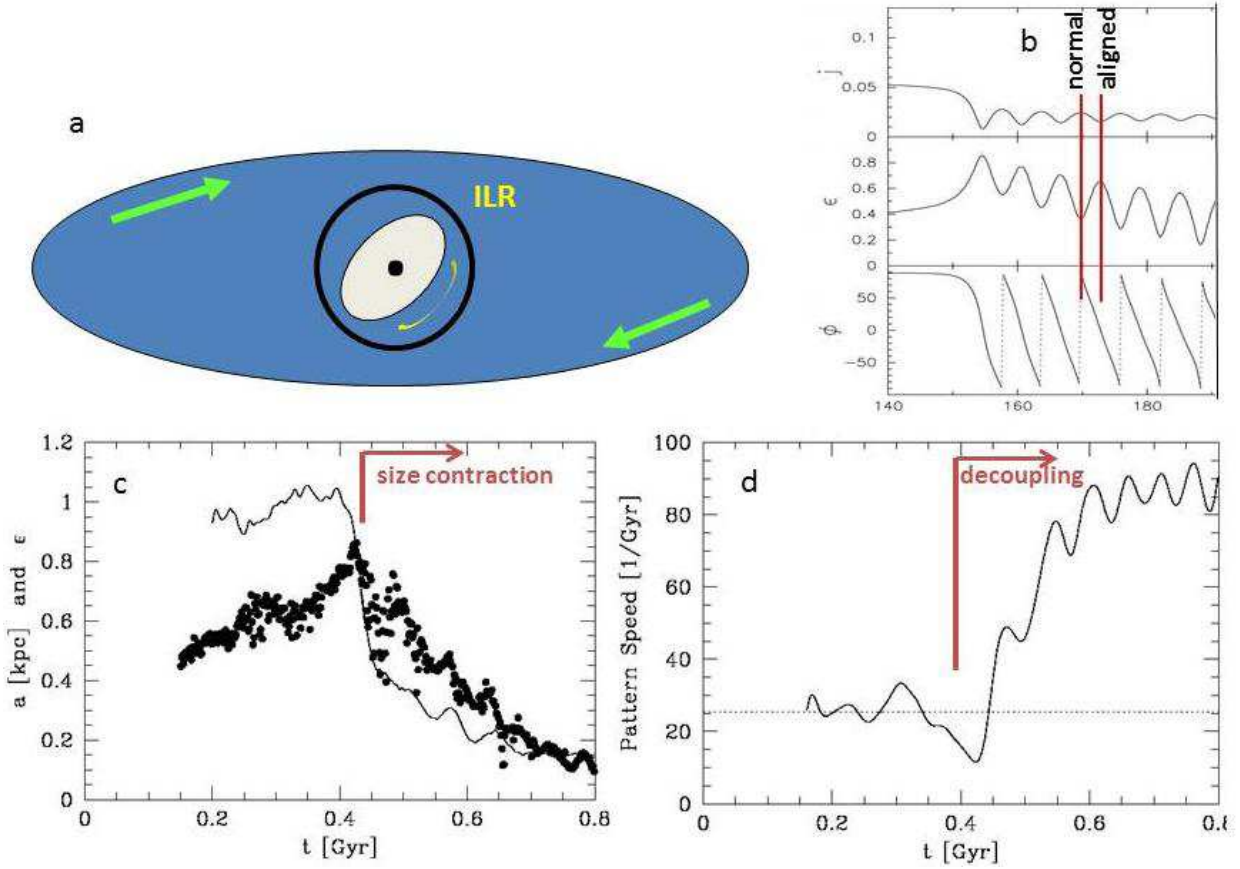}
\end{center}
\caption{\small Formation and evolution of nested bars with dissipation (Englmaier \&
Shlosman 2004): (a) Schematic structure of a nested bar system; (b) Specific
angular momentum,  ellipticity, and angle between the primary and secondary bars
(top to bottom). Note the correlation between the shape and the angle between
the bars; (c) Evolution of ellipticity (dotted line) and semimajor axis (thick
dots) of the secondary (nuclear) bar during decoupling; (d) Pattern speed of the
primary (dotted line) and secondary (solid line) bars during decoupling, in the
inertial frames. }
\label{fig:fig18}
\end{figure}

The accumulation of gas within the ILR can and probably does lead to a nuclear
starburst in nuclear rings. If, however, the gas inflow rate across the ILR is
high enough, or the star formation is dampened by a local turbulent field, the
stellar/gaseous fluid can trigger the formation of a nuclear bar (Shlosman {\it
et al.} 1989), when the gas becomes dynamically important and triggers an
additional responce from the stellar fluid, as confirmed by numerical
simulations (Friedli \& Martinet 1993; Englmaier \& Shlosman 2004). In this
respect, the large-scale bar is a primary and the nuclear bar is a secondary
feature (Fig.\,\ref{fig:fig18}a). This is the so-called bars-in-bars mechanism.

>From a dynamical point of view, it is challenging to explain how two
gravitational quadrupoles tumble with different pattern speeds, $\Omega_{\rm p}$
(primary) and $\Omega_{\rm s}$ (secondary), without exerting a braking effect on each other.
Such a system can serve as an astrophysical counterpart of a system of coupled
oscillators which is a familiar tool to study non-linear behaviour. Essentially,
three dynamical states exist for such a system, but only two of them appear
dynamically long-lived. The first state involves two bars with equal pattern
speeds -- this is a stable state and the bars stay perpendicular to each other,
which is energetically advantageous. The second state is made out of two bars
which tumble with different pattern speeds, but their ratio is fixed,
$\Omega_{\rm s}/\Omega_{\rm p}$\,$\sim$\,constant.  In a way, the first state is a
special case of the more general second one, and is also long-lived. The third
case consists of two bars which tumble with different pattern speeds, where
their ratio is time-dependent and evolves on a dynamical timescale -- this is a
transition state between two stable states with the fixed ratio of pattern
speeds. Naturally, this state is a short-lived one. We call the first two states
{\it coupled}\footnote{Note that various definitions of {\it coupled/decoupled}
states exist in the literature, and are refined occasionally by the same
authors, including in our own work.}, and the third one {\it decoupled}. 

The explanation for the coupled state is based on a non-linear {\it mode
coupling}: these modes exchange energies and angular momentum and, therefore,
support their pattern speeds which otherwise would decay exponentially due to
the gravitational torques. When the bars are locked in $\Omega_{\rm
s}/\Omega_{\rm p}$\,$\sim$\,constant, a local minimum should exist in the
efficiency of energy transfer between the bars. A strong resonance which can
`capture' the bar is necessary. One would expect the low resonances to play a
major role in the locking process, especially the ILR (e.g., Lichtenberg \&
Lieberman 1995; Tagger {\it et al.} 1987; Shlosman 2005).

Numerical simulations enable one to follow the decoupling process in nested bars
(Englmaier \& Shlosman 2004). Figures\,\ref{fig:fig18}c,d provide an example of
this process when a secondary bar, which forms within the ILR of the primary bar
and obeys $\Omega_{\rm s}/\Omega_{\rm p}$\,$\sim$\,1, speeds up  in a short
time, until its corotation radius moves to the ILR position of the primary bar,
and $\Omega_{\rm s}/\Omega_{\rm p}$\,$\rightarrow$\,3.6. The shape of the
nuclear bar depends on their mutual orientation (Fig.~\ref{fig:fig18}b), and is
dynamically important -- the bar axial ratio is one of the measures of its
strength and, therefore, determines the fraction of chaotic orbits there, which
is a measure of stochasticity within the bar and its possible demise. Another
example in the cosmological setting describes the evolution of a nested bar
system which is locked in two different coupled states and transits between them
(Heller {\it et al.} 2007a).

Different methods have been developed to quantify nested bar systems, especially
to measure the amount and the effect of multi-periodic and chaotic orbits. As we
are interested in the secular evolution of these systems, we only mention that
the orbital structure associated with long-lived nested bars has been
investigated. A counterpart of periodic orbits in single bars is based on the
fruitful concepts of a \textit{loop} (Maciejewski \& Sparke 2000) and on that of
the Liapunov exponents (El-Zant \& Shlosman 2003). Orbit analysis based on these
two concepts has demonstrated that orbits are dominated by the potentials of
single bars with the possibility of migration from bar to bar across the
interface between the bars (El-Zant \& Shlosman 2003; Maciejewski \&
Athanassoula 2007).

The gas dynamics at the bar-bar interface is determined by the already irregular
gas flow perturbed by the secondary bar with $\Omega_{\rm s}/\Omega_{\rm
p}$\,$>$\,1. For such bars, the inflow is chaotic and dominated by shocks.
Strong dissipation in the gas will not allow it to settle on stable orbits
there. Instead the gas will fall to smaller radii (e.g., Shlosman \& Heller
2002). The immediate corollary is that one should not expect starbursts
throughout secondary bars, except in the central regions. Shocks and shear
within the bar would slash molecular clouds reducing the SFRs.
In particular, the mode where star formation occurs in massive stellar clusters
should be absent in secondary bars, except (maybe) in the central region,
although even there it can be dampened if the gas can be maintained in a
strongly turbulent state.

The situation is very different for bars with $\Omega_{\rm s}/\Omega_{\rm p}$\,=\,1. 
In this case one should observe a relaxed flow, and a `grand-design' shock
system, but no random shocks. The dissipation is decreased compared to other
cases. 

\subsubsection{Nested bars: evolutionary corollaries}
\label{sec:corol}

Bars are known to channel the gas to the central kpc. Over a Hubble time, bars
are capable of affecting the angular momentum profile in the stellar component
as well. This process acts slowly but relentlessly in changing the mass
distribution in galaxies. Occasionally, during various instabilities, bars are
capable of increasing the central mass concentration in galaxies on short
dynamical timescales (e.g., Dubinski {\it et al.} 2009). Nested bars are the
result of this evolution when the amount of gas moved by the primary bar is able to
change the stellar dynamics inside the central kpc. Generally, when the gas has
reached $\sim$10\%  of the mass fraction in the central regions, it can affect
the stellar dynamics there. When a secondary bar forms, the local conditions for
star formation will be altered as well. The decrease in the SFR will eliminate
the ISM sink and facilitate further radial infall of gas. Further surveys of gas
kinematics in the central region should answer the question of whether these
flows fuel the AGN.

The fuelling of AGN is of course one of the outstanding issues in galaxy
evolution. Are they fuelled locally, say by a `neighbourhood' stellar cluster or
by the main body of the host galaxy? Diverging views prevail on this subject.
The duty cycle of AGN is not known at present. Does it depend on the class of
AGN, say QSOs versus Seyferts? If the AGN are fuelled by the bars-in-bars
mechanism, that characteristic timescale of the process can be as short as
$\sim$10$^6$\,yr, as gaseous bars are short-lived. What fraction of the gas
ends up in the accretion disk around the SMBH? 

What is clear is that preferring local mechanisms (e.g., star clusters) in
fuelling the AGN does not solve the issue, as it begs the question of what
fuelled the formation of local stars. Rather, one can argue that star formation
in the vicinity of the SMBH is a by-product of the overall gas inflow to the
galactic centre. Such an inflow will always be associated with compression and
star formation along the AGN fuelling process. The following scenario can
actually lead to an anti-correlation of gaseous bars with AGN activity: if the
gaseous bar activates the AGN cycle and as a by-product the local star
formation, the AGN and stellar feedback would disperse the gas in the next
stage.

What is the fate of gaseous bars? The central feedback can drive the local
azimuthal mixing of the `barred' gas component if the energy is deposited in the
turbulent motions in the gas, and ultimately contribute to the growth of the
stellar bulge, disky or classical. Strong winds driven by the nuclear starburst
can be a by-product. Alternatively, these winds can be driven via hydromagnetic
winds, as discussed earlier. Such extensive outflows from the centres of AGN
host galaxies have recently been detected. 

The study of nuclear bars, especially gaseous ones, will proceed quickly when
ALMA comes online. We have omitted interesting options which can in fact be
detected by upcoming observations of these objects. One of these is the
occasional injection of the gaseous component in the nuclear
stellar bars. How
does this influence the evolution of the system, and especially the gas inflow
toward the central SMBH?

\subsection{\textit{\textbf{The origin of SMBHs: the by-product of galaxy evolution?}}}
\label{sec:origin}

QSOs have been detected so far up to $z$\,$\sim$\,7, when the age of the
Universe was substantially less than 1\,Gyr. Even more intriguing is the
inferred mass of their SMBHs, $\gtorder 10^9\,M_\odot$. Within the framework of
hierarchical buildup of mass, these objects must originate in rare, highly
over-dense regions of the Universe. 

Indeed, recent studies of high-$z$ QSOs have indicated that they reside in very
rare over-dense regions at $z$\,$\sim$\,6 with a comoving space density of
$\sim$(2.2\,$\pm$\,0.73)$h^3\,{\rm Gpc^{-3}}$. Numerical simulations give a
similar comoving density of massive DM haloes, of a few\,$\times
10^{12}\,M_\odot$. The caveat, however, is that this similarity depends on the
QSO duty cycle -- the fraction of time the QSO is actually active (Romano-D\'iaz
{\it et al.} 2011a). For a duty cycle which is less than unity, QSOs will appear
less rare and reside in less massive haloes. On the other hand, these QSOs are
metal-rich and are therefore plausibly located at the centres of massive
galaxies. 

The causal connection between AGN and their host galaxies is a long-debated
issue, and substantial evidence has accumulated in favour of this relation.
SMBHs are ubiquitous. If the formation and growth of SMBHs is somehow correlated
with their host galaxy growth within the cosmological framework, what are the
possible constraints on the formation time of these objects? What are the
possible seeds of SMBHs?

\subsubsection{SMBH seeds: population III versus direct collapse}
\label{sec:versus}

Population III (Pop III) stars can form BH seeds of
$M_\bullet$\,$\sim$\,10$^2\,M_\odot$ (e.g., Bromm \& Loeb 2003).  An alternative
to this is the so-called `monolithic' (or direct) collapse to a SMBH (Begelman
{\it et al.} 2006). A gas with a primordial composition, which has a cooling
floor of $T_{\rm gas}$\,$\sim$\,10$^4$\,K, can collapse into DM haloes with a
virial temperature of $T_{\rm vir}$\,$\gtorder$\,$T_{\rm
gas}$\,$\sim$\,10$^4$\,K. This corresponds roughly to $M_{\rm
vir}$\,$\sim$\,10$^8\,M_\odot$. In a WMAP7 (\textit{Wilkinson Microwave Anisotropy
Probe}) universe, this can involve about $\sim$2\,$\times$\,10$^7\,M_\odot$
baryons. If the gas has been enriched previously by Pop III stars, the halo mass
can be smaller and the amount of baryons involved can be smaller as well. 

The Pop III SMBH seeds will be required to grow from $\sim$10$^2\,M_\odot$
to $\gtorder$\,10$^9\,M_\odot$ in less than $\sim$6\,$\times$\,10$^8$\,yr, i.e.,
from $z$\,$\sim$\,20 to $z$\,$\sim$\,7. The e-folding time for SMBH growth is
(Salpeter 1964):
\begin{equation}
t_{\rm e}\sim \epsilon \frac{c\sigma_{\rm T}}{4\pi Gm_{\rm p}}
    \left(\frac{L_{\rm E}}{L}\right) \sim 4.4\times 10^8\epsilon 
    \left(\frac{L_{\rm E}}{L}\right)\,{\rm yr},
\label{eq:timee}
\end{equation}
where $L_{\rm E}$ is the Eddington luminosity and $\epsilon$\,$\sim$\,0.1 is the
accretion efficiency. The growth time to the SMBH masses estimated for high-$z$
QSOs is
\begin{equation}
t_{\rm growth}\sim 3.1\times 10^9\epsilon 
    \left(\frac{L_{\rm E}}{L}\right)\,{\rm yr},
\label{eq:timegro}
\end{equation}
which is uncomfortably long and close to the age of the Universe. If the SMBH
growth is via mergers, this requires frequent merger events. When the merging
rate is too high, SMBHs can be ejected via slingshots, which should limit the
efficiency of this process. If the dominant mode is accretion, it will be
limited by $\sim L_{\rm E}$ and maybe by the feedback, which again would limit
the efficiency, or even cut off the accretion process.

Alternatively, monolithic (direct) collapse will require a growth from
$\sim$10$^{6-7}\,M_\odot$ to $\sim$10$^9\,M_\odot$, from $z$\,$\sim$\,12--20
to $z$\,$\sim$\,6--7. This will involve fewer mergers, but growth by accretion
may lead to fragmentation of the accretion flows, star formation and depletion
of the gas supply for the seed SMBH.

The collapse rates can be estimated from $\dot M$\,$\sim$\,$v^3/G$ (Shlosman \&
Begelman 1989), where $v$ is the characteristic infall velocity, which results
in mass accretion rates of $\sim$10$^{-4}$ to 10$^{-5}\,M_\odot\,{\rm
yr^{-1}}$ for a stellar collapse with virial temperatures $\sim$10$^{1-2}$\,K,
$\sim$10$^{-2}$ to 10$^{-4}\,M_\odot\,{\rm yr^{-1}}$ for a Pop III collapse
with $\sim$10$^{2-3}$\,K, and $\gtorder$\,10$^{-1}\,M_\odot\,{\rm yr^{-1}}$
for a direct collapse to the seed SMBH with $\gtorder$\,10$^{4}$\,K. 

However, a spherical collapse over decades in $r$ is improbable -- the angular
momentum barrier will stop it sufficiently quickly. Numerical simulations of a
baryon collapse into 10$^8\,M_\odot$ DM haloes have shown explicitly that the
angular momentum plays an important role (e.g., Wise {\it et al.} 2008), as
expected of course. The baryon collapse in the presence of a typical halo
angular momentum, $\lambda$\,$\sim$\,0.05, will develop virial turbulent
velocities, driven by the gravitational potential energy. The turbulence is
supersonic. The collapse can proceed until the angular momentum barrier stops
it. If the system reaches equilibrium, the turbulent velocities will decay and
the disk will fragment forming stars. However, the ability of the flow to
fragment depends on its equation of state, and should not be taken for granted
(e.g., Paczynski 1978; Shlosman {\it et al.} 1990). Begelman \& Shlosman (2009)
find that such a decay in the turbulent support will trigger a bar instability
in the gaseous disk, {\it before} it fragments. In other words, the global
instability sets in before the local instability develops. The bar instability
will create intrinsic shocks in the gas and the collapse will resume -- again
pumping energy into turbulent motions. This is the same bars-in-bars runaway
scenario as discussed in the previous section. 

The turbulent support is especially helpful in suppressing fragmentation in
$\sim$10$^8\,M_\odot$ DM haloes when the baryons are metal-rich. Previously,
it was suggested that the fragmentation in metal-rich flows in such haloes would
suppress the formation of a SMBH, and lead to star formation instead.

Supersonic turbulence requires continuous driving. A reliable diagnostic of this
flow is the density PDF, discussed in
Section~\ref{sec:feedback}. The log-normal PDF depends on the density
fluctuations around some average $\rho_0$, or $x$\,$\equiv$\,$\rho/\rho_0$ when
normalised (Fig.~\ref{fig:fig19}). The supersonic turbulence developing during
the overall collapse will generate gas clumps. Two spatial scales characterise
this turbulence -- the Jeans scale and the so-called transition scale, below
which the flow becomes supersonic, i.e., in essence the clump size.

\begin{figure}[ht!!!]  
\begin{center}
\includegraphics[angle=0,scale=0.84]{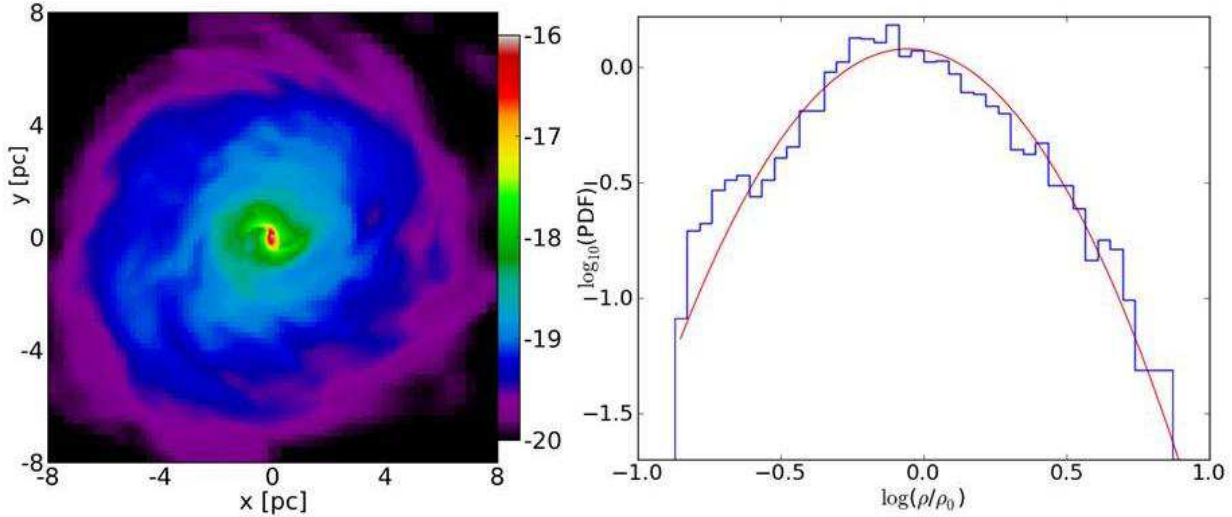}
\end{center}
\caption{\small Simulation of direct collapse to the SMBH (from Choi {\it et al.}
2012). Left: the projected gas density along the rotation axis at an
intermediate time of $t$\,$\sim$\,4.6\,Myr shows the central disk-like
configuration. The frame captures the central runaway collapse forming one of
the gaseous bars on a scale of $\sim$1\,pc. The collapse proceeds from the
initial scale of $\sim$3\,kpc -- the DM halo virial radius. The overall gas
density profile at this time is $\propto$\,$r^{-2}$. Right: a representative
log(PDF) of the gas density field in the left frame (blue histogram line). The
red solid line is the analytical counterpart of the log-normal PDF (see text).
The close match between the two distributions confirms the fully developed
turbulence field in the collapsing gas. }
\label{fig:fig19}
\end{figure}

A fraction of these clumps will be Jeans-unstable. However, only the fraction of
these clumps that contract on a timescale shorter than the free-fall time for
the overall collapse can actually contribute to the star formation. These
`active' clumps should have $x$\,$\gtorder$\,$x_{\rm crit}$\,$\sim$\,$(v_{\rm
turb}/v_{\rm K})^2 \mathcal{M}^2$, where $v_{\rm turb}$ is the typical turbulent
velocity, $v_{\rm K}$ is the Keplerian velocity and $\mathcal{M}$ is the Mach
number of the supersonic flow. Other clumps, even when Jeans-unstable, will be
swept away by the next crossing shock. To estimate the fraction of star-forming
clumps, one should integrate the PDF from $x_{\rm crit}$ to $\infty$.  For
$\mathcal{M}$\,$\gtorder$\,3, this fraction is $\ltorder$\,0.01. Hence in a flow
with a mildly supersonic turbulence the SFR is heavily dampened.

\subsubsection{Direct collapse: two alternatives}
\label{sec:alternate}

Present numerical simulations of the direct collapse to the SMBH cannot answer
the ultimate question about how and when the SMBH will form. At small spatial
scales new physical processes should become important, such as radiative
transfer, as the matter becomes opaque. This can be estimated to happen at
$\sim$10$^{13}\alpha\sigma_{10}^3(\kappa/\kappa_{\rm es})$\,cm, where
$\sigma_{10}$\,$\equiv$\,$\sigma/10\,{\rm km\,s^{-1}}$, and $\kappa_{\rm es}$ is
the electron scattering opacity. At what stage will the bars-in-bars cycle be
broken? What are the intermediate configurations that lead to the SMBH? 

The fate of the direct collapse has been analysed by Begelman {\it et al.}
(2006) under the assumption that the angular momentum is unimportant. Under
these conditions, there is no preferential channel for energy release. The
photon trapping in the collapsing matter results in the formation of a single
accreting massive object in a pressure equilibrium, termed a {\it quasistar}.
The follow-up evolution leads to thermonuclear reactions within the object for
$\sim$10$^6$\,yr. At some point, neutrino cooling becomes important, which
defines the quasistar core, $\sim$10\,$M_\odot$. Neutrino-cooled core collapse
will determine the seed SMBH mass. The subsequent rapid super-Eddington growth
of the seed will result in $M_\bullet$\,$\sim$\,10$^{4-6}\,M_\odot$.

What is the possible alternative to this scenario? Choi {\it et al.} (2012) have
assumed that the angular momentum dominates at some spatial scale of the inflow
and it does indeed define the preferential channel for energy release. Under
these conditions, photon trapping may not be important, and an optically-thick
disk/torus will form instead of a quasistar. This is a way to bypass the
thermonuclear reaction stage as well, but at the price of additional dynamical
instabilities. The SMBH seed which forms will be more massive than in the
previous case, $M_\bullet$\,$\sim$\,10$^{6-7}\,M_\odot$. 

%
%

\section{Summary and future prospects}
\label{sec:summary}
 
The cosmological evolution of galaxies is a fascinating subject which has
experienced explosive growth lately due to the incredible rate of new
observational data and the development of new methods and codes in this
observationally  and computationally intensive research field. In this review,
we have discussed various aspects of galaxy evolution. The original paradigm of
this evolution is being replaced slowly but persistently by a modified view -- a
process which is driven by a long list of recent discoveries. 

Probably, nowhere is this more obvious than in our understanding of
the main factor(s) behind galaxy growth. The cold accretion scenario has
successfully challenged the merger-only picture. The new approach raised a
number of challenging questions. How important is the accretion shock? What
fraction of the gas is actually processed by the shock? How does the accretion
flow join the growing disk? Via shocks or smoothly assembling in the outer disk?
These outstanding questions will be answered shortly via high-resolution
numerical simulations. On the other hand, observations must answer questions
about the redshift evolution of cold accretion flows. Furthermore, the issue of
the actual detection of flows in cosmological filaments is an open one and will
help to resolve at least partly the problem of the missing baryons. And of
course the increasing number of detected galaxies at redshifts corresponding to
the re-ionisation epoch must answer the fundamental questions about the
morphological types of the first galaxies, their mode of growth and the other
scaling relations established at low redshifts, e.g., the morphology-density
relation. 

The subject of stellar and AGN feedback on galaxy evolution is truly a Pandora's
 box. The inclusion of feedback has clearly solved the overcooling problem.
However, the current subgrid physics used to quantify SN feedback, and feedback
from OB stellar winds, AGN and galactic winds has too many parameters to be
fine-tuned. For AGN feedback, mechanisms more sophisticated than considering the
AGN as a giant O star are needed. Does it quench the star formation and clean
the galaxies of their ISM, or is this feedback much more anisotropic, producing
much less disturbance for the rest of the host galaxy. At high redshifts,
$z$\,$\sim$\,6--10, does the QSO feedback induce or dampen galaxy
formation? 

How do galactic winds form? They are clearly detected in observations, but the
underlying theory requires much more work. In fact, almost any model of the wind
from any astrophysical object experiences difficulties in describing the physics
of wind {\it initiation}.

There are too many bulgeless disks in the Universe. This comes as a surprise and
as another challenge. One approach attempted successfully has related this
phenomenon to the feedback problem. Indeed, back-of-the-envelope estimates show
convincingly that the SNe can expel the ISM and therefore eliminate the bulge
buildup in smaller galaxies. Additional work is required to understand the
efficiency of this process. Could it be too efficient? Massive disks provide
another challenge. In a number of cases cold gas has been detected in the
central regions, but no indication of intensive star formation, at least not in
the mode of massive clusters. How can one suppress the star formation while
leaving the gas intact?  One possibility lies in pumping the energy into
turbulent motions in the gas. Observations will resolve this issue soon. 

Where do stars form? This simple question has an interesting twist. Do stars
form as a result of Jeans instability in the molecular gas? Or does the
molecular gas itself form as a result of the Jeans instability in the neutral
gas? This issue has immediate consequences for numerical simulations of galaxy
evolution. Because we still cannot resolve the star-forming regions, what should
be used as a density threshold for star formation? 

The growing list of high-redshift galaxies presently extending to
$z$\,$\sim$\,10.4, and protoclusters extending to $z$\,$\sim$\,8, will soon
increase dramatically. We have already mentioned the importance of measuring the
scaling relations at these redshifts, but even more basic parameters, like the
galaxy LF, or the rate of star formation, must be measured as
well. Is the evolution of the {\it specific} SFR so flat even at
redshifts beyond 7? 

Different challenges are expected to resolve the kinematics of cold gas in
galactic centres. A high-resolution survey of gas morphology will probably be
produced by ALMA. What fraction of this gas is in a relaxed orbital motion, and
what fraction is in a `barred' state? What is the relation between the AGN host
galaxy and the central SMBH? We know that Seyfert activity is probably not
triggered by galaxy interactions or mergers. We also know that the brightest of
AGN, the QSOs, can be fuelled by this process. Overall, a number of factors can
fuel the accretion onto the SMBH, but it would be interesting to understand the
statistics of local versus non-local sources of fuelling. If both are involved,
where is the boundary between them?

The bulge buildup is partly a dissipative process, but it appears now that
stellar-dynamical instabilities play an important role as well. The relation
between the origin of peanut/boxy bulges and vertical buckling in stellar bars
has been now established. But this instability is {\it recurrent}, and what
determines the timescale of the onset of the next stage in this instability is
not yet clear. 

The fate of nuclear bars is really `lost' in the darkness. While purely stellar
nested bars can live indefinitely, the gas can still make a difference,
especially in gas-rich nuclear bars. What is their relation to the bulge
build-up? And to AGN fuelling? 

Finally, an outstanding issue is when and where the SMBHs form and how massive
their seeds are.  Are they formed at very high redshifts from Pop~III remnants,
or at lower redshifts in a direct collapse inside DM minihaloes of
$\sim$10$^8\,M_\odot$? The separate problems in galaxy evolution we have
discussed above join together to complicate this process. Does the accretion
flow fragment or does the induced turbulence dampen the Jeans instability and
does the seed SMBH form more massive? Is the angular momentum problem resolved
in this case by the bars-in-bars mechanism?  Observers can hope to detect the
quasistars -- the last stage before the horizon is formed. Or does the energy
escape along the preferred channel without interacting and stopping the inflow?
In this case, the evolution may bypass the quasistar stage. The detection of
intermediate-mass black holes can help here.

I would like to complete these lecture notes with my favourite question asked by
Stanislaw Lem in his novel {\it Ananke}: `Where is that order and whence came
this mocking illusion?'

%
%

\section*{Acknowledgments}
I am grateful to the organisers of the XXIII Winter School, Johan Knapen and
Jes\'us Falc\'on-Barroso, for their financial support and patience, and for
bringing together this highly enjoyable meeting. I thank Mitch Begelman,
Jun-Hwan Choi and Michele Trenti for insightful discussions on some of the
topics presented here. As this is not a full-fledged review, but encompasses a
broad range of topics on galaxy evolution, I apologise if some references have
been left out. My research is supported by NSF, NASA and STScI grants.

\end{document}